\documentclass[12pt]{article}
\usepackage{amsmath,amssymb,amsfonts,amsthm}
\usepackage[dvips]{graphicx} 
\usepackage{epic}

\title  {
        The scaling limit of the incipient infinite cluster \\ 
        in high-dimensional percolation. \\
	I. Critical exponents 
        }

\author {Takashi Hara\thanks{
        Department of Mathematics,
        Tokyo Institute of Technology,
        Oh-Okayama, Meguro-ku, Tokyo 152-8551,
        Japan,
        {\tt hara@ap.titech.ac.jp}.} 
        \; and 
        Gordon Slade\thanks{
        Department of Mathematics and Statistics, 
        McMaster University, 
        Hamilton, ON, Canada L8S 4K1,
        {\tt slade@math.mcmaster.ca}.
	Address after July 1, 1999:  Department of Mathematics, University
	of British Columbia, Vancouver, BC, Canada V6T 1Z2,
	{\tt slade@math.ubc.ca}.}
        }

\date{March 18, 1999}

\def\UseSection{%%
        \numberwithin{equation}{section}
	\theoremstyle{plain}% default theorem style 
        \newtheorem{theorem}    {Theorem}[section]
        \DefineTheorems % Use this to define other environments to be 
}

\def\DefineTheorems{%%
	\newtheorem{lemma}      [theorem] {Lemma}
	\newtheorem{prop}       [theorem] {Proposition}
	\newtheorem{cor}        [theorem] {Corollary}

	\theoremstyle{definition}% ``defn'' theorem style 
	\newtheorem{defn}       [theorem] {Definition}

	\newtheorem{example}       [theorem] {Example}
	\theoremstyle{definition}% ``remark'' theorem style 

}

\newcommand{\bt}   {\begin{theorem}}
\newcommand{\et}   {\end  {theorem}}
\newcommand{\bl}   {\begin{lemma}}
\newcommand{\el}   {\end  {lemma}}
\newcommand{\bp}   {\begin{prop}}
\newcommand{\ep}   {\end  {prop}}
\newcommand{\bc}   {\begin{cor}}
\newcommand{\ec}   {\end  {cor}}
\newcommand{\bd}   {\begin{defn}}
\newcommand{\ed}   {\end  {defn}}

\newcommand{\ba}   {\begin{array}}
\newcommand{\ea}   {\end  {array}}
\newcommand{\be}   {\begin{enumerate}}
\newcommand{\ee}   {\end  {enumerate}}
\newcommand{\bi}   {\begin{itemize}}
\newcommand{\ei}   {\end  {itemize}}

\def\eq#1\en{\begin{equation}#1\end{equation}}  
\def\eqsplit#1\ensplit{
	\begin{equation}\begin{split}#1\end{split}\end{equation}
	}
\def\eqalign#1\enalign{
	\begin{align}#1\end{align}
	}
\newcommand{\eqarrstar} {\begin{eqnarray*}} 
\newcommand{\enarrstar} {\end{eqnarray*}} 
\newcommand{\eqarray}   {\begin{eqnarray}} 
\newcommand{\enarray}   {\end{eqnarray}} 
\newcommand{\nnb}	{\nonumber \\} 

\newcommand{\lbeq}[1]  {\label{eq:#1}}
\newcommand{\refeq}[1] {\eqref{eq:#1}}    % AMS-LaTeX trick!
\newcommand{\Ebold} {{\mathbb E}}

\newcommand{\Rbold} {{\mathbb R}}

\newcommand{\Zbold} {{\mathbb Z}}

\newcommand{\Bcal}   {\mathcal{B}}

\newcommand{\Ecal}   {\mathcal{E}} 
\newcommand{\Fcal}   {\mathcal{F}} 
\newcommand{\Gcal}   {\mathcal{G}} 
\newcommand{\Hcal}   {\mathcal{H}}

\newcommand{\Mcal}   {\mathcal{M}} 
\newcommand{\Ncal}   {\mathcal{N}} 
 
\newcommand{\Pcal}   {\mathcal{P}}

\newcommand{\Rcal}   {\mathcal{R}}

\newcommand{\Ucal}   {\mathcal{U}} 
\newcommand{\Vcal}   {\mathcal{V}}

\newcommand{\Dhat} {{\hat{D} }}  
\newcommand{\Rhat} {{\hat{R} }} 
\newcommand{\tauhat}{{ \hat{\tau}  }}
  
\newcommand{\Rd}    {{ {\Rbold}^d}}
\newcommand{\Zd}    {{ {\Zbold}^d }}

\newcommand{\cupd} {\stackrel{\cdot}{\cup}}

\newcommand{\Ind}  { {\rm I} } 

\newcommand{\nin}  {{ \not\in }}

\newcommand{\expec}[1]	{\left \langle #1 \right \rangle} 

\UseSection  %Necessary to define Numbering scheme for Theorem, etc. 

\newcommand{\AND}       {\;\&\;}
\newcommand{\IN}        {\textnormal{ in }}
\newcommand{\INSIDE}    {\textnormal{ occurs in }}
\newcommand{\ON}        {\textnormal{ occurs on }}

\newcommand{\conn}      {\longleftrightarrow }
\newcommand{\nc}        { \conn  {\hspace{-3.0ex} /} \hspace{1.8ex}   }
\newcommand{\ct}[1]     { \stackrel{#1} {\conn} }
\newcommand{\dbc}{\Longleftrightarrow}

\renewcommand{\to}      {\rightarrow}

\newcommand{\prob}[1]   {  {  P ( #1 ) }  } 
 
\newcommand{\event}[1] {\left \{ #1 \right \} }

\newcommand{\Ctilde}    {\tilde{C}}

\newcommand{\Proof}	{\noindent \textbf{Proof}. \hspace{2mm}}

\newcommand{\oneOd}   {\lambda}
\newcommand{\pol}[1]  {{\sf P}^{(#1)}}
\newcommand{\wpol}[1] {{\sf W}^{(#1)}} 
\newcommand{\Phihat}  {{\hat{\Phi}}}
\newcommand{\ok}      {o_k(1)}
\newcommand{\oh}      {o_h(1)}
\newcommand{\ou}      {o_u(1)}
\newcommand{\okone}   {o_k(1)}
\newcommand{\ohone}   {o_h(1)}

\newcommand{\degIR}[1][\cdot]	{\underline{\textnormal{deg}}(#1)}

\newcommand{\setlengthmine}{\setlength{\unitlength}{0.4pt}} 
\setlengthmine
\newcommand{\raiseEqnAmount} {-18pt}
\newcommand{\myeqnpic}[3] {%
	\raisebox{\raiseEqnAmount}{\setlengthmine 
	\begin{picture}#1#2
	    #3
	\end{picture}
	} % end of raisebox
} % end of defn of \myeqnpic

\newcommand{\pictril}  % triangle on the left.  
	{\begin{picture}(0,0)%  %
	    \drawline(0,0)(-30, 30)(0,60)(0,0) %
	 \end{picture} %
	}
\newcommand{\pictrir}  % triangle on the right.  
	{\begin{picture}(0,0)%
	    \drawline(0,0)(30, 30)(0,60)(0,0) %
	 \end{picture}%
	}
\newcommand{\pictrim} % triangle in the middle.  (tri at bottom) 
	{\begin{picture}(0,0)%
	    \drawline(0,60)(0,30)(-15,0)(15,0)(0,30) %
	 \end{picture} %
	}
\newcommand{\pictrimu} % triangle in the middle.  (tri at top) 
	{\begin{picture}(0,0)%
	    \drawline(0,0)(0,30)(-15,60)(15,60)(0,30) %
	 \end{picture} %
	}
\newcommand{\picpivone}  % pivotal bond, #1
	{\begin{picture}(0,0) \thicklines% 
	    \multiput( -4, 6)(0.5, -1.0){13}{\circle*{2}}%
	    \multiput( -3, 6)(0.5, -1.0){13}{\circle*{2}}%
	    \multiput( -2, 6)(0.5, -1.0){13}{\circle*{2}}%
	    \multiput(  6, 6)(0.5, -1.0){13}{\circle*{2}}%
	    \multiput(  7, 6)(0.5, -1.0){13}{\circle*{2}}%
	    \multiput(  8, 6)(0.5, -1.0){13}{\circle*{2}}%
	 \end{picture}%
	}
\newcommand{\picpivtwo} % pivotal bond, #2
	{\begin{picture}(0,0) \thicklines%
	    \multiput( -4,-6)(0.5, 1.0){13}{\circle*{2}}%
	    \multiput( -3,-6)(0.5, 1.0){13}{\circle*{2}}%
	    \multiput( -2,-6)(0.5, 1.0){13}{\circle*{2}}%
	    \multiput(  6,-6)(0.5, 1.0){13}{\circle*{2}}%
	    \multiput(  7,-6)(0.5, 1.0){13}{\circle*{2}}%
	    \multiput(  8,-6)(0.5, 1.0){13}{\circle*{2}}%
	 \end{picture}%
	}

\newcommand{\picFonepNconn} %
	{\myeqnpic{(130,100)}{(-35,0)}{ %begin my-eqn-pic %%%%%
		\drawline( 0,80)(50,80)
		\put( 50,20){\pictrir} 
    		\put(-30,90){$v_{n-1}$}
    		\put( 40,90){$v_{n}'$}
    		\put(  0, 5){$w_{n-1}$}
    		\put( 85,45){$x$}
	 } % end of my-eqn-pic  %%%%%%%%%%%%%%%%%%%%%%%%%%%%%%%%
	}
	
\newcommand{\picFonepNbd} %
	{\myeqnpic{(130,100)}{(-35,0)}{ %begin my-eqn-pic %%%%%%
	    \thicklines
    		\drawline( 0,80)(50,80)
		\put( 50, 20){\pictrir} 
    		\put(-30,90){$v_{n-1}$}
    		\put( 25, 5){$w_{n-1}$}
    		\put( 85,45){$x$}
	 } % end of my-eqn-pic  %%%%%%%%%%%%%%%%%%%%%%%%%%%%%%%%
	}

\newcommand{\picFoneppwNMONEa} %
	{\myeqnpic{(200,100)}{(-105,0)}{ %begin my-eqn-pic %%%%%%
		\drawline(-50,80)(50,80)%%
		\put(   0, 20){\pictrim}%
    		\put(-100, 90){$v_{n-2}$}%
    		\put(  40, 90){$w_{n-1}$}%
    		\put( -60,  0){$w_{n-2}$}%
    		\put(  30,  0){$u_{n-1}$}%
	 } % end of my-eqn-pic  %%%%%%%%%%%%%%%%%%%%%%%%%%%%%%%%%
	}
\newcommand{\picFoneppwNMONEb} %
	{\myeqnpic{(240,100)}{(-55, 0)}{ %begin my-eqn-pic %%%%%
    		\drawline(  0,80)(150,80)
   		\drawline( 50,20)(100,20)
    		\drawline( 50,20)( 50,80) 
    		\drawline(100,20)(100,80) 
    		\put(-50,90){$v_{n-2}$}
    		\put(130,90){$w_{n-1}$}
    		\put(  0, 0){$w_{n-2}$}
    		\put(100, 0){$u_{n-1}$}
	 } % end of my-eqn-pic  %%%%%%%%%%%%%%%%%%%%%%%%%%%%%%%%
	}

\newcommand{\picFzeroppw} %
	{\myeqnpic{(130,100)}{(-55, 0)}{ %begin my-eqn-pic %%%%%
    		\drawline( 0,20)(50,20)
		\put(  0,20){\pictril} 
    		\put(-10,90){$u_{0}$}
    		\put( 50, 0){$w_{0}$}
    		\put(-50,40){$0$}
	 } % end of my-eqn-pic %%%%%%%%%%%%%%%%%%%%%%%%%%%%%%%%
	}

\newcommand{\picFoneppwGfreeNMONEa} %
	{\myeqnpic{(200,100)}{(-105,0)}{ %begin my-eqn-pic %%%%%%
		\thicklines
		\drawline(-50,80)( 0,80)
		\dottedline{6}(  0,80)(50,80)
		\put(   0, 20){\pictrim}
    		\put(-100, 90){$v_{n-2}$}
    		\put(  40, 90){$w_{n-1}$}
    		\put( -60,  0){$w_{n-2}$}
    		\put(  30,  0){$u_{n-1}$}
	 } % end of my-eqn-pic  %%%%%%%%%%%%%%%%%%%%%%%%%%%%%%%%%
	}
\newcommand{\picFoneppwGfreeNMONEb} %
	{\myeqnpic{(240,100)}{(-55, 0)}{ %begin my-eqn-pic %%%%%
    		\drawline(  0,80)(100,80)
		\dottedline{6}(100,80)(150,80)
   		\drawline( 50,20)(100,20)
    		\drawline( 50,20)( 50,80) 
    		\drawline(100,20)(100,80) 
    		\put(-50,90){$v_{n-2}$}
    		\put(130,90){$w_{n-1}$}
    		\put(  0, 0){$w_{n-2}$}
    		\put(100, 0){$u_{n-1}$}
	 } % end of my-eqn-pic  %%%%%%%%%%%%%%%%%%%%%%%%%%%%%%%%
	}

\begin{document}

\maketitle

%\begin{center}
%       {\large \bf IN PROGRESS: NOT FOR DISTRIBUTION}
%\end{center}

\begin{abstract}
This is the first of two papers on the critical behaviour
of bond percolation models in high dimensions.  In this paper,
we obtain strong joint control of the critical exponents
$\eta$ and $\delta$, for the nearest-neighbour model in very
high dimensions $d \gg 6$ and for sufficiently spread-out models
in all dimensions $d>6$.  The exponent
$\eta$ describes the low frequency behaviour of 
the Fourier transform of the critical two-point connectivity function,
while $\delta$ describes the behaviour of the magnetization at 
the critical point.  Our main result is an asymptotic relation
showing that, in a joint sense, $\eta = 0$ and $\delta = 2$.
The proof uses a
major extension of our earlier expansion method for percolation.
This result provides evidence that the scaling limit of the incipient
infinite cluster is the random probability measure on $\Rd$ 
known as integrated
super-Brownian excursion (ISE), in dimensions above $6$.
In the sequel to this paper, we extend our methods to
prove that the scaling limits
of the incipient infinite cluster's two-point and three-point functions 
are those 
of ISE for the nearest-neighbour model in dimensions $d \gg 6$.
\end{abstract}

\tableofcontents

%%%%%%%%%%%%%%%%%%%%%%%%%%%%%%%%%%%%%%%%%%%%%%%%%%%%%%%%%%%%%%%%%%%%%%%%%%%%%%%
%%%%%%%%%%%%%%%%%%%%%%%%%%%%%%%%%%%%%%%%%%%%%%%%%%%%%%%%%%%%%%%%%%%%%%%%%%%%%%%
%%%%%%%%%%%%%%%%%%%%%%%%%%%%%%%%%%%%%%%%%%%%%%%%%%%%%%%%%%%%%%%%%%%%%%%%%%%%%%%
%%%%%%%%%%%%%%%%%%%%%%%%%%%%%%%%%%%%%%%%%%%%%%%%%%%%%%%%%%%%%%%%%%%%%%%%%%%%%%%
\section{Introduction}

%%%%%%%%%%%%%%%%%%%%%%%%%%%%%%%%%%%%%%%%%%%%%%%%%%%%%%%%%%%%%%%%%%%%%%%%%%%%%%%
%%%%%%%%%%%%%%%%%%%%%%%%%%%%%%%%%%%%%%%%%%%%%%%%%%%%%%%%%%%%%%%%%%%%%%%%%%%%%%%
\subsection{Critical exponents}

We consider two models of independent bond percolation on $\Zd$.
For the nearest-neighbour model, a bond is a pair
$\{x,y\}$ of distinct sites in $\Zd$, separated by unit Euclidean distance.
For the spread-out model, a bond is a pair $\{x,y\}$ of distinct sites in
$\Zd$, with $0< \| x -y \|_\infty \leq L$.
We will consider the case of large, but finite, $L$.
In either model, we associate to each bond an independent 
Bernoulli random variable $n_{\{x,y\}}$ taking 
the value 1 with probability $p$, and the value
0 with probability $1-p$.  A bond $\{x,y\}$ 
is said to be \emph{occupied} if $n_{\{x,y\}}=1$, and \emph{vacant} if
$n_{\{x,y\}}=0$. We say that sites $u,v \in \Zd$ are \emph{connected},
denoted $x \conn y$, 
if there is a lattice path from $u$ to $v$ consisting of occupied bonds. 
If $x$ and $y$ are not connected, we write $x \nc y$.
 
For both the nearest-neighbour and spread-out models, 
a phase transition occurs for $d \geq 2$, 
in the sense that there is a critical value $p_c \in (0,1)$,
such that for $p<p_c$ there is with probability 1 no infinite connected
cluster of occupied bonds, whereas for $p>p_c$ there is with probability
1 a unique infinite connected cluster of occupied bonds (percolation occurs).
It is an unproven prediction of the hypothesis of universality
that, in any dimension $d$, the behaviours of the 
nearest-neighbour and spread-out
models (for any $L$) in the vicinity of the critical point
are identical in all important aspects.

Much of this important behaviour can be described in terms of
critical exponents.  At present, the existence of critical
exponents has been proved only in high dimensions, where
the critical behaviour is the same as that on a tree, using
the triangle condition.  Aizenman and Newman 
\cite{AN84} introduced the triangle condition as a sufficient
condition for the existence of the critical exponent $\gamma$
for the susceptibility (expected cluster size of the origin),
with the value $\gamma=1$.  Subsequently Barsky and Aizenman 
\cite{BA91} showed that the triangle condition implied existence
of the exponents $\delta$ for the magnetization and $\beta$ for
the percolation probability, with $\delta=2$ and $\beta=1$.  
Nguyen \cite{Nguy87} showed
that the triangle condition implied existence of the gap exponent
$\Delta$, with $\Delta = 2$.  
Implications of the triangle condition for differentiability of
the number of clusters per site were explored in \cite{YZ92}.
In the above results, existence of
critical exponents is in the form of upper and lower bounds with
different constants.  For example, for the susceptibility $\chi(p)$,
the consequence of the triangle condition is that $c_1(p_c-p)^{-1} \leq
\chi(p) \leq c_2(p_c-p)^{-1}$ for $p \in [0,p_c)$.
In \cite{HS90a}, an infra-red bound
was proved and used to show that the triangle
condition holds for the nearest-neighbour model in sufficiently high
dimensions and for the spread-out model for $d>6$ and $L$ sufficiently
large.  We subsequently showed that $d \geq 19$ is large enough for
the nearest-neighbour model \cite{HS94}.
Thus the above critical exponents are known to exist, and to take on
the corresponding values for a tree, in these contexts.
In addition, it was shown in \cite{Hara90} that the critical exponent
$\nu$ for the correlation length is equal to $\frac{1}{2}$,
in the sense of upper and lower bounds with different constants,
for the nearest-neighbour model in sufficiently high dimensions
and for sufficiently spread-out models for $d>6$.

In this paper, we extend some of the above results in two ways.
Firstly, we obtain power law \emph{asymptotic} 
behaviour of the Fourier transform
of the two-point function in the presence of a magnetic field,
for small values of the magnetic field and the frequency variable.
Secondly, this asymptotic behaviour is \emph{joint}, as a function
of two variables.  In addition to any intrinsic interest, 
this joint behaviour turns out to be relevant
for the identification of the scaling limit of the incipient infinite
cluster as integrated super-Brownian excursion, or ISE (see 
\cite{Aldo93,BCHS99,LeGa99b} for discussions of ISE).  
We will return
to this point below, and it will be the main topic of the sequel \cite{HS98c} 
to this paper, hereafter referred to as II.

Our method of proof involves a major extension of the expansion
for percolation introduced in 
\cite{HS90a}.  Moreover, a double expansion
will be used here.  
Our analysis is based in part on the corresponding analysis
for lattice trees, for which a double lace expansion was performed in
\cite{HS92c}, and 
for which a proof that the scaling limit is ISE in high dimensions
was given in \cite{DS97,DS98}.
We will also make use
of the infra-red bound proved in 
\cite{HS90a}, and of its consequence that,
for example, the triangle condition of 
\cite{AN84} holds in high dimensions.

The results obtained in this paper were announced in \cite{HS98a}.
A survey of the occurrence of ISE as a scaling limit for lattice trees
and percolation is given in \cite{Slad99}.

%%%%%%%%%%%%%%%%%%%%%%%%%%%%%%%%%%%%%%%%%%%%%%%%%%%%%%%%%%%%%%%%%%%%%%%%%%%%%%%
%%%%%%%%%%%%%%%%%%%%%%%%%%%%%%%%%%%%%%%%%%%%%%%%%%%%%%%%%%%%%%%%%%%%%%%%%%%%%%%
\subsection{The main result}
\label{sec-results}

Consider nearest-neighbour or spread-out independent bond percolation  
on $\Zd$, with bond density $p \in [0,1]$.  
Let $C(0)$ denote the random set of sites connected to $0$,
and let $|C(0)|$ denote its cardinality.  Let
\eq
	\tau_p(0,x; n) = P_p \left( C(0) \ni x , |C(0)|=n \right)
\en 
denote the probability  
that the origin is connected to $x$ by a cluster containing $n$ sites.
For $h \geq 0$, we define the generating function
\eq
\lbeq{Mzdef}
	\tau_{h,p}(0,x) = \sum_{n=1}^\infty \tau_p(0,x;n) e^{-hn}.
\en 
The generating function \refeq{Mzdef} converges for $h \geq 0$.

We will work with Fourier transforms, and for an absolutely summable function
$f$ on $\Zd$ define
\eq
	\hat{f}(k) = \sum_{x \in \Zd} f(x) e^{ik\cdot x}, \quad
	k = (k_1,\ldots,k_d) \in [-\pi,\pi]^d,
\en
with $k \cdot x = \sum_{j=1}^d k_jx_j$.
For $h>0$ and any $p \in [0,1]$, the Fourier transform $\hat{\tau}_{h,p}(k)$
exists since
\eq
\lbeq{tauhatexists}
	\sum_x \sum_{n=1}^\infty \tau_p(0,x;n) e^{-hn} = 
	\sum_{n=1}^\infty n P_p(|C(0)|=n) e^{-hn} 
	\leq \sum_{n=1}^\infty n e^{-hn} < \infty.
\en 
A similar estimate shows that the Fourier transform $\hat{\tau}_{h,p}(k)$
exists also for $h=0$ when $p<p_c$, using the fact that $P_p(|C(0)|=n)$
decays exponentially in the subcritical regime.  Our main object of
study will be $\hat{\tau}_{h,p_c}(k)$.

There is a convenient and well-known
probabilistic interpretation for the
generating function \refeq{Mzdef}, upon which we will rely heavily.
For this,
we introduce a probability distribution on the lattice sites by declaring 
a site to be ``green'' with probability $1-z$ and ``not green'' with
probability $z$.  These site variables are independent, and are independent
of the bond occupation variables.  
We use $G$ to denote the random set of green sites.
In this framework, $\tau_{h,p}(0,x)$ 
is the probability that the origin is connected
to $x$ by a cluster of any finite size, but containing no green sites, i.e.,
\eq
\lbeq{taugreen}
	\tau_{h,p}(0,x) = \sum_{n=1}^\infty P_p(0 \conn x, |C(0)|=n) e^{-hn}
	= P_p(0 \conn x, C(0) \cap G = \emptyset, |C(0)|<\infty ).
\en
Assuming there is no infinite cluster at $p_c$, 
$\tau_{0,p}(0,x)$ is the probability that $0$ is connected to $x$,
for any $p \leq p_c$.  It is a consequence of the results of
\cite{BA91,HS90a} that
there is no infinite cluster at $p_c$ for the high-dimensional systems
relevant in this paper.

For $h \geq 0$, $p \in [0,1]$, we define the magnetization
\eq
	M_{h,p} = P_p(C(0)\cap G \neq \emptyset) 
	= 1 - \sum_{n=1}^\infty P_p(|C(0)|=n) e^{-hn}
\en
and the susceptibility 
\eq
\lbeq{chidef}
	\chi_{h,p} =  \frac{\partial}{\partial h} M_{h,p}
	= \sum_{n=1}^\infty n P_p(|C(0)|=n) e^{-hn} 
	= \Ebold \left[ |C(0)| I[C(0) \cap G = \emptyset ] \right] 
	= \hat{\tau}_{h,p}(0) .
\en
Here $\Ebold$ denotes expectation and $I$ denotes an indicator function.

For $k  \in \Rd$, we write $k^2 = k \cdot k$
and $|k| = (k \cdot k)^{1/2}$.
The conventional definitions of the critical exponents $\eta$ and $\delta$ 
(see \cite[Section~7.1]{Grim89}) suggest that
\eq
\lbeq{2relns}
	\hat{\tau}_{0,p_c}(k) \sim \mbox{const.}\frac{1}{|k|^{2-\eta}}, 
	\;\; \mbox{as} \;
	k \to 0; \quad \quad
	\hat{\tau}_{h,p_c}(0) = \chi_h
	\sim \mbox{const.}\frac{1}{h^{1-1/\delta}}, \;\; 
	\mbox{as} \; h \downarrow 0.
\en 
We use `$\sim$' to denote an asymptotic formula, in which the ratio of
left and right sides tends to $1$ in the limit.  The above 
asymptotic relations
go beyond what has been proved previously, even in high dimensions.

The closest proven analogue of the first relation of
\refeq{2relns} is the infrared bound
\eq
\lbeq{irbd}
	0 \leq \hat{\tau}_{1,p}(k) \leq \frac{c}{k^2}
	\quad \quad (p \in [0,p_c), \; k \in [-\pi,\pi]^d),
\en
valid for sufficiently spread-out models for $d>6$ and for the
nearest-neighbour model for $d \geq 19$ \cite{HS90a,HS94}.
The constant $c$ in \refeq{irbd} is uniform in $p<p_c$ and $k \in [-\pi,\pi]^d$.
The triangle condition, which states that the triangle
diagram defined by
\eq
	\nabla(p) = \sum_{x,y} \tau_{0,p}(0,x)\tau_{0,p}(x,y)\tau_{0,p}(y,0)
	= \int_{[-\pi,\pi]^d} \hat{\tau}_{0,p}(k)^3 \frac{d^dk}{(2\pi)^d}
\en
is finite for $p=p_c$, is implied by \refeq{irbd}, if $d>6$.

For the second relation of \refeq{2relns}, Barsky and Aizenman (1991)
\nocite{BA91} proved that, under the triangle condition, $M_h$ is bounded
above and below by constant multiples of $h^{1/2}$.  This is weaker than
the second relation in two senses:  no \emph{asymptotic} bound was
obtained, and a relation for $\chi_h$ is a stronger statement about the
\emph{derivative} of $M_h$.

Using the mean-field values $\eta =0$ and $\delta = 2$
above six dimensions, the simplest possible combination of the relations
\refeq{2relns} for $d>6$ would be
\eq
\lbeq{tauE}
	\hat{\tau}_{h,p_c}(k) 
	= \frac{1}{C_1k^{2} + C_2h^{1/2}} + \mbox{error},
\en 
where $C_1$ and $C_2$ are constants and the error term is lower order
in some suitable sense in the limit $(k,h) \to (0,0)$.
\emph{A priori}, we cannot rule out the possibility of cross terms
such as $|k|h^{1/4}$, and some such cross terms could possibly occur for $d<6$.
The following theorem shows that the simple combination \refeq{tauE}
is what actually
occurs in high dimensions, and provides \emph{joint} control of the
\emph{asymptotic} behaviour in the limits $h \to 0$, $k \to 0$.
In its statement, we denote by $\okone$ a function  
of $k$ that goes 
to zero as $k$ approaches $0$.
Similarly, $\ohone$ denotes a function of $h$ that goes to zero as $h$
approaches $0$.  The factor $2^{3/2}$ in the statement of the theorem
is introduced to agree with our convention in II.

\begin{theorem}
\label{thm-tauasymp}
For nearest-neighbour bond percolation with $d$ sufficiently large,
and for spread-out bond percolation with 
$d>6$ and $L$ sufficiently large (depending on $d$), 
there are positive constants $C, D^2$, depending on $d,L$,
and a bounded function $\epsilon(h,k)$, such that for 
all $k \in [-\pi,\pi]^d$ and $h>0$,
\eq
\lbeq{tauasy}
	\hat{\tau}_{h,p_c}(k)
 	= \frac{C}{D^2 k^2 + 2^{3/2} h^{1/2} } \left[ 1 + \epsilon(h,k) \right],
\en 
with
\eq
	|\epsilon(h,k)| \leq  \okone + \ohone  
\en
as $h \to 0$ and/or $k \to 0$.  In addition, the limit
$\hat{\tau}_{0,p_c}(k) = \lim_{h \downarrow 0} \hat{\tau}_{h,p_c}(k)$ exists
and is finite for $k \neq 0$, and obeys 
\eq
\lbeq{tauirpc}
	\hat{\tau}_{0,p_c}(k) = \frac{C}{D^2 k^2 } \left[ 1 + \okone \right].
\en 
\end{theorem} 

Assuming that universality holds, Theorem~\ref{thm-tauasymp} would indicate 
that \refeq{tauasy} and \refeq{tauirpc} should actually 
be valid for the nearest-neighbour model for all $d>6$.
Setting $k=0$ in \refeq{tauasy} gives
\eq
	\chi_{h,p_c} = \hat{\tau}_{h,p_c}(0) =  h^{-1/2}
	\left[ 2^{-3/2}C + \ohone \right],
\en
which gives the second statement of \refeq{2relns}.
Consequently, since $M_{0,p_c}=0$,
\eq
\lbeq{Mhformula}
	M_{h,p_c} = \int_0^h \chi_{t,p_c} dt = h^{1/2}
	\left[ 2^{-1/2}C + \ohone \right],
\en
which is a statement that $\delta =2$, where in general 
it is expected that $M_{h,p_c} \sim \mbox{const.} h^{1/\delta}$.

Note that   
$\tau_{0,p_c}(0,x)$ is not summable if it decays like $|x|^{2-d}$, 
as expected for $d>6$.  Therefore its Fourier transform is not well-defined
without some interpretation.  We use the interpretation
$\hat{\tau}_{0,p_c}(k) = \lim_{h \downarrow 0} \hat{\tau}_{h,p_c}(k)$
because $\tau_{0,p_c}(0,x)$ is then the inverse
Fourier transform of $\hat{\tau}_{0,p_c}(k)$.  
In fact, using \refeq{tauasy} and the dominated convergence theorem
in the last step, we have
\eq
	\tau_{0,p_c}(0,x) = \lim_{h \downarrow 0}\tau_{h,p_c}(0,x)
	= \lim_{h \downarrow 0}
	\int_{[-\pi,\pi]^d}\hat{\tau}_{h,p_c}(k) 
	e^{-ik\cdot x} \frac{d^dk}{(2\pi)^d}
	=
	\int_{[-\pi,\pi]^d}\hat{\tau}_{0,p_c}(k)
	e^{-ik\cdot x}  \frac{d^dk}{(2\pi)^d}.
\en
Equation~\refeq{tauirpc} is a statement that $\eta = 0$.  It does not
immediately imply that $\tau_{0,p_c}(0,x)$ behaves like $|x|^{2-d}$ as
$x \to \infty$, but we intend to return to this matter in a future publication.

If we write $z=e^{-h}$, then the leading behaviour on the right
side of \refeq{tauasy} can be rewritten as 
$C(D^2k^2 + 2^{3/2}\sqrt{1-z})^{-1}$.
This generating function has been identified as a signal for the occurrence
of ISE as a scaling limit \cite{DS97,Slad99}, and this led us to
conjecture in \cite{HS98a} (see also \cite{DS97,Slad99}) that above the upper
critical dimension the scaling limit of the incipient infinite cluster
is ISE.  

The incipient infinite cluster is a concept admitting various
interpretations.  In \cite{CCD87,Kest86}, an incipient infinite cluster
is constructed in 2-dimensional percolation models as an infinite cluster at
the critical point.  Such constructions are necessarily singular with
respect to the original percolation model, which has no infinite cluster
at $p_c$.  Our point of view is to regard the incipient infinite cluster
as a cluster in $\Rd$ arising in the scaling limit.  More precisely,
we condition the size of the cluster of the origin to be $n$,
scale space by a multiple of $n^{-1/4}$, and examine the
cluster in the limit $n \to \infty$.  In II, we obtain strong evidence
that this scaling limit is ISE for $d>6$.  ISE can be regarded as the law
of a random probability measure on $\Rd$, but in addition it contains
more detailed information including the structure of all paths joining
pairs of points in the cluster.  This is consistent with the 
recent approach of \cite{Aize96,Aize97,Aize97a,AB98} to the scaling limit, 
although here our focus is on a single
percolation cluster, rather than on many clusters.  ISE is almost surely
supported on a compact subset of $\Rd$, but on the scale of the lattice,
this corresponds to an infinite cluster.  Thus we regard the limiting
object as the scaling limit of the incipient infinite cluster.

To relate the scaling limit of the incipient infinite cluster
to ISE, we will prove in II that for the nearest-neighbour model in 
sufficiently
high dimensions, \refeq{tauasy} can be promoted to a statement for
\emph{complex} $z=e^{-h}$ in the unit disk $|z|<1$, with uniform error 
estimates.  
Let
\eq
	\hat{A}(k) = \int_0^\infty t e^{-t^2/2} e^{-k^2t/2} dt
\en
denote the Fourier transform of the ISE two-point function  
(see \cite{Aldo93,BCHS99,DS97,LeGa93,Slad99}).
For the nearest-neighbour model in high dimensions, contour integration
can then be used to show that, as $n \to \infty$, 
\eq
	\hat{\tau}_{p_c}(kD^{-1}n^{-1/4};n) = C(8\pi n)^{-1/2}\hat{A}(k)
	[1+ O(n^{-\epsilon})],
\en 
for any $\epsilon \in (0, \frac{1}{2})$.
In particular, 
\eq
	P_{p_c}(|C(0)|=n) = \frac{1}{n}\hat{\tau}_{p_c}(0;n) 
	= C(8\pi)^{-1/2} n^{-3/2} [1+ O(n^{-\epsilon})],
\en
which is stronger than \refeq{Mhformula}.
Analogous results will be obtained
for the three-point function.  However, as we will explain in II,
for technical reasons we are unable to obtain these results for sufficiently
spread-out models in all dimensions $d>6$.

It has been argued since \cite{Toul74} 
that the upper critical
dimension of percolation is equal to $6$, i.e., that critical
exponents depend on the dimension for $d \leq 6$ but not for $d>6$.
Our proof provides an understanding of the critical dimension as
arising as $6=4+2$.   To explain this, we first introduce the
notion of \emph{backbone}.  Given a cluster containing $x$ and $y$,
the backbone joining $x$ to $y$ is defined to
consist of those sites $u \in C(x)$ for
which there are edge-disjoint paths 
consisting of occupied bonds from $x$ to $u$ and from $u$ to $y$.  
The backbone can be depicted as consisting of all connections between
$x$ and $y$, with all ``dangling ends'' removed.
An ISE cluster is 4-dimensional for $d \geq 4$
\cite{DP97,LeGa99a}, and distinct points in
the cluster are joined by a 2-dimensional Brownian path.
In our expansion,
the leading behaviour corresponds to neglecting intersections between
a backbone and a percolation cluster.  Considering the 
percolation cluster to scale
like an ISE cluster, intersections will generically not occur
above $4+2=6$ dimensions.  This points to $d=6$ as the upper critical
dimension.

%%%%%%%%%%%%%%%%%%%%%%%%%%%%%%%%%%%%%%%%%%%%%%%%%%%%%%%%%%%%%%%%%%%%%%%%%%%%%%%
%%%%%%%%%%%%%%%%%%%%%%%%%%%%%%%%%%%%%%%%%%%%%%%%%%%%%%%%%%%%%%%%%%%%%%%%%%%%%%%
\subsection{Organization}

This paper is organized as follows.  The proof of 
Theorem~\ref{thm-tauasymp} makes use of a double expansion.
The first expansion is described in Section~\ref{sec-expansion}.  
It is based on the expansion of 
\cite{HS90a}, but requires major adaptation
to deal with the presence of the 
magnetic field $h$.  Two versions of this expansion will
be presented in Section~\ref{sec-expansion}: a simpler version
which we call the ``one-$M$ scheme,'' and a more extensive expansion
which we call the ``two-$M$ scheme.''  The one-$M$ scheme is used
in Section~\ref{sec-part.exp.bd} to prove a weaker version of
Theorem~\ref{thm-tauasymp} that involves only upper and lower bounds.
The two-$M$ scheme is used to refine these bounds to an asymptotic
relation.  The $k^2$ term in \refeq{tauasy} is extracted in 
Section~\ref{sec-2M.exp.err}, where existence of the limit
$\lim_{h \downarrow 0}\hat{\tau}_{h,p_c}(k)$ is established  
and \refeq{tauirpc} is proved.  The more
difficult $h^{1/2}$ term involves a 
second expansion, derived in Section~\ref{sec-Psidef}, which
is used to complete the proof of Theorem~\ref{thm-tauasymp}.

Our method involves bounding terms in an expansion by Feynman diagrams.
To estimate these Feynman diagrams, we will at times employ the
method of power counting.  In Appendix~\ref{app}, we recall some
power counting results of Reisz \cite{Reis88b,Reis88a} that we will use.

This paper can be read independently of \cite{HS90a}, apart from the
fact that we will make use of the infrared bound and techniques
of diagrammatic estimation from \cite{HS90a}.

%%%%%%%%%%%%%%%%%%%%%%%%%%%%%%%%%%%%%%%%%%%%%%%%%%%%%%%%%%%%%%%%%%%%%%%%%%%%%%%
%%%%%%%%%%%%%%%%%%%%%%%%%%%%%%%%%%%%%%%%%%%%%%%%%%%%%%%%%%%%%%%%%%%%%%%%%%%%%%%
%%%%%%%%%%%%%%%%%%%%%%%%%%%%%%%%%%%%%%%%%%%%%%%%%%%%%%%%%%%%%%%%%%%%%%%%%%%%%%%
%%%%%%%%%%%%%%%%%%%%%%%%%%%%%%%%%%%%%%%%%%%%%%%%%%%%%%%%%%%%%%%%%%%%%%%%%%%%%%%
\section{The first expansion}
\label{sec-expansion}

Our method makes use of a double expansion.  In this section,
we derive the first of the two expansions, to finite order.  
We will derive two versions of the expansion in this section,
a ``one-$M$'' scheme and ``two-$M$'' scheme.  
For $p<p_c$ and $h=0$, these two expansions are the
same, and are essentially the expansion of \cite{HS90a}.   
Additional terms arise, however, for $h>0$.  Dealing with these
new terms poses new difficulties that must be overcome.  
The derivation of
the expansion applies equally well to the nearest-neighbour 
and spread-out models, and we treat the two cases simultaneously.

The derivation is
based on probabilistic arguments requiring 
$p\leq p_c$ and $h \geq 0$, which we henceforth assume.
We also assume henceforth that there is almost surely no infinite cluster
at the critical point, which is known to be the case under the assumptions  
of Theorem~\ref{thm-tauasymp} \cite{BA91,HS90a}.
We will first derive the expansions to 
finite order, and then prove that they can be extended  
to infinite order, for $h \geq 0$ when $p<p_c$, and for
$h>0$ when $p=p_c$.

Our starting point is
\refeq{taugreen}.  For $p \leq p_c$ and $h \geq 0$, \refeq{taugreen} reduces 
under the above assumptions to
\eq
	\tau_{h,p}(0,x) = \sum_{n=1}^\infty P_p(0 \conn x, |C(0)|=n) e^{-hn}
	= P_p(0 \conn x, C(0) \cap G = \emptyset).
\en
This is the quantity for which we want an expansion.  
Before beginning the derivation of the expansion, 
we first 
introduce some definitions and
prove a basic lemma that is at the heart of the expansion
method.

%%%%%%%%%%%%%%%%%%%%%%%%%%%%%%%%%%%%%%%%%%%%%%%%%%%%%%%%%%%%%%%%%%%%%%%%%%%%%%%
%%%%%%%%%%%%%%%%%%%%%%%%%%%%%%%%%%%%%%%%%%%%%%%%%%%%%%%%%%%%%%%%%%%%%%%%%%%%%%%
\subsection{Definitions and basic lemmas}
\label{sec-2ndexpdefs}

The following definitions will be used repeatedly throughout the paper.

\begin{defn}
\label{def-percterms}
(a)  
Define $\Omega = \{ x \in \Zd: \|x\|_1 =1 \}$ for the nearest-neighbour model
and $\Omega = \{ x \in \Zd : 0 < \|x\|_\infty \leq L \}$ for the
spread-out model.
A \emph{bond} is an unordered pair of distinct sites $\{x,y\}$ with
$y-x \in \Omega$.  A \emph{directed bond} is an ordered pair $(x,y)$ of 
distinct sites with $y-x \in \Omega$.  A \emph{path}
from $x$ to $y$ is a self-avoiding walk from $x$ to $y$, considered to be
a set of bonds.  Two paths are \emph{disjoint} if they have no bonds in
common (they may have common sites).  
Given a bond configuration, an \emph{occupied path} is a path
consisting of occupied bonds.
\newline
(b)  Given a bond configuration, two sites $x$ and $y$ are \emph{connected},
denoted $x \conn y$,
if there is an occupied path from $x$ to $y$ or if $x=y$. 
We write $x \nc y$ when it is not the case that $x \conn y$. 
We denote by $C(x)$ the random set of sites
which are connected to $x$.  Two sites $x$ and $y$ are 
\emph{doubly-connected}, denoted $x \dbc y$,
if there are at least two disjoint occupied paths from $x$ to $y$ or
if $x=y$.  
Given a bond $b = \{u,v\}$ and a bond configuration, we define 
$\tilde{C}^b(x)$ 
to be the set of sites which remain connected to $x$ in the new configuration
obtained by setting $n_b=0$.  Given a set of sites $A$, we say $x \conn A$
if $x \conn y$ for some $y \in A$, and we define 
$C(A) = \cup_{x \in A} C(x)$ and
$\tilde{C}^b(A) = \cup_{x \in A} \tilde{C}^b(x)$.
\newline
(c)  Given a set of sites $A \subset \Zd$ and a bond configuration, 
we say
$x \conn y$  \emph{in} $A$ if there is an occupied path from 
$x$ to $y$ having all of its sites in $A$ (so in particular 
$x,y \in A$), or if $x=y \in A$.
Two sites $x$ and $y$ are \emph{connected
through} $A$, denoted $x \ct{A} y$, if they are connected  
in such a way that every occupied path from $x$ to $y$ has at least one bond
with an endpoint in $A$, or if $x=y \in A$.
\newline
(d)  Recall that site variables were introduced above \refeq{taugreen}.
Given a bond/site configuration $\omega$ and a bond $b$, let $\omega^b$
be the configuration that agrees with $\omega$ everywhere except possibly
in the occupation status of $b$, which is occupied in $\omega^b$.
Similarly, $\omega_b$ is defined to be the configuration 
that agrees with $\omega$ everywhere except possibly
in the occupation status of $b$, which is vacant in $\omega_b$.
Given an event $E$ and a bond/site configuration $\omega$, 
a bond $b$ (occupied or not) is called
\emph{pivotal}
for $E$ if $\omega^b \in E$ and  
$\omega_b \nin E$.  We say that 
a directed bond $(u,v)$ is pivotal for the connection from $x$ to
$y$ if $x \in \tilde{C}^{\{u,v\}}(u)$, $y\in \tilde{C}^{\{u,v\}}(v)$ and 
$y \, \nin \, \tilde{C}^{\{u,v\}}(x)$.
If $x \conn A$ then there is a natural order to the set of
occupied pivotal bonds for the connection from $x$ to $A$ (assuming there
is at least one occupied pivotal bond), and each of these pivotal bonds is
directed in a natural way, as follows.  The \emph{first pivotal bond from}
$x$ \emph{to} $A$ is the directed occupied pivotal bond $(u,v)$ such that
$u$ is doubly-connected to $x$.  If $(u,v)$ is the first pivotal bond
for the connection from $x$ to $A$, then the second pivotal bond is the
first pivotal bond for the connection from $v$ to $A$, and so on.
\newline
(e)  
We say that an event $E$ is {\em increasing}\/ if, given a bond/site
configuration $\omega \in E$, and a configuration $\omega'$ having the
same site configuration as $\omega$ and for which each occupied bond in
$\omega$ is also occupied in $\omega'$, then $\omega' \in E$.
\end{defn}

\begin{defn}
\label{def-event-on}
(a)
Given a set of sites $S$, we refer to bonds with both endpoints in $S$
as \emph{bonds in $S$}.  
A bond having at least one endpoint in $S$ 
is referred to as a \emph{bond touching $S$}.  
We say that a site $x \in S$ is \emph{in $S$} or \emph{touching $S$}.
We denote by $S_{I}$ the set of bonds and sites in $S$.
We denote by $S_{T}$ the set of bonds and sites touching $S$.
\newline
(b)
Given a bond/site configuration $\omega$ and a set of sites $S$,
we denote by $\omega|_{S_I}$ the bond/site configuration which agrees
with $\omega$ for all bonds and sites in $S$, and which has all
other bonds vacant and all other sites non-green.  
Similarly,
we denote by $\omega|_{S_T}$ the bond/site configuration which agrees
with $\omega$ for all bonds and sites touching $S$, and which has all
other bonds vacant and all other sites non-green.
Given an event $E$ and a deterministic set of sites $S$, the event
$\{E$ \emph{occurs in} $S\}$ is defined to 
consist of those configurations $\omega$ for which
$\omega|_{S_I} \in E$.
Similarly, we define the event $\{E$ \emph{occurs on} $S\}$ 
to consist of those configurations $\omega$ for which
$\omega|_{S_T} \in E$.
Thus we distinguish between ``occurs on'' and ``occurs in.''
\newline 
(c) 
The above definitions will now be extended to certain random sets 
of sites.  Suppose $A \subset \Zd$.
For $S=C(A)$ or $S=\Zd \backslash C(A)$, 
we have $\omega |_{S_T} = \omega |_{S_I}$,
since bonds touching but not in $C(A)$ are automatically vacant.  
For such an $S$, we therefore
define $\{ E$ occurs on $S\} = \{ E$ occurs in $S\} = \{ \omega :
\omega |_{S_T} \in E\}$.
For $S= \tilde{C}^{\{u,v\}}(A)$ (see Definition~\ref{def-percterms}(b))
or $S= \Zd \backslash \tilde{C}^{\{u,v\}}(A)$, 
we define $\tilde{S}_T = S_T \backslash \{u,v\}$
and $\tilde{S}_I = S_I \backslash \{u,v\}$, and denote by
$\omega|_{\tilde{S}_T}$ and $\omega |_{\tilde{S}_I}$ 
the configurations obtained by setting $\{u,v\}$ vacant in 
$\omega|_{S_T}$ and $\omega |_{S_I}$ respectively.
Then $\omega |_{\tilde{S}_T} = \omega |_{\tilde{S}_I}$ for these two
choices of $S$, and we define 
$\{ E$ occurs on $S\} 
= \{ E$ occurs in $S\} = \{ \omega : \omega |_{\tilde{S}_T} \in E\}$.
\end{defn}

The above definition of $\{E$ occurs on $S\}$ is intended to capture
the notion that if we restrict attention to the status of
bonds and sites touching $S$, then $E$ is seen to occur.
A kind of asymmetry has been introduced, intentionally, by our
setting bonds and sites not touching $S$ to be respectively vacant
and non-green, as a kind of ``default'' status.  Some
examples are: 
(1) $\{v \conn x$ occurs in $S\}$,
for which Definitions~\ref{def-percterms}(c) and \ref{def-event-on}(b) agree,
(2) $\{ x \conn G$ occurs on $S\} = \{ x \in S\} \cap
\{C(x) \cap S \cap G \neq \emptyset\}$,
and (3) $\{ x \nc G$ occurs on $S\} = \{ x \nin S\} \cup
\{C(x) \cap S \cap G = \emptyset\}$.  

The following lemma is an immediate consequence of
Definition~\ref{def-event-on}, and shows that the notions of ``occurs on''
and ``occurs in'' preserve the basic operations of set theory.
The first statement of the lemma is illustrated by examples (2) and (3) above.

\begin{lemma}
\label{lem-occurson}
For any events $E,F$ and for random or deterministic sets $S,T$ of sites,
\eqalign
    \{E \textnormal{ occurs on }S\}^c 
    	& =  \{E^c \textnormal{ occurs on } S\}, 
	\nnb 
    \{(E \cup F) \textnormal{ occurs on } S\} 
    	& =  \{E \textnormal{ occurs on } S\} 
	\cup \{F \textnormal{ occurs on } S\}, 
	\nnb 
    \{\{E \textnormal{ occurs on } S\} \textnormal{ occurs on } T\} 
	& =  \{E \textnormal{ occurs on } S\cap T\}. 
	\nonumber 
\enalign
The corresponding identities with ``occurs in'' are also valid.
\end{lemma}

We are now able to prove our basic factorization lemma.
An erroneous lemma of this sort was given in \cite[Lemma~2.1]{HS90a}. 
Corrected versions
can be found in \cite{HS94} or \cite{MS93}.
We use angular brackets to denote the joint expectation
with respect to the bond and site variables.

\begin{lemma}
\label{lem-cond.0}
Let $p \leq p_c$.  For $p=p_c$, assume there is no infinite cluster.
Given a bond $\{u, v\}$, a finite set of sites $A$, and events $E$, $F$,
we have
\eqarray 
        && \Bigl \langle \Ind 
        \left[ E \textnormal{ occurs on } \tilde{C}^{\{u, v\}}(A) \AND 
        F \textnormal{ occurs in } \Zd \backslash \tilde{C}^{\{u, v\}}(A)
        \AND \{u,v\} \textnormal{ occupied} \right] 
                \Bigr \rangle 
        \nnb
        && 
\lbeq{cond.0lem}
	\qquad = p
        \Bigl \langle \Ind [ 
        E \textnormal{ occurs on } \tilde{C}^{\{u, v\}}(A) ]  
        \langle 
        \Ind [ F \textnormal{ occurs in } 
        \Zd \backslash \tilde{C}^{\{u, v\}}(A) ] 
        \rangle 
        \Bigr \rangle ,
\enarray 
where, in the second line, $\tilde{C}^{\{u, v\}}(A)$
is a random set associated with the outer expectation. 
In addition, the analogue of \refeq{cond.0lem}, in which ``$\{u,v\}$ occupied''
is removed from the left side and ``$p$'' is removed from the right side,
also holds.
\end{lemma}

\Proof
The proof is by conditioning on
the \emph{bond} cluster of $A$ which remains after
setting $n_{\{u,v\}}=0$, which we denote $\tilde{C}^{\{u,v\}}(A)_b$.
This cluster is finite with probability 1.
We do not condition on the status of the sites in this bond cluster.
Let $\Bcal$ denote the set of all finite bond 
clusters of $A$.  Given $B \in \Bcal$, we denote the set of 
\emph{sites} in $B$ by $B_s$.  
Conditioning on $\tilde{C}^{\{u,v\}}(A)_b$, we have 
\eqarray
        &&\Bigl \langle \Ind \left[ E  \ON  
                        \tilde{C}^{\{u, v\}}(A) \AND 
                F \INSIDE \Zd \backslash \tilde{C}^{\{u, v\}}(A) 
        \AND \{u,v\} \textnormal{ occupied} \right] 
                \Bigr \rangle 
        \nnb
\lbeq{condonB}
        && \quad = \sum_{B \in \Bcal}
        \Bigl \langle \Ind \left[\tilde{C}^{\{u,v\}}(A)_b = B 
                \AND E \ON \tilde{B}_{s}  \AND 
                F \INSIDE \Zd \backslash \tilde{B}_{s} 
        \AND \{u,v\} \textnormal{ occupied} \right] 
                \Bigr \rangle ,  \hspace{1cm}
\enarray
where $\tilde{B}_s$ signifies the 
vacancy of $\{u,v\}$, as
described in Definition~\ref{def-event-on}(c).

Since the first two 
of the four events on the right side of \refeq{condonB}
depend only on bonds/sites touching $B_s$
(according to Definition~\ref{def-event-on}(c), excluding $\{u,v\}$),
while the third event depends only on bonds/sites which do not touch $B_s$
(again, excluding $\{u,v\}$), and the fourth event depends only on $\{u,v\}$,
this independence allows us to write \refeq{condonB} as
\eqarray
        && p
        \sum_{B \in \Bcal}
        \Bigl \langle \Ind \left[\tilde{C}^{\{u,v\}}(A)_b = B 
        \AND E\ON \tilde{B}_{s} \right]
        \Bigr \rangle \; 
        \Bigl \langle 
        \Ind \left[ F \INSIDE \Zd \backslash \tilde{B}_s \right] 
        \Bigr \rangle
        \nnb 
        && = \quad p
        \Bigl \langle \Ind 
        [ E \ON \tilde{C}^{\{u, v\}}(A)  ] 
        \langle 
        \Ind [ 
        F \INSIDE \Zd \backslash \tilde{C}^{\{u,v\}}(A)
        ] 
        \rangle 
        \Bigr \rangle .
\enarray
The random set $\tilde{C}^{\{u,v\}}(A)$ in the inner expectation corresponds
to the outer expectation.
This completes the proof of \refeq{cond.0lem}.  The analogue stated in
the lemma holds by the same proof. 
\qed

In Sections~\ref{sec-exp.rem} and \ref{subsub-exp.no-rem}, we will  
apply Lemma~\ref{lem-cond.0} several times.  Further applications
will occur in Section~\ref{sec-Psidef}.  As an example of a situation
in which an event of the type appearing on the left side of \refeq{cond.0lem}
arises, we have the following lemma.

\begin{lemma}
\label{lem-pivotal2}
Given a deterministic set $A \subset \Zd$, a directed bond $(a',a)$, 
and a site $y \nin A$, the event $E$ defined by
\eq
	E= \{(a',a) \textnormal{  is a pivotal bond for } y \to A \} 
\en
is equal to the event $F$ defined by
\eq
	F= \bigl \{ a \conn A \ON \tilde{C}^{\{a, a'\}}(A) 
        \AND 
        y \conn a' \INSIDE \Zd\backslash \tilde{C}^{\{a, a'\}}(A) 
	\bigr \} .
\en
\end{lemma}

\Proof
First we show that $E \subset F$.  Suppose $E$ occurs, so
we have a configuration
for which $(a',a)$ is pivotal for the connection from $y$ to $A$.
Then $a \in \tilde{C}^{\{a',a\}}(A)$ and hence $a \conn A$ occurs on
$\tilde{C}^{\{a',a\}}(A)$.  Also, $y \in \tilde{C}^{\{a',a\}}(a')$,
and hence $y \conn a'$ occurs in $\tilde{C}^{\{a',a\}}(a')$.  But since
$(a',a)$ is pivotal, $\tilde{C}^{\{a',a\}}(a') \subset \Zd \backslash
\tilde{C}^{\{a',a\}}(A)$ and hence $y \conn a'$ occurs in 
$\Zd \backslash \tilde{C}^{\{a',a\}}(A)$.  Thus $F$ occurs.

Now we show that $F \subset E$.  Suppose $F$ occurs.
It suffices to show that (1) $y \conn A$ when $(a',a)$
is occupied, and (2) $y \nc A$ and $y \conn a'$ when $(a',a)$ is vacant.  
We see this as follows.
(1) If $(a',a)$ is
occupied, then it is clear from the definition of $F$ that $y \conn A$.
(2) If $(a',a)$ is vacant, then $\tilde{C}^{\{a',a\}}(A) =C(A)$.
Since $y \conn a'$ in $\Zd \backslash \tilde{C}^{\{a',a\}}(A)$, 
we have $y \conn a'$.  Also, it 
follows that $y \nin \tilde{C}^{\{a',a\}}(A)$.  Thus $y \nin C(A)$.
\qed

%%%%%%%%%%%%%%%%%%%%%%%%%%%%%%%%%%%%%%%%%%%%%%%%%%%%%%%%%%%%%%%%%%%%%%%%%%%%%%%
%%%%%%%%%%%%%%%%%%%%%%%%%%%%%%%%%%%%%%%%%%%%%%%%%%%%%%%%%%%%%%%%%%%%%%%%%%%%%%%
\subsection{The first expansion: one-$M$ scheme} 
\label{sec-exp.rem}

In this section, we generate an expansion that will be used
to prove upper and lower bounds on the two-point function, 
as an initial step in the proof of Theorem~\ref{thm-tauasymp}.  
The expansion will produce a convolution 
equation for $\tau_{h,p}(0,x)$, for $h,p$ such that $h \geq 0$ and $p < p_c$
or $h >0$ and $p=p_c$.
We refer to this expansion as the one-$M$ scheme, because remainder
terms in the expansion will be bounded 
in Section~\ref{sec-part.exp.bd} using a single factor of the
magnetization $M_{h,p}$.

The starting point for the expansion is to regard a cluster contributing
to $\tau_{h,p}(0,x) = P(0 \conn x, 0 \nc G)$ as a string of sausages
joining $0$ to $x$ and not connected to $G$.  In this picture,
the ``string'' corresponds to the pivotal bonds for the connection
from $0$ to $x$, and the sausages are the connected components of 
$C(0)$ that remain when these pivotal bonds are made vacant.  
Suppose the pivotal bonds for the connection from $0$ to $x$ are given, in
order, by $(u_i,v_i)$, $i=1,\ldots, n$.  Let $v_0=0$ and $u_{n+1}=x$.
Then the $j^{\rm th}$ sausage is defined to be the connected cluster
of $v_{j-1}$ after setting $\{u_{j-1},v_{j-1}\}$ and $\{u_{j},v_{j}\}$ vacant
($j=1,\ldots, n+1$), omitting reference to the undefined bonds
$\{u_0,v_0\}$ and $\{u_{n+1},v_{n+1}\}$ when $j=1$ or $j=n+1$.
By definition, the $j^{\rm th}$ sausage is 
doubly connected between $v_{j-1}$
and $u_j$, which we refer to respectively as the
{\em left}\/ and {\em right endpoints}\/ of the $j^{\rm th}$ sausage.
We regard the
sausages as interacting with each other, in the sense that they
cannot intersect each other.  In high dimensions, the interaction
should be weak, and our goal is to make an approximation in which
the sausages are treated as independent.  The approximation will
introduce error terms which are represented as higher order terms in
the expansion, and these can be controlled in high dimensions. 

We begin by defining some events.  
Given a bond $\{ u', v' \}$, let 
\eqalign
        E_0(0,x) & =  \left\{ 0 \conn x \AND 0 \nc G \right\} ,
        \\
        E_0'(0, x) & =  
        \left \{ 0 \dbc x \AND 0 \nc G \right \} ,
        \\
        E_0''(0, u', v') & =  E_0'(0, u') \ON \Ctilde^{\{u', v'\}}(0)  ,
        \\
        E_0(0,u',v',x) & =  E_0'(0, u') \cap \left \{  
        (u',v') \mbox{ is occupied and pivotal for } 0 \conn x \right \} .
\enalign
Given a set of sites $A \subset \Zd$, we also define
\eq
        \tau^A_{h,p}(0,x) = \langle I[(0 \conn x \AND 0 \nc G) \INSIDE
        \Zd \backslash A] \rangle.
\en 

The first step in the expansion is to write
\eq
        \tau_{h,p}(0,x) = \langle I[E_0(0, x)] \rangle
        = \langle I[E_0'(0, x)] \rangle
        + \sum_{(u_0,v_0)} \langle I[E_0(0, u_0, v_0, x)] \rangle,
\en 
where the sum is over directed bonds $(u_0,v_0)$.
We now wish to apply Lemma~\ref{lem-cond.0} to factor the expectation
in the last term on the right side.  Arguing as in the proof of 
Lemma~\ref{lem-pivotal2}, $E_0(0, u_0,v_0,x)$ can be written as
the intersection of the events that
$E'_{0}(0, u_0)$ occurs on $\Ctilde^{\{u_0, v_0\}}(0)$, 
that $\{u_0,v_0\}$ is occupied, and 
that $(v_0 \conn x \AND v_0 \nc G) \INSIDE
\Zd \backslash \Ctilde^{\{u_0, v_0\}}(0)$. Applying
Lemma~\ref{lem-cond.0} then gives
\eq
        \langle I[E_0(0, u_0, v_0, x)] \rangle =
        p \langle I[E_0''(0, u_0, v_0)] 
        \tau^{\tilde{C}^{\{u_0,v_0\}}(0)}_{h,p}(v_0,x) \rangle .
\en 
Therefore,
\eq
\lbeq{expan0}
        \tau_{h,p}(0,x) = \langle I[E_0'(0, x)] \rangle
        + p \sum_{(u_0,v_0)} \langle I[E_0''(0, u_0, v_0)] 
        \tau^{\tilde{C}^{\{u_0,v_0\}}(0)}_{h,p}(v_0,x) \rangle.
\en 

Before proceeding with the expansion, we give a brief perspective on
where we are heading.
To leading order, we would like to replace 
$\tau^{\tilde{C}^{\{u_0,v_0\}}(0)}_{h,p}(v_0,x)$ by $\tau_{h,p}(v_0,x)$, which
would produce a simple convolution equation for $\tau_{h,p}$ 
and would effectively treat the first sausage in the cluster joining
$0$ to $x$ as independent of the other sausages.
Such a replacement
should create a small error provided the backbone 
(see Section~\ref{sec-results}) joining
$v_0$ to $x$ typically does not intersect the cluster 
$\tilde{C}^{\{u_0,v_0\}}(0)$.  Above the upper critical dimension, where
we expect the backbone to have the character of Brownian motion and the
cluster $\tilde{C}^{\{u_0,v_0\}}(0)$ to have the character of an ISE
cluster, this lack of intersection demands the mutual avoidance of a
2-dimensional backbone and a 4-dimensional cluster.  This is a weak
demand when $d>6$, and this leads to the interpretation of the
critical dimension 6 as $4+2$.  As was pointed out in \cite{AN84},
and as we will show in Section~\ref{sec-part.exp.bd}, 
bounding errors in the above replacement leads
to the triangle diagram, whose convergence at the critical point is also
naturally associated with $d>6$.  When $h=0$, the diagrams that emerge
in estimating the expansion can be bounded in terms of the triangle diagram,
as was done in \cite{HS90a}, but for $h \neq 0$ other diagrams,
including the square, will also arise.  However,
square diagrams arise only in conjunction with factors of the
magnetization $M_{h,p} = P(0 \conn G)$ that vanish in the limit $h \to 0$ 
more rapidly than the divergence of the square diagram as a function of $h$.
These terms therefore make no contribution in the limit.

We now return to the derivation of the expansion.
Let $A$ be a set of sites.
To effect the replacement described in the previous paragraph, we write
\eq
\lbeq{tautauA}
        \tau^{A}_{h,p}(v,x)  
	=  \tau_{h,p}(v,x) - [ \tau_{h,p}(v,x) - \tau^{A}_{h,p}(v,x) ]     
\en 
and proceed to derive an expression for the difference in square brackets
on the right side.  
Recall the notation $v \ct{A} x$ of Definition~\ref{def-percterms}(c).  
Similarly, we denote by $v \ct{A} G$
the event that every occupied path from $v$ to any green site must contain
a site in $A$, or that $v \in G \cap A$.
The quantity in square brackets in \refeq{tautauA} is then given by
\eqalign
        \tau_{h,p}(v,x) - \tau^{A}_{h,p}(v,x) & = 
        \langle I[ v \conn x \AND v \nc G] \rangle
        - \langle I[(v \conn x \AND v \nc G) \INSIDE
        \Zd \backslash A] \rangle
        \nnb
        & =  \langle I[ v \conn x \AND v \nc G] \rangle
        - \langle I[v \conn x \INSIDE
        \Zd \backslash A \AND v \nc G ] \rangle
        \nnb 
        & \quad  + \langle I[v \conn x \INSIDE
        \Zd \backslash A \AND v \nc G ] \rangle
        \nnb 
	& \quad 
        - \langle I[(v \conn x \AND v \nc G) \INSIDE
        \Zd \backslash A] \rangle
        \nnb
        & =  \langle I[ v \ct{A} x \AND v \nc G] \rangle
        - \langle I[v \conn x \mbox{ in }
        \Zd \backslash A \AND v \ct{A} G ] \rangle   .
\lbeq{F14exp}
\enalign
Defining
\eqalign
\lbeq{F1def}
        F_1(v, x;A) & =  \left \{ v \ct{A} x  \AND v \nc G \right \}  ,
        \\ 
        F_2 (v, x; A) & = \left \{v \conn x \mbox{ in }
        \Zd \backslash A \AND v \ct{A} G \right \} ,
\enalign
this gives
\eq
\lbeq{tAF12}
        \tau_{h,p}(v,x) - \tau^{A}_{h,p}(v,x) =
        \langle I[F_1(v, x;A)] \rangle 
        - \langle I[F_2 (v, x; A)] \rangle.
\en 

We define events associated with the event $F_1(v,x;A)$ by
\eqalign
%        F_1(v, x;A)  & =   \left \{ v \ct{A} x  \AND v \nc G \right \},
%        \\
        F'_{1}(v, x; A) 
	& =  F_1(v, x;A) \cap
        \left \{ \textrm{ $\nexists$ pivotal
        $(u',v')$ for } v \conn x \textrm{ such that } v \ct{A} u'  \right \} ,
        \\ 
        F''_{1}(v, u', v'; A) 
	& = 
        F'_{1}(v, u'; A) \ON \Ctilde^{\{u', v'\}}(v)  ,
        \\
        F_1(v,u',v',x;A) & =  F_1'(v,u';A) \cap 
        \left \{ (u',v') \textrm{ is occupied and pivotal for } 
        v \conn x 
        \right\} .
\enalign
For $n \geq 0$, let 
\eq
\lbeq{Cnabbr}
        \tilde{C}_n = \tilde{C}^{\{u_n,v_n\}}(v_{n-1}),
\en 
with $v_{-1}=0$.  The random set $C_n$ is associated to an expectation,
and we will sometimes 
emphasize this association by using a subscript $n$ for
the corresponding expectation.  
Using \refeq{expan0}, \refeq{tautauA}, \refeq{tAF12} and
\refeq{Cnabbr}, we have
\eqalign
        \tau_{h,p}(0,x) & =  \langle I[E_0'(0, x)] \rangle
        + p \sum_{(u_0,v_0)} \langle I[E_0''(0, u_0, v_0)] 
        \rangle \tau_{h,p}(v,x)
        \nnb 
	& \quad 
        - p \sum_{(u_0,v_0)} \langle I[E_0''(0, u_0, v_0)] 
        \langle 
	I[F_1(v_0, x;\tilde{C}_0)] 	
	\rangle \rangle
        \nnb 
	& \quad 
\lbeq{taueq14}
        +  p \sum_{(u_0,v_0)} \langle I[E_0''(0, u_0, v_0)] 
        \langle 
        I[F_2(v_0, x;\tilde{C}_0)] 
        \rangle \rangle.
\enalign
Here, we have tacitly assumed that the sums on the right side converge.
We will continue to make this kind of assumption in what follows, and
return to this issue at the end of Section~\ref{sec-exp.rem}.

In the one-$M$ scheme, we will expand terms involving $F_1$,
but not expand those involving $F_2$.
For the $F_1$ terms, by definition we have 
\eq
\lbeq{F1exp}
        \langle I[F_1(v_{n-1},x;\tilde{C}_{n-1})] \rangle_n 
        = \langle I[F_1'(v_{n-1},x;\tilde{C}_{n-1})] \rangle_n
        + \sum_{(u_{n},v_{n})} 
	\langle I[F_1(v_{n-1},u_{n},v_{n},x;\tilde{C}_{n-1})] 
        \rangle_n .
\en 
Arguing as in the proof of Lemma~\ref{lem-pivotal2},
the event $F_1(v_{n-1},u_{n},v_{n},x;\tilde{C}_{n-1})$ 
is the intersection of the events 
that $F'_{1}(v_{n-1}, u_{n}; \tilde{C}_{n-1})$ occurs on 
$\Ctilde^{\{u_{n}, v_{n}\}}_{n}(v_{n-1})$, 
that $\{u_{n},v_{n}\}$ is occupied, and 
that $(v_{n} \conn x \AND v_{n} \nc G)$ occurs in 
$\Zd \backslash \Ctilde^{\{u_{n}, v_{n}\}}_{n}(v_{n-1})$.  
Therefore, applying Lemma~\ref{lem-cond.0}, we have
\eq
\lbeq{Cut1}
        \langle I[F_1(v_{n-1},u_{n},v_{n},x;\tilde{C}_{n-1})] \rangle_n 
	= 
        p \langle I[F_1''(v_{n-1},u_{n},v_{n};\tilde{C}_{n-1})] 
	\tau^{\tilde{C}_{n}}_{h,p}(v_{n},x)
        \rangle_n .
\en 
Using \refeq{tAF12}, substitution of \refeq{Cut1} into \refeq{F1exp} leads
to
\eqalign 
        \langle I[F_1(v_{n-1},x;\tilde{C}_{n-1})] \rangle_n 
        & = 
	\langle I[F_1'(v_{n-1},x;\tilde{C}_{n-1})] \rangle_n
        + p \sum_{(u_{n},v_{n})}
	\langle I[F_1''(v_{n-1},u_{n},v_{n};\tilde{C}_{n-1})] 
        \rangle_n \tau_{h,p}(v_{n},x)
	\nnb 
	& \quad 
	- p \sum_{(u_{n},v_{n})}
	\langle I[F_1''(v_{n-1},u_{n},v_{n};\tilde{C}_{n-1})] 
        \langle I[F_1(v_{n},x;\tilde{C}_{n})] \rangle_{n+1}
	\rangle_n
	\nnb 
	& \quad
\lbeq{F1expa}
	+ p \sum_{(u_{n},v_{n})}
	\langle I[F_1''(v_{n-1},u_{n},v_{n};\tilde{C}_{n-1})] 
        \langle I[F_2(v_{n},x;\tilde{C}_n)] \rangle_{n+1}
	\rangle_n .
\enalign

To abbreviate the notation, we define
\eqalign
\lbeq{Yndef}
	Y_n   & =  I[F_1(v_{n-1},x;\tilde{C}_{n-1})]  , \\
	Y_n'  & =  I[F_1'(v_{n-1},x;\tilde{C}_{n-1})] , \\
	Y_n'' & =  I[F_1''(v_{n-1},u_{n},v_{n};\tilde{C}_{n-1})].
\enalign
Then \refeq{F1expa} can be rewritten as
\eq
\lbeq{Yniterate}
	\langle Y_n \rangle_n = \langle Y_n' \rangle_n 
	+ \langle Y_n'' \rangle_n \tau 
	- \langle Y_n'' \langle Y_{n+1} \rangle_{n+1} \rangle_n
	+ \langle Y_n'' \langle I[F_2(v_{n},x;\tilde{C}_n)]
	\rangle_{n+1} \rangle_n,
\en
where we further abbreviate the notation by omitting 
$p\sum_{(u_{n},v_{n})}$ from the last three terms on the right side.
Substitution of \refeq{Yniterate}, with $n=1$,
into the third term of \refeq{taueq14} gives
\eqalign
\lbeq{N1partial}
	\tau_{h,p}(0,x) 
	& =  \Big( \langle I[E_0'] \rangle_0
	- \langle I[E_0''] \langle Y_1' \rangle_1 \rangle_0 \Big)
	+ \Big( \langle I[E_0''] \rangle_0
	- \langle I[E_0''] \langle Y_1'' \rangle_1 \rangle_0 \Big) \tau_{h,p}
	\nnb 
	& \quad 
	+ \langle I[E_0''] \langle Y_1'' \langle Y_2 
	\rangle_2 \rangle_1 \rangle_0 
	+ \langle I[E_0''] \langle I[F_2] \rangle_1 \rangle_0
	+ \langle I[E_0''] \langle Y_1'' \langle I[F_2] 
	\rangle_2 \rangle_1 \rangle_0 .
\enalign
The expansion can be iterated by applying \refeq{Yniterate}
to the term on the right involving $\langle Y_2 \rangle_2$,
and so on.

To express the result of this iteration compactly, we introduce
the following notation.  In place of $\langle \cdot \rangle_n$, we write 
$\Ebold_n$.  For $n \geq 1$, let
\eqalign
	\phi_{h,p}^{(0)}(0,x) 
	& = \Ebold_0 I[E_0'(0,x)] , 
	\lbeq{f0def}
	\\
	\phi_{h,p}^{(n)}(0,x) 
	& =  
	(-1)^n \Ebold_0 I[E_0''] \Ebold_1 Y_1'' 
	\cdots \Ebold_{n-1} Y_{n-1}'' \Ebold_n Y_{n}'  , 
	\lbeq{fndef}
	\\
	\Phi_{h,p}^{(0)}(0,v_0) 
	& = 
	p \sum_{u_0 \in v_0 -\Omega} \Ebold_0 I[E_0''(0,u_0,v_0)] , 
	\lbeq{Phi0def}
	\\
	\Phi_{h,p}^{(n)}(0,v_n) 
	& =  (-1)^n
	p \sum_{u_n \in v_n -\Omega}  \Ebold_0 I[E_0''] 
	\Ebold_1 Y_1'' 
	\cdots \Ebold_{n-1} Y_{n-1}'' \Ebold_n Y_{n}''  , 
	\lbeq{Phindef}
	\\
	r_{h,p}^{(n)}(0,x) 
	& =  (-1)^n 
	\Ebold_0 I[E_0''] \Ebold_1 Y_1'' 
	\cdots \Ebold_{n-1} Y_{n-1}'' \Ebold_n Y_{n}  ,
	\\
	R_{h,p}^{(n)}(0,x) 
	& =  (-1)^{n-1} 
	\Ebold_0 I[E_0''] \Ebold_1 Y_1'' 
	\cdots \Ebold_{n-1} Y_{n-1}'' 
	\Ebold_n I[F_2 (v_{n-1}, x; \tilde{C}_{n-1})] . 
	\lbeq{Rndef}
\enalign
In the above, the notation continues to omit sums and factors of $p$
associated with each product.  For each $N \geq 0$,
the iteration indicated in the previous paragraph then gives
\eq
\lbeq{tauhHR}
	\tau_{h,p}(0,x) = \sum_{n=0}^N \phi_{h,p}^{(n)}(0,x) + 
	\sum_{n=0}^N \sum_{v_{n}}
	\Phi_{h,p}^{(n)}(0,v_{n}) \tau_{h,p}(v_{n},x) 
	+ \sum_{n = 1}^{N+1} R_{h,p}^{(n)}(0,x)
	+ r_{h,p}^{(N+1)}(0,x).
\en
The cases $N=0$ and $N=1$ are given explictly above in \refeq{taueq14}
and \refeq{N1partial}.
For $p < p_c$, $h \geq 0$, or for $p=p_c$, $h>0$, it was argued below
\refeq{tauhatexists} that the Fourier transform $\hat{\tau}_{h,p}(k)$ exists.
The bounds of Lemmas~\ref{lem-Phinbd} and \ref{lem-Rnbd} below
will show that the Fourier transform of each of the quantities on the
right side of \refeq{tauhHR} also exists, under the hypotheses of 
Theorem~\ref{thm-tauasymp}.  These bounds will also imply convergence
of the various summations arising in the course of deriving the expansion.
For each $N \geq 0$, this leads to 
\eq
\lbeq{tauhHRk}
	\hat{\tau}_{h,p}(k) = 
	\frac{\sum_{n=0}^N \hat{\phi}_{h,p}^{(n)}(k)  
	+ \sum_{n = 1}^{N+1} \hat{R}_{h,p}^{(n)}(k) 
	+ \hat{r}_{h,p}^{(N+1)}(k)}
	{1- \sum_{n=0}^N \hat{\Phi}_{h,p}^{(n)}(k)}.
\en

In this one-$M$ scheme for the expansion, $\phi_{h,p}$ and $\Phi_{h,p}$
are given by the same diagrams as in \cite{HS90a},
but now there is a $G$-free condition on the connections in 
each of the nested expectations defining the diagrams.  If we set $h=0$,
the $G$-free condition becomes vacuous, the terms involving $F_2$ in 
the remainder vanish, and we recover the expansion of \cite{HS90a}.

%%%%%%%%%%%%%%%%%%%%%%%%%%%%%%%%%%%%%%%%%%%%%%%%%%%%%%%%%%%%%%%%%%%%%%%%%%%%%%%
%%%%%%%%%%%%%%%%%%%%%%%%%%%%%%%%%%%%%%%%%%%%%%%%%%%%%%%%%%%%%%%%%%%%%%%%%%%%%%%
\subsection{The first expansion: two-$M$ scheme}
\label{subsub-exp.no-rem}

For the proof of Theorem~\ref{thm-tauasymp}, we 
require a more complete expansion, in which bounds on remainder
terms will involve two factors of the magnetization $M_{h,p}$.
We therefore refer to this new expansion as the two-$M$ scheme.
The expansion proceeds by further expanding the $F_2$ 
that was left unexpanded in the one-$M$ scheme, in $R_{h,p}^{(n)}(0,x)$
of \refeq{Rndef}. 

We begin by decomposing $F_{2}$ into several events.  Using 
the notion of ``sausage'' defined at the beginning of 
Section~\ref{sec-exp.rem}, we introduce the following definitions: 
\begin{description}
    \item  
    $F_{3}(v, x; A)$ is the event that $v \conn x$, $v \ct{A} G$, 
    exactly one sausage is connected to $G$, and the right endpoint 
    of the sausage which is connected to $G$ is connected to $v$ in 
    $\Zd\backslash A$.    
    \item 
    $F_{4}(v, x; A)$ is the event that $v \conn x$, $v \ct{A} G$, 
    two or more sausages are connected to $G$, and the right 
    endpoints of all sausages which are connected to $G$ are 
    connected to $v$ in $\Zd\backslash A$.  
    \item 
    $F_{5}(v, x; A)$ is the event that $v \ct{A} x$, $v \ct{A} G$,
    and the right 
    endpoints of all sausages which are connected to $G$ are 
    connected to $v$ in $\Zd\backslash A$.
\end{description}
The event $F_4$ involves two disjoint connections to $G$ and will
lead to a bound involving $M_{h,p}^2$.  It does not require further 
expansion.
The events $F_3$, $F_4$, $F_5$ are related to $F_2$ in the following lemma. 
In the lemma, $\cupd$ denotes disjoint union.

\begin{lemma}
    \label{lem-F2dcmp}
For $v, x \in \Zd$ and $A \subset \Zd$, 
	\eq
	\lbeq{F2dcmp}
		F_{2}(v, x; A) 
		= \{ F_{3}(v, x; A) \cupd F_{4}(v, x; A) \} \, 
		\backslash  F_{5}(v, x; A) .
	\en 
\end{lemma}

\Proof
Since $F_{2}$ and $F_{5}$ are disjoint,  
$F_{2}(v, x; A) =  \{ F_{2}(v, x; A)  \cupd F_{5}(v, x; A) \} \, 
\backslash F_{5}(v, x; A)$. 
Thus it suffices to show that 
\eq
\lbeq{F2534iden}
	F_{2}(v, x; A)  \cupd F_{5}(v, x; A) 
	= F_{3}(v, x; A)  \cupd F_{4}(v, x; A) . 
\en 
By definition, the left side is the event that
$v \conn x$, $v \ct{A} G$, and
the right endpoints of all sausages which are connected to $G$ are 
connected to $v$ in $\Zd\backslash A$.  The desired identity \refeq{F2534iden}
then follows, since
$F_{3}$ and $F_{4}$ provide a 
disjoint decompositions of the above event, 
according to the number of sausages connected to $G$. 
\qed

Now we define the events  
\eqalign
        F_3'(v, x; A) 
	& =  F_3 (v, x; A) \cap
        \bigl \{ (\textrm{last sausage of } v \conn x) \ct{A} G \bigr \} ,
	\\
        F_4'(v, x; A) 
	& = F_4 (v, x; A) \cap
        \bigl \{ (\textrm{last sausage of } v \conn x) \ct{A} G \bigr \} ,
	\\
        F_5'(v, x; A) 
	& =  F_5 (v, x; A) \cap
        \bigl \{\textrm{$\nexists$ pivotal } (u',v') \textrm{ for } v \conn x 
	\textrm{ such that } v \ct{A} u' \bigr \} ,
\intertext{and for $j = 3,4,5$}
	F_j''(v, u', v'; A) 
	& = F_j'(v, u'; A) \ON \Ctilde^{\{u'v'\}}(v) ,
        \\
        F_j(v,u',v',x;A) 
	& = F_j'(v,u';A) \cap 
        \bigl \{ (u',v') \textrm{ is occupied and pivotal for } v \conn x 
	\bigr \}
        \nnb 
	& \quad  \cap \bigl \{ 
        \tilde{C}^{\{u',v'\}}(x) \cap G = \emptyset  \bigr \} . 
\enalign  
These events obey the identity of the following lemma.  

\begin{lemma}
\label{lem-F345cut}
For $j =3,4,5$, 
\eq
\lbeq{Ej0}
        \langle I[F_{j}(v,x;A)] \rangle 
        = \langle I[F_{j}'(v,x;A)] \rangle
        + p \sum_{(u',v')} 
	\langle I[F_{j}''(v,u',v';A)] 
	\tau^{\tilde{C}^{\{u',v'\}}(v)}_{h,p}(v',x)
        \rangle.
\en 
\end{lemma}

\Proof 
Let $j =3,4,5$.  We first observe that 
\eq
\lbeq{Ej4}
        \langle I[F_j(v,x;A)] \rangle 
        = \langle I[F_j'(v,x;A)] \rangle
        + \sum_{(u',v')} 
	\langle I[F_j(v,u',v', x; A)] 
        \rangle.
\en 
Arguing as in the proof of Lemma~\ref{lem-pivotal2},
each $F_{j}(v, u',v',x;A)$ can be written as 
the intersection of the events that
$F'_{j}(v, u'; A) \ON \Ctilde^{\{u', v'\}}(v)$, 
that $(v' \conn x \AND v' \nc G) \INSIDE
\Zd \backslash \Ctilde^{\{u', v'\}}(v)$,
and that $\{u',v'\}$ is occupied. Hence
Lemma~\ref{lem-cond.0} can be applied to conclude 
\eq
\lbeq{Ej2}
        \langle I[F_j(v, u',v',x;A)] \rangle = 
        p \langle I[F_j''(v,u',v';A)] 
        \tau^{\tilde{C}^{\{u',v'\}}(v)}_{h,p}(v',x)
        \rangle .
\en 
Combined with \refeq{Ej4}, this gives \refeq{Ej0}.
\qed

We can now begin the expansion of the $F_2$ term.
Using Lemma~\ref{lem-F2dcmp}, and Lemma~\ref{lem-F345cut} for $F_{3}$ and 
$F_{5}$, we obtain 
\eqalign 
	\expec{I[F_2(v, x; A)]} 
	& = 
	\expec{I[F_{3}(v, x; A)]} - \expec{I[F_5(v, x; A)]} 
	+ \expec{I[F_{4}(v, x; A)]}
	\nnb	
	& = 
	\expec{I[F_3'(v, x; A)]} - \expec{I[F_5'(v, x; A)]} 
	+ \expec{I[F_4(v, x; A)]} 
	\nnb
\lbeq{F2-rewrite1}
	& \quad 
	+ p \sum_{(u', v')} 
	\expec{
	\{ I[F_3''(v, u', v'; A)] - I[F_5''(v, u', v'; A)] \} 
	 \, \tau_{h,p}^{\tilde{C}^{\{u',v'\}} (u')  } (v', x) }
	.
\enalign 
The $F_4$ term is not expanded further.
For the last term, we use \refeq{tautauA} and \refeq{tAF12}.  This gives 
\eqarray 
	\expec{I[F_2(v, x; A)]} 
	& = &
	\expec{I[F_3'(v, x; A)]} - \expec{I[F_5'(v, x; A)]} 
	+ \expec{I[F_4(v, x; A)]} 
	\nonumber \\
\lbeq{F2-rec1}
	&&
	+ p \sum_{(u', v')} 
	\expec{I[F_3''(v, u', v'; A)] - I[F_5''(v, u', v'; A)] }  
	 \, \tau_{h,p}(v', x) 
	\\ \nonumber
	&& - p \sum_{(u', v')} 
	\expec{
	\{ I[F_3''(v, u', v'; A)] - I[F_5''(v, u', v'; A)] \} 
	 \, 
	\langle I[F_1(v',x; \tilde{C}^{\{u',v'\}} (u') )] \rangle  
	}
	\\ \nonumber
	&&
	+ p \sum_{(u', v')} 
	\expec{
	\{ I[F_3''(v, u', v'; A)] - I[F_5''(v, u', v'; A)] \} 
	 \, 
	\langle I[F_2(v',x; \tilde{C}^{\{u',v'\}} (u') )] \rangle  
	}  .
\enarray 

We are now in a position to generate the expansion.
First, we introduce some abbreviated notation.  Let
\eqalign
\lbeq{alndef}
	W_n & =    I[F_3  (v_{n-1},x; \tilde{C}_{n-1})] 
		 -  I[F_5  (v_{n-1},x; \tilde{C}_{n-1})] , \\
\lbeq{alnpdef}
	W_n' & =   I[F_3' (v_{n-1},x; \tilde{C}_{n-1})] 
		 -  I[F_5' (v_{n-1},x; \tilde{C}_{n-1})] , \\
\lbeq{alnppdef}
	W_n'' & =  I[F_3''(v_{n-1}, u_{n}, v_{n}; \tilde{C}_{n-1})] 
		  - I[F_5''(v_{n-1}, u_{n}, v_{n}; \tilde{C}_{n-1})] , \\
\lbeq{Fnabdef}
	(F_2)_n & =  I[F_2(v_{n-1}, x; \tilde{C}_{n-1})] , 
	\qquad 
	(F_4)_n =  I[F_4(v_{n-1}, x; \tilde{C}_{n-1})] . 
\enalign
To further abbreviate the notation, in generating the expansion we omit
all arguments related to sites and omit the summations
$p \sum_{(u_{n}, v_{n})}$ that are associated with
each product.
Then, recalling \refeq{Yndef}, \refeq{F2-rec1} can be written more compactly
as 
\eq
\lbeq{alit}
        \expec{ (F_2)_n }_{n}
	= \expec{ W'_{n} }_{n} 
	+ \expec{ (F_4)_n }_{n}
	+ 
	 \expec{ W''_{n} }_n \tau 
	+ 
	 \expec{ W''_{n} \, \langle (F_2)_{n+1} \rangle_{n+1} }_{n} 
	- 
	 \expec{ W''_{n} \, \langle Y_{n+1} \rangle_{n+1} }_{n} .
\en 
An expansion can now be generated by recursively substituting
\refeq{Yniterate}, which now reads 
\eq
\lbeq{Yniterate3}
	\langle Y_n \rangle_n = \langle Y_n' \rangle_n 
	+ \langle Y_n'' \rangle_n \tau 
	- \langle Y_n'' \langle Y_{n+1} \rangle_{n+1} \rangle_n
	+ \langle Y_n'' \langle (F_2)_{n+1}
	\rangle_{n+1} \rangle_n,
\en
into the last term on the right side of \refeq{alit}. 
The first iteration yields
\eqalign 
\lbeq{F2-rec2}
	\expec{ (F_2)_n }_{n}
	& 
	 = \expec{ W'_{n} }_{n} 
	+ \expec{ (F_4)_n }_{n}
	+ 
	 \expec{ W''_{n} }_n \tau
	+ 
	 \expec{ W''_{n} \, \langle (F_2)_{n+1} \rangle_{n+1} }_{n} 
	\nnb 
	& \quad
	- \expec{ W''_{n} \, \langle Y'_{n+1} \rangle_{n+1} }_{n} 
	- \expec{ W''_{n} \, \langle Y''_{n+1} \rangle_{n+1} }_{n} \tau 
	\nnb 
	& \quad
	- \expec{ W''_{n} \, \langle Y''_{n+1} 
		\langle (F_2)_{n+2} \rangle_{n+2} \rangle_{n+1} }_{n} 
	+ \expec{ W''_{n} \, \langle Y''_{n+1} 
		\langle Y_{n+2} \rangle_{n+2} \rangle_{n+1} }_{n} .
\enalign
We then apply \refeq{Yniterate3} to $Y_{n+2}$, and so on. 
We halt the expansion in any term in which an $F_4$ appears, or when
an $F_2$ appears in a term already containing a $W''$.
The result is substituted into the formula 
for $R_{h,p}^{(n)}$ of \refeq{Rndef}.  

To organize the resulting terms, we introduce the following
quantities, for $n \geq 1, m \geq 1$.  
\eqarray
\lbeq{xin0def}
	 \xi^{(n, 0)}_{h,p} (0,x) & = & (-1)^{n-1} 
	{\mathbb E}_{0} I[E_0''] {\mathbb E}_{1} Y_1'' 
	\cdots {\mathbb E}_{n-1} Y_{n-1}'' 
	{\mathbb E}_{n} W_n' ,
	\\
	\xi^{(n, m)}_{h,p} (0,x) & = &  (-1)^{n+m-1} 
	{\mathbb E}_{0} I[E_0''] {\mathbb E}_{1} Y_1'' 
	\cdots {\mathbb E}_{n-1} Y_{n-1}'' 
	{\mathbb E}_{n} W_n'' 
	\nonumber \\
	&& \hspace{10mm} \times
	{\mathbb E}_{n+1} Y_{n+1}''  \cdots 
	{\mathbb E}_{n+m-1} Y_{n+m-1}'' 
	{\mathbb E}_{n+m} Y_{n+m}'  ,
	\\
	\Xi^{(n, 0)}_{h,p} (0,v_n) & = &  (-1)^{n-1}
	{\mathbb E}_{0} I[E_0''] {\mathbb E}_{1} Y_1'' 
	\cdots {\mathbb E}_{n-1} Y_{n-1}'' 
	{\mathbb E}_{n} W_n''  ,
	\\
	\Xi^{(n, m)}_{h,p} (0,v_{n+m}) & = &  (-1)^{n+m-1}
	{\mathbb E}_{0} I[E_0''] {\mathbb E}_{1} Y_1'' 
	\cdots {\mathbb E}_{n-1} Y_{n-1}'' 
	{\mathbb E}_{n} W_n'' 
	\nonumber \\ && \hspace{10mm} \times 
	{\mathbb E}_{n+1} Y_{n+1}''  \cdots {\mathbb E}_{n+m-1} Y_{n+m-1}'' 
	{\mathbb E}_{n+m} Y_{n+m}''  ,
	\\
\lbeq{Rtilndef}
	S^{(n)}_{h,p} (0,x) & = &  (-1)^{n}
	{\mathbb E}_{0} I[E_0''] {\mathbb E}_{1} Y_1'' 
	\cdots {\mathbb E}_{n-1} Y_{n-1}'' 
	{\mathbb E}_{n} (F_4)_{n}  ,
	\\
	U^{(n, m)}_{h,p} (0,x) & = &  (-1)^{n+m}
	{\mathbb E}_{0} I[E_0''] {\mathbb E}_{1} Y_1'' 
	\cdots {\mathbb E}_{n-1} Y_{n-1}'' 
	{\mathbb E}_{n} W_n'' 
	\nonumber \\ && \hspace{10mm} \times
	{\mathbb E}_{n+1} Y_{n+1}''  \cdots {\mathbb E}_{n+m-1} Y_{n+m-1}'' 
	{\mathbb E}_{n+m} (F_2)_{n+m}  ,
	\\ 
	u_{h,p}^{(n, m)} (0,x) & = &  (-1)^{n+m-1}
	{\mathbb E}_{0} I[E_0''] {\mathbb E}_{1} Y_1'' 
	\cdots {\mathbb E}_{n-1} Y_{n-1}'' 
	{\mathbb E}_{n} W_n'' 
	\nonumber \\ && \hspace{10mm} \times
\lbeq{UNdef}
	{\mathbb E}_{n+1} Y_{n+1}''  \cdots {\mathbb E}_{n+m-1} Y_{n+m-1}'' 
	{\mathbb E}_{n} Y_{n+m}  . 
\enarray
In the above, the notation continues to omit sums and factors of $p$
associated with each product.
We substitute the result of the expansion into the term $R_{h,p}^{(n)}$ of
\refeq{tauhHR}.   
Define
\eqarray
	A_{h,p}^{(M,N)}(0,x) & = & 
	\sum_{n=0}^N \phi_{h,p}^{(n)}(0,x) 
	+ \sum_{n=1}^N \sum_{m=0}^M \xi^{(n, m)}_{h,p}(0,x) 
	+ \sum_{n=1}^N \sum_{m=1}^{M+1} U^{(n, m)}_{h,p}(0,x) 
	+ \sum_{n=1}^{N} S^{(n)}_{h,p}(0,x) 
	\nonumber \\
	&& 
	+  \sum_{n=1}^N u_{h,p}^{(n, M+1)}(0,x) 
	+ r_{h,p}^{(N+1)} (0,x),
	\\
	B_{h,p}^{(M,N)}(0,x) & = &
	\sum_{n=0}^N \Phi_{h,p}^{(n)}(0, v_n) 
	+ \sum_{n=1}^N \sum_{m=0}^M \Xi^{(n, m)}_{h,p}(0,v_n).
	\enarray
For each $N, M \geq 1$, the result of the expansion is then
\eq
\lbeq{taugPiU}
	\tau_{h,p}(0,x) 
	= A_{h,p}^{(M,N)}(0,x)
	+ \sum_{v_n} 
	B_{h,p}^{(M,N)}(0,v_n) \tau_{h,p}(v_n, x) .
\en 
Under the high dimension assumptions of Theorem~\ref{thm-tauasymp},
existence of the Fourier transforms of $A_{h,p}^{(M,N)}(0,x)$ and
$B_{h,p}^{(M,N)}(0,x)$ will follow from Lemmas~\ref{lem-Phinbd},
\ref{lem-Rnbd} and \ref{lem-xiXiURbds} below, leading to the conclusion that
for $p < p_c$, $h \geq 0$, or for
$p=p_c$, $h>0$,
\eq
\lbeq{taugPiUk}
	\hat{\tau}_{h,p}(k) = 
	\frac{\hat{A}_{h,p}^{(M,N)}(k)}
	{1- \hat{B}^{(M,N)}_{h,p}(k) 
	}.
\en

In Section~\ref{sub-2Mbd.infinite}, we will take the limits $M,N \to \infty$
in \refeq{taugPiUk}, and in this limit, the terms 
$\sum_{n=1}^N \hat{u}_{h,p}^{(n, M+1)}(k)$
and $\hat{r}^{(N+1)}_{h,p}(k)$ in $\hat{A}_{h,p}^{(M,N)}(k)$ vanish.
For $h=0$, the set $G$ of green sites is empty, and
the events $F_{3}$, $F_{4}$ and $F_5$, 
which require connection to $G$, cannot occur.
Therefore the terms involving $\xi$, $\Xi$, $S$, $U$, and $u$ all
vanish for $h=0$, and 
\refeq{taugPiU} reduces to the expansion of \cite{HS90a}.

%%%%%%%%%%%%%%%%%%%%%%%%%%%%%%%%%%%%%%%%%%%%%%%%%%%%%%%%%%%%%%%%%%%%%%%%%%%%%%%
%%%%%%%%%%%%%%%%%%%%%%%%%%%%%%%%%%%%%%%%%%%%%%%%%%%%%%%%%%%%%%%%%%%%%%%%%%%%%%%
%%%%%%%%%%%%%%%%%%%%%%%%%%%%%%%%%%%%%%%%%%%%%%%%%%%%%%%%%%%%%%%%%%%%%%%%%%%%%%%
%%%%%%%%%%%%%%%%%%%%%%%%%%%%%%%%%%%%%%%%%%%%%%%%%%%%%%%%%%%%%%%%%%%%%%%%%%%%%%%
\section{Bounds on the two-point function via the one-$M$ scheme}
\label{sec-part.exp.bd}
\setcounter{equation}{0}

In this section, we use the one-$M$ scheme of the expansion to
prove upper and lower 
bounds on $\hat{\tau}_{h,p_c}(k)$.  The bounds involve the function
\eq
	\hat{D}(k) = \frac{1}{|\Omega|} \sum_{x \in \Omega} e^{ik \cdot x},
\en
where $|\Omega|$ denotes the cardinality of the set $\Omega$ of neighbours
of the origin.
We will frequently write simply $\Omega$, rather than $|\Omega|$.
For the nearest-neighbour model, we have simply 
$\hat{D}(k) = d^{-1}\sum_{j=1}^d \cos k_j$, 
and for both the nearest-neighbour and spread-out models, 
$1-\hat{D}(k)$ is asymptotic to an $\Omega$-dependent 
multiple of $k^2$ as $k \to 0$.
Useful bounds on $\hat{D}(k)$ can be found in \cite[Appendix~A]{MS93}.

\begin{prop}
\label{prop-taubd}
For the nearest-neighbour model with $d$ sufficiently large, or
for the spread-out model with 
$d>6$ and $L$ sufficiently large (depending on $d$), there are 
positive constants $K_1$ and $K_2$ (independent of $L,d$), such that 
for $h>0$ and $k \in [-\pi,\pi]^d$, 
\eq
\lbeq{tauwant}
        \frac{K_1 e^{-h}}{[1-\Dhat(k)] + \sqrt{1-e^{-h}}} 
	\leq \tauhat_{h,p_c}(k)
        \leq \frac{K_2 e^{-h}}{[1-\Dhat(k)] + \sqrt{1-e^{-h}}}  .
\en 
\end{prop}

We treat the nearest-neighbour and spread-out models simultaneously
in this section.
To facilitate this,
we will use $\oneOd$ to denote a function of $L$ or of $d$ 
which goes to zero as $L \to \infty$ or $d \to \infty$.
We will use
$O(\oneOd^n)$ to denote a quantity bounded by $(K \oneOd)^n$, 
with $K$ independent 
of $h$, $p$, $n$ and of $L$ or $d$.  
We assume without further mention that henceforth
$d \gg 6$ for the nearest-neighbour model, and $d>6$
and $L \gg 1$ for the spread-out model.

Our starting point for proving \refeq{tauwant} is \refeq{tauhHRk}.
Introducing the notation
\eq
\lbeq{rhodef}
	\hat{\rho}_{h,p}^{(N+1)}(k) = \sum_{n = 1}^{N+1} \hat{R}_{h,p}^{(n)}(k) 
	+ \hat{r}_{h,p}^{(N+1)}(k),
\en
\refeq{tauhHRk} states that for any $N \geq 0$, $h>0$,
\eq
\lbeq{oneF4.tauhat1}
        \tauhat_{h,p_c}(k) = 
        \frac{\sum_{n=0}^N \hat{\phi}_{h,p_c}^{(n)}(k) 
	+ \hat{\rho}_{h,p_c}^{(N+1)}(k) }
	{ 1 - \sum_{n=0}^N \Phihat_{h,p_c}^{(n)}(k)} .
\en 
It will be a consequence of what follows that the limit $N \to \infty$
can be taken in \refeq{oneF4.tauhat1}.
The proof of \refeq{tauwant} 
is organized as follows.  In Section~\ref{subsub-tau.main},
we will extract the leading terms from \refeq{oneF4.tauhat1}.
The denominator of \refeq{oneF4.tauhat1}, and the 
contribution $\sum_{n=0}^N \hat{\phi}_{h,p_c}^{(n)}(k)$ to the numerator,
will be bounded in Sec.~\ref{sub-bound.basic}.  
The remainder term $\hat{\rho}_{h,p_c}^{(N)}(k)$ will be bounded
in Sec.~\ref{sub-bd.rem}.  At this point, we will be able to take the
limit $N \to \infty$.  The remainder term will be bounded using
Lemma~\ref{lem-tail.tau1}, the ``cut-the-tail'' lemma,
whose proof is deferred to 
Section~\ref{sec-cut-the-tail}.  The cut-the-tail lemma will also be
used in Sections~\ref{sec-2M.exp.err} and \ref{sec-Psidef}, and in II.
In Section~\ref{sub-taubd}, 
we combine the bounds obtained thus far, and prove \refeq{tauwant}. 

In this section, we will use the infra-red bound 
\refeq{irbd} and the bound 
\eq
\lbeq{pc}
	1 \leq p_c \Omega \leq 1 + O(\Omega^{-1}),
\en 
both of which are due to \cite{HS90a}.
For the nearest-neighbour model, \refeq{pc} was improved in \cite{HS95}.

%%%%%%%%%%%%%%%%%%%%%%%%%%%%%%%%%%%%%%%%%%%%%%%%%%%%%%%%%%%%%%%%%%%%%%%%%%%%%%%
\subsection{The main contribution} 
\label{subsub-tau.main}

We rewrite the $n=0$ terms of \refeq{oneF4.tauhat1} as 
\eq
\lbeq{phi0def}
	\hat{\phi}^{(0)}_{h,p} (k) 
	= \hat{\phi}^{(00)}_{h,p} (k) + \hat{\phi}^{(01)}_{h,p} (k), 
	\qquad
	\Phihat^{(0)}_{h,p} (k) 
	= \Phihat^{(00)}_{h,p} (k) + \Phihat^{(01)}_{h,p} (k), 
\en 
with 
\begin{align}
	\hat{\phi}^{(00)}_{h,p} (k) 
	& = \langle I[E_0'(0,0)] \rangle, 
	&
	\hat{\phi}^{(01)}_{h,p} (k) 
	& = \sum_{x \neq 0} \langle I[E_0'(0,x)] \rangle e^{i kx}, 
	\\
\lbeq{Phi00def}
	\Phihat^{(00)}_{h,p} (k) 
	& = 
	p \sum_{(0, v_0)} \expec{I[E_0''(0,0, v_0)]} e^{i k v_0}, 
	& \Phihat^{(01)}_{h,p} (k) 
	& = p \sum_{(u_0, v_0): u_0 \neq 0} 
	\expec{I[E_0''(0, u_0, v_0)]} e^{i k v_0}.  \hspace{1cm}
\end{align}
The terms $\hat{\phi}^{(00)}_{h,p} (k)$ and $\Phihat^{(00)}_{h,p} (k)$ 
are the leading ones.
The former is given simply by
\eq
\lbeq{f00bd.0}
	 \hat{\phi}_{h,p}^{(00)}(k) = \phi_{h,p}^{(0)} (0,0) 
	 = \expec{I[0 \nc G]} = 1 - M_{h,p} .
\en 
For the latter, we have the following lemma.

\begin{lemma}
\label{lem-Pi00bd}
For $p\leq p_c$, $h \geq 0$, and $k \in [-\pi,\pi]^d$,
\eqalign 
	\Phihat_{h,p}^{(00)}(k) & =  p \Omega 
	\left [ (1 - M_{h,p}) \Dhat(k) + O(\oneOd) M_{h,p}  \right ] ,
	\lbeq{Phi00bd.0}
	\\ 
	\Phihat_{h,p}^{(00)}(0) - \Phihat_{h,p}^{(00)}(k) 
	& =  
	p \Omega [ 1- \Dhat(k)] 
	 \left[ 1 - M_{h,p} 
	+ O(\oneOd) M_{h,p}   
	\right].
	\lbeq{Phi00bd.1}
\enalign
\end{lemma}

\Proof
We first note that \refeq{Phi00bd.0} would follow immediately from 
\eq
	\expec{I[E_0''(0,0,v_0)]}  
	= 1 - M_{h,p}  + O(\Omega^{-1})  M_{h,p}.
	\lbeq{E00bd.0}
\en 
To prove \refeq{E00bd.0}, we begin by observing that
\begin{align}
        \expec{I[E_0''(0,0,v_0)]} & = 
        \prob{\Ctilde^{\{0,v_{0}\}}(0) \cap G = \emptyset} 
        \nonumber \\ 
        & =  \prob{C(0) \cap G  = \emptyset} + 
        \left \{   
        \prob{\Ctilde^{\{0,v_{0}\}}(0) \cap G = \emptyset} 
        - \prob{C(0) \cap G  = \emptyset}
        \right \} .
\lbeq{pi0.1}    
\end{align}
The first term on the right side equals $1-M_{h,p}$.  The second term is
the probability that $\{0,v_0\}$ is occupied and pivotal for the event
$\{ 0 \conn G\}$, and is bounded by
$pP(v_0 \conn G) =pM_{h,p}$.
With \refeq{pc}, this proves \refeq{E00bd.0}. 

Finally, \refeq{Phi00bd.1} follows from substitution of \refeq{E00bd.0} into 
\eq
	\Phihat_{h,p}^{(00)}(0) - \Phihat_{h,p}^{(00)}(k) 
	= p \sum_{(0, v_0)} \expec{I[E_0''(0,0, v_0)]} 
	[ 1 - \cos( k \cdot v_0) ] .
\en  
\qed

%%%%%%%%%%%%%%%%%%%%%%%%%%%%%%%%%%%%%%%%%%%%%%%%%%%%%%%%%%%%%%%%%%%%%%%%%%%%%%%
%%%%%%%%%%%%%%%%%%%%%%%%%%%%%%%%%%%%%%%%%%%%%%%%%%%%%%%%%%%%%%%%%%%%%%%%%%%%%%%
\subsection{Standard diagrammatic estimates}
\label{sub-bound.basic}
% previously \label{sub-bound.rem}

In this section, we obtain bounds on the subdominant terms
$\hat{\phi}^{(n)}_h(k)$ and $\hat{\Phi}^{(n)}_h(k)$, for $n \geq 1$
and $n=01$.  
The bounds are standard, in the sense that they do not require methods
beyond those used in \cite{HS90a}.  They are based on bounds for simple
polygonal diagrams, and we begin by reviewing these bounds.

For $p \in [0,p_c]$ and $h \geq 0$, we define
the polygon and weighted polygon diagrams:
\begin{align}
	\pol{m}_{h, p} (x) 
	& =  
	\sum_{y_1, y_2, \ldots, y_{m-1} \in \Zd} 
	\tau_{h,p}(0, y_1) \tau_{h,p}(y_1, y_2) 
	\cdots \tau_{h,p}(y_{m-1}, x)
	- \delta_{0,x} \{\tau_{h,p}(0, 0)\}^{m} ,
	\\
\lbeq{wpoldef}
	\wpol{m}_{h, p}(x) 
	& =  
	\sum_{y_1, y_2, \ldots, y_{m-1} \in \Zd} 
	| y_1 |^2 
	\tau_{h,p}(0, y_1) \tau_{h,p}(y_1, y_2) 
	\cdots \tau_{h,p}(y_{m-1}, x) .
\end{align}
The second term of 
$\pol{m}$ just subtracts the $y_{1} = y_{2} = \cdots = y_{m-1} = 
x =0$ term
from the sum, and thus $\pol{m}$ can be rewritten as a sum of products of 
$\tau_{h,p}$'s, with positive coefficients.
The following lemma gives bounds on these quantities.

\begin{lemma}
\label{lem-triangles}
For $p \in [0, p_c]$, $h \geq 0$,  and $\lambda$ sufficiently small,
\begin{align}
	\sup_x \pol{m}_{h,p}(x) & \leq O(\oneOd) \qquad {\rm for} \quad d > 2m,
\\
	\sup_x  \wpol{m}_{h,p}(x)  & \leq  O(\oneOd) 
	\qquad {\rm for} \quad d > 2m +2 .
\end{align}
\end{lemma}

\Proof
For $h \geq 0$, by \refeq{Mzdef} we have 
$0 \leq \tau_{h,p}(0, x) \leq \tau_{0,p}(0, x)$.   
Therefore, $\pol{m}_{h,p}(x)$ and $\wpol{m}_{h,p}(x)$ are 
dominated by their values at $h=0$.   
Also, $\pol{m}_{0,p}(x)$ and $\wpol{m}_{0,p}(x)$ are 
monotone nondecreasing in $p$, since $\tau_{0,p}(0,x)$ is.  Thus
we need only bound their values at $h=0$ by $O(\oneOd)$, 
uniformly in $p<p_c$ and in $x$, to establish the lemma. 

It was shown in \cite{HS90a} that 
$\pol{3}_{h,p}(x)$ and $\wpol{2}_{h,p}(x)$ are $O(\oneOd)$ 
for $h=0$, uniformly in $p < p_c$ and in $x$.  
The method involved writing these quantities
in terms of the Fourier transform of the two-point function and using
the infra-red bound \refeq{irbd}.  The same method can be
used for general $m$, yielding the lemma.
\qed

We now turn to bounds on $\hat{\phi}_{h,p}^{(n)}(k)$ and 
$\hat\Phi_{h,p}^{(n)}(k)$.
To discuss the cases $n=01$ and $n \geq 1$ simultaneously, we introduce
the notation
\eq
\lbeq{nbardef}
	\bar{n} = \left\{ \begin{array}{ll} 
	1 & n = 01 , \\
	n & n \geq 1 .
	\end{array} \right. 
\en
The following lemma gives bounds on the subdominant 
$\hat{\phi}^{(n)}_{h,p}(k)$ and $\hat{\Phi}^{(n)}_{h,p}(k)$ corresponding
to these values of $n$.

\begin{lemma}
\label{lem-Phinbd}
For $h \geq 0$ and $p \in [0, p_c]$, and for 
$n=01$ or $n \geq 1$, we have 
\begin{align} 
	|  \hat{\phi}^{(n)}_{h,p}(k) | 
	& \leq O(\oneOd^{\bar{n}}) e^{-h(\bar{n}+1)} , 
	& 
        | \hat{\phi}^{(n)}_{h,p}(0) - \hat{\phi}^{(n)}_{h,p}(k) | 
        & \leq  O(\oneOd^{\bar{n}}) e^{-h(\bar{n}+1)}\, [1-\Dhat(k)] , 
	\lbeq{fnbd}   \\  
	|  \hat{\Phi}^{(n)}_{h,p}(k) | 
	& \leq p \Omega O(\oneOd^{\bar{n}}) e^{-h(\bar{n}+1)} , 
 	& | \hat{\Phi}^{(n)}_{h,p}(0) - \hat{\Phi}^{(n)}_{h,p}(k) | 
                & \leq p \Omega O(\oneOd^{\bar{n}}) e^{-h(\bar{n}+1)} 
	\, [1-\Dhat(k)] , 
	\lbeq{Phinbd}
\end{align}
and
\eq
\lbeq{Phinbd.2}
	| \hat{\Phi}^{(n)}_{0,p}(0) - \hat{\Phi}^{(n)}_{h,p}(0) | 
                \leq p \Omega O(\oneOd^{\bar{n}} ) M_h .
\en 
\end{lemma}

The remainder of Section~\ref{sub-bound.basic} is devoted to the proof of 
Lemma~\ref{lem-Phinbd}.  The method of proof illustrates our basic
strategy for bounding diagrams.  Because the proof is lengthy, 
we present it in several steps.

%%%%%%%%%%%%%%%%%%%%%%%%%%%%%%%%%%%%%%%%%%%%%%%%%%%%%%%%%%%%%%%%%%%%%%%%%%%%%%%
\subsubsection{Explicit $h$-dependence}

We begin by making explicit the $h$-dependence of quantities of interest.
For this purpose, we define auxiliary events which only depend on 
\emph{bond}\/ 
variables, with no $G$-dependence:  
\eqalign 
	E_{0, b}' (0, x) & = 
	\bigl \{ 0 \dbc x \bigr \}  
	\\
	E_{0, b}'' (0, u', v') & = 
	\bigl \{ 0 \dbc u' \bigr \} \ON \tilde{C}^{\{u', v'\}}(0) 
	\\
	F_{1, b}' (v, x; A) & = 
	\bigl \{ v \ct{A} x \bigr \} \cap 
	\bigl \{\nexists \mbox{ pivotal $(u', v')$ for $v \conn x$ s.t. }
	v \ct{A} u' \bigr \} 
	\\ 
	F_{1, b}'' (v, u', v'; A) & = 
	F_{1, b}'(v, u'; A)  \ON \tilde{C}^{\{u',v'\}}(v) .
\enalign
These are the events occuring in \cite{HS90a}. 
We denote by $\expec{\cdot}_{s}$ or ${\mathbb E}_s$
the expectation with respect to the
site variables alone.  
Also we use $\expec{\cdot}_{b}$  or ${\mathbb E}_{b}$
to denote expectation with respect to the 
bond variables.  The joint expectation is then given by 
$\expec{\expec{\cdot}_{s}}_{b}$

By definition,
\eqalign 
	\expec{I[E_0'(0, u)]}_{s} & = 
	I[E_{0, b}'(0, u)] \, e^{-h|C(0)|} 
	\\
	\expec{I[E_0''(0, u', v')]}_{s} & = 
	I[E_{0, b}''(0, u', v')] \, e^{-h|\tilde{C}^{\{u', v'\}}(0)|} 
	\\
	\expec{I[F_1'(v, x)]}_{s} & = 
	I[F_{1, b}'(v, x)] \, e^{-h|C(v)|} 
	\\
	\expec{I[F_1''(v, u', v')]}_{s} & = 
	I[F_{1, b}''(v, u', v')] \, e^{-h|\tilde{C}^{\{u', v'\}}(v)|} . 
\enalign
Recalling \refeq{Cnabbr}, we introduce the abbreviations
\eq
	Y_{n,b}   =  I[F_{1, b}  (v_{n-1}, x; \tilde{C}_{n-1})], \quad  
	Y_{n,b}'  =  I[F_{1, b}' (v_{n-1}, x; \tilde{C}_{n-1})],  \quad 
	Y_{n,b}'' =  I[F_{1, b}''(v_{n-1}, u_{n}, v_{n}; \tilde{C}_{n-1})].
\en 
We also write $C_n = C(v_{n-1})$.  
Then we have 
\begin{align} 
	\phi_{h,p}^{(0)} (0,x) & =   
	{\mathbb E}_{0,b}[I[E_{0, b}'(0,x)] e^{-h|C_0|} ] ,
	\lbeq{f0-bond}
	\\
	\phi_{h,p}^{(n)} (0,x) & =  (-1)^{n} 
	{\mathbb E}_{0,b} I[E_{0, b}''] e^{-h|\tilde{C}_0|} \, 
	{\mathbb E}_{1,b} Y_{1, b}'' e^{-h|\tilde{C}_1|} \, 
	\cdots 
	{\mathbb E}_{n-1,b}Y_{n-1, b}'' e^{-h|\tilde{C}_{n-1}|} \, 
	{\mathbb E}_{n,b} Y_{n, b}' e^{-h|C_{n}|}  ,
	\lbeq{fn-bond}
	\\
	\Phi_{h,p}^{(0)} (0, v_0) & = 
	p \sum_{u_{0}} 
	{\mathbb E}_{0,b}[I[E_{0, b}''(0, u_0, v_0)] e^{-h|\tilde{C}_0|} ] ,
	\lbeq{Phi0-bond}
	\\
	\Phi_{h,p}^{(n)} (0, v_{n}) & =  (-1)^{n} \, 
	p \sum_{u_{n}} 
	{\mathbb E}_{0,b} I[E_{0, b}''] e^{-h|\tilde{C}_0|} \, 
	{\mathbb E}_{1,b} Y_{1, b}'' e^{-h|\tilde{C}_1|} \, 
	\cdots 
	{\mathbb E}_{n-1,b} Y_{n-1, b}'' e^{-h|\tilde{C}_{n-1}|} \, 
	{\mathbb E}_{n,b} Y_{n, b}'' e^{-h|\tilde{C}_{n}|}  . 
	\lbeq{Phin-bond}
\end{align}

%%%%%%%%%%%%%%%%%%%%%%%%%%%%%%%%%%%%%%%%%%%%%%%%%%%%%%%%%%%%%%%%%%%%%%%%%%%%%%%
\subsubsection{Bounds involving $\hat{\phi}^{(n)}$}
\label{subsub-bounds.phihat}

We begin with the simplest case $n=01$.
For $x \neq 0$ we have
\eq
	\phi_{h,p}^{(0)} (0,x) = {\mathbb E}_{0,b}[I[0 \dbc x] e^{-h|C(0)|}] 
	\leq e^{-2h} {\mathbb E}_{0,b}I[0 \dbc x]  
	\leq e^{-2h} \tau_{0,p}(0,x)^{2} ,
	\lbeq{f00-bd0}
\en 
using the BK inequality and $|C(0)| \geq 2$.  We thus have 
\eq
	\left | \hat{\phi}_{h,p}^{(01)} (k) \right | 
	\leq  
	\sum_{x \neq 0} \phi_{h,p}^{(0)} (0,x) 
	\leq \sum_{x \neq 0} e^{-2h} \tau_{0,p}(0,x)^{2}
	= e^{-2h} \pol{2}_{0,p}(0) = O(\oneOd) e^{-2h} .
	\lbeq{f00-bd}
\en 
Similarly, using the lattice symmetry and \refeq{f00-bd0} we obtain
\eqsplit 
	\hat{\phi}_{h,p}^{(01)} (0) 	- \hat{\phi}_{h,p}^{(01)} (k)  
	& =  
	\sum_{x \neq 0} \phi_{h,p}^{(0)} (0,x) [ 1 - \cos k \cdot x ]  
	\\ 
	& \leq  \sum_{x \neq 0} \phi_{h,p}^{(0)} (0,x) \frac{k^2 x^2}{2d}
	\leq  e^{-2h} \frac{k^2}{2d} \wpol{2}_{0,p}(0) 
	= O(\oneOd) e^{-2h} [1 - \Dhat(k)]  .
\ensplit 

For $n \geq 1$, each expectation in $|\hat{\phi}^{(n)}_{h,p}(k)|$
involves at least one 
factor of $e^{-h}$, since $C$ or $\tilde{C}$ cannot be empty. 
Bounding each of these using $e^{-h|\tilde{C}|} \leq e^{-h}$, we obtain 
\eq
	\phi_{h,p}^{(n)} (0,x) \leq e^{-h(n+1)} 
	{\mathbb E}_{0,b} I[E_{0, b}''] \,   
	{\mathbb E}_{1,b} Y_{1, b}'' \, 
	\cdots 
	{\mathbb E}_{n-1,b} Y_{n-1, b}''  \, 
	{\mathbb E}_{n,b} Y_{n, b}' .
\en 
The resulting bond expectation was treated in 
\cite{HS90a}, and can be bounded using the critical
triangle diagram $\pol{3}_{0,p_c}$, yielding 
\eq
        |  \hat{\phi}^{(n)}_{h,p}(k) | \leq O(\oneOd^{n} ) e^{-h(n+1)}  
	\quad (n \geq 1) .
	\lbeq{fhatbd.2}
\en  
Because similar 
diagrammatic estimates will be required 
repeatedly in the rest of the paper, we recall the main ideas entering
into the proof of \refeq{fhatbd.2}.  Further details can be found
in \cite{HS90a}.  There are two main steps:
(1) We first bound the nested expectation in terms of 
    	$\tau_{h,p}$, from right to left. 
	The original nested expectation is thus bounded by a sum of 
    	products of $\tau_{h,p}$, which can be represented by 
    	diagrams.  
(2) We estimate the resulting expression by decomposing it into triangles. 

\medskip
\noindent\emph{Step~1: Bounds on building blocks.}

\nopagebreak 

We bound the nested expectation from right to left, starting 
with $\expec{Y_{n,b}'}_{n,b}$.  For this expectation,
we first note that   
\eq
	F_{1,b}'(v_{n-1}, x; \tilde{C}_{n-1})_{n} 
	\subset 
	\bigl \{v_{n-1}\conn x 
	\AND v_{n}' \ct{\tilde{C}_{n-1}} x ] \bigr \}_{n} , 
\en
where we used the subscript $n$ to emphasize we are considering level-$n$ 
connections, and 
$(u_{n}', v_{n}')$ denotes the last pivotal bond for the connection 
$v_{n-1} \to x$ (if it does not exist, we set $v_{n}' = v_{n-1}$). 
This is a subset of the event 
\eq  
	\bigcup_{w_{n-1}, v_{n}' \in \Zd} 
	\left[ 
	\bar{F}_{1,b}'(v_{n-1}, x, w_{n-1}, v_{n}')_{n} 
	\cap \bigl \{w_{n-1} \in \tilde{C}_{n-1} \bigr \}
	\right] ,
	\lbeq{F1pnbd.1}
\en  
where 
\eqalign 
	\bar{F}_{1,b}'(v_{n-1}, x, w_{n-1}, v_{n}') 
	& =  
	\bigl \{
	(v_{n-1} \conn v_{n}') \circ (v_{n}' \conn w_{n-1}) \circ 
	(w_{n-1} \conn x) \circ (v_{n}' \conn x) 
	\bigr \} 
	\nnb 
	& = 
	\biggl \{ \picFonepNconn
	\biggr \}
	\lbeq{F1bar-def}. 
\enalign
In \refeq{F1bar-def}, we have introduced a suggestive diagrammatic notation 
for events, in which thin lines represent disjoint connections 
between vertices.  

Now we continue to estimate \refeq{F1pnbd.1}, 
using the BK inequality.  We have 
\eqalign 
	&
	\expec{I[F_{1,b}'(v_{n-1}, x; \tilde{C}_{n-1})_{n}]}_{n} 
	\leq 
	\sum_{w_{n-1} \in \Zd} I[ w_{n-1} \in \tilde{C}_{n-1} ]\, 
	\expec{I[\bar{F}_{1,b}'
		(v_{n-1}, u_{n}, w_{n-1}, v_{n}')]}_{n}
	\nnb 
	& \quad 
	\leq 
	\sum_{w_{n-1} \in \tilde{C}_{n-1}}  
	\sum_{v_{n}'} 
	\tau_{0,p}(v_{n-1}, v_{n}') \, \tau_{0,p}(v_{n}', w_{n-1})  \, 
	\tau_{0,p}(w_{n-1}, x) \, \tau_{0,p}(v_{n}', x) 
	= 
	\sum_{w_{n-1} \in \tilde{C}_{n-1}}  
	\Biggl [ \picFonepNbd
	\Biggr ] 
	, 
	\lbeq{F1pbd.2}
\enalign 
where on the right side, thick lines represent 
factors of $\tau_{0,p}$, and summation over $\Zd$ 
is implicit over the unlabelled vertex.  This is the desired bound on the 
level-$n$ expectation. 

Next, we consider the expectation at level-$(n-1)$.  Here we 
have two conditions: the event $F_{1,b}''$ coming from $Y_{n-1,b}''$, and 
the requirement 
$w_{n-1}\in \tilde{C}_{n-1}$ which has just been produced in the 
process of  bounding the level-$n$ expectation.  
Our goal is to bound the right side of 
\eqalign 
	& \Bigl \langle 
	Y_{n-1,b}'' \, I[w_{n-1} \in  \tilde{C}_{n-1}]
	\Bigr \rangle_{n-1,b} 
	\leq 
	\Bigl \langle  
	I[v_{n-2}\conn u_{n-1} 
	\AND v_{n-1}' \ct{\tilde{C}_{n-2}} u_{n-1}
	\AND w_{n-1} \in  \tilde{C}_{n-1}]
	\Bigr \rangle_{n-1,b} . 
	\nnb 
\intertext{%%%%%  intertext begins
This can be further bounded by the following 
(essentially, 
we add a connection $v_{n-2} \conn w_{n-1}$ to the diagram 
of \refeq{F1bar-def}): 
} %%%%% intertext ends
	& \hspace{15mm}
	\sum_{w_{n-2} \in \tilde{C}_{n-2}}
	\biggl \langle I \biggl [ \picFoneppwNMONEa
	\biggr ] 
	\; 
	+ 
	\; 
	I 
	\biggl [ \picFoneppwNMONEb
	\biggr ] 
	\biggr \rangle_{n-1, b}
	\lbeq{F1ppbd.event} . 
	\\ 
\intertext{By the BK inequality, this is bounded above by} 
	& \hspace{15mm}	
	\sum_{w_{n-2} \in \tilde{C}_{n-2}}
	\Biggl \{ \thicklines \picFoneppwNMONEa
	\; 
	+ 
	\; \picFoneppwNMONEb
	\Biggr \} .  
	\lbeq{F1ppbd.2}
\enalign 
This is the desired bound for level-$(n-1)$. 

The remaining expectations are bounded in a similar fashion, until
we reach level-$0$. Arguing as above, it is bounded by 
\eq
	\expec{I[E_{0}''] \, I[w_{0} \in \tilde{C}_{0}]}_{0,b} 
	\leq 
	\Biggl \langle I \biggl [ 
	\thinlines \picFzeroppw
	\biggr ]
	\Biggr \rangle_{0,b}
	\leq 
	\thicklines \picFzeroppw 
	.
\en 
Combining the above, we can bound $\hat{\phi}^{(n)}_{h,p}(0)$ for any $n$.   
For example, $\hat{\phi}^{(2)}_{h,p}(0)$ is bounded by the sum of two terms: 
\eq
\lbeq{f2bd.res}
	\sum_{x} \phi^{(2)}_{h,p} (x) \leq 
	e^{-3h}
	\sum_{w_{0}, w_{1}, x} \, 
	\myeqnpic{(200,100)}{(-50, 0)}{ %begin my-eqn-pic %%%%
	    \thicklines
		\put(  0,20){\pictril} 
		\put( 50,20){\pictrim}
		\put(100,20){\pictrir}
		\put(  0,80){\picpivone}
		\put( 65,20){\picpivtwo}
    		\drawline(  0,20)( 35,20)
    		\drawline( 10,80)(100,80)
    		\drawline( 75,20)(100,20)
    		\put(-50,40){$0$}
    		\put( 90, 90){$w_{1}$}
    		\put(-10, 0){$w_{0}$}
   		\put(140,40){$x$}
	} % end of my-eqn-pic %%%%%%%%%%%%%%%%%%%%%%%%%%%%%%%%
	\, + \, e^{-3h}
	\sum_{w_{0}, w_{1}, x} \, 
	\myeqnpic{(250,100)}{(-50, 0)}{ %begin my-eqn-pic %%%%
	    \thicklines
		\put(  0,20){\pictril} 
		\put(150,20){\pictrir}
		\put(  0,80){\picpivone}
		\put(100,20){\picpivtwo}
    		\drawline(  0,20)(100,20)
    		\drawline(110,20)(150,20)
    		\drawline( 10,80)(150,80)
    		\drawline( 50,20)( 50,80)
    		\drawline(100,20)(100,80)
    		\put(-50,40){$0$}
    		\put(140,90){$w_{1}$}
    		\put(-10,-5){$w_{0}$}
   		\put(190,40){$x$}
	} % end of my-eqn-pic %%%%%%%%%%%%%%%%%%%%%%%%%%%%%%%% 
	.
\en 
In the above, a pair of thick lines represents a (pivotal) bond, and summation 
over all unlabelled vertices, including pivotal bonds, is understood. 
Each pivotal bond also carries a factor $p$.

\medskip
\noindent\emph{Step~2: Decomposition of the diagrams.} 

\nopagebreak 

For $n=2$, we illustrate the method for estimating diagrams 
via a decomposition into triangles.  The basic tool is the simple inequality 
\eq
	\sum_{x} f(x) g(x) \leq \bigl [ \sup_{x} f(x) \bigr ] 
	\, \sum_{x} g(x) ,
	\qquad \textrm{for } f(x),  \, g(x) \geq 0 . 
	\lbeq{basic.ineq}
\en 
Applying \refeq{basic.ineq} and translation invariance, 
the first diagram (including the summation) 
of \refeq{f2bd.res} is bounded by
\eqalign
	& 
	\Biggl [ \sup_{w_{1}} 
	\myeqnpic{(150,100)}{(-35, 0)}{ %begin my-eqn-pic %%%%
	    \thicklines
		\put(  0,20){\pictril} 
		\put( 50,20){\pictrim}
		\put(  0,80){\picpivone}
		\put( 65,20){\picpivtwo}
    		\drawline(  0,20)( 35,20)
    		\drawline( 10,80)(100,80)
    		\drawline( 75,20)(100,20)
    		\put( 90, 90){$w_{1}$}
    		\put(100,  0){$0$}
	} % end of my-eqn-pic %%%%%%%%%%%%%%%%%%%%%%%%%%%%%%%%
	\Biggr ] \, \, 
	\Biggl [ 
	\sum_{w_{1}} 
	\myeqnpic{(50,100)}{(-10, 0)}{ %begin my-eqn-pic %%%%
	    \thicklines
		\put(  0,20){\pictrir}
    		\put(-10,90){$w_{1}$}
    		\put(  0,-5){$0$}
	} % end of my-eqn-pic %%%%%%%%%%%%%%%%%%%%%%%%%%%%%%%%
	\Biggr ] 
	\leq 
	\Biggl [ 
	\sup_{w_{3}}
	\myeqnpic{(60,100)}{(-35, 0)}{ %begin my-eqn-pic %%%%
	    \thicklines
		\drawline(0,20)(-30,50)(0, 80)
     		\put(-10,90){$w_{3}$}
   		\put(  0, 0){$0$}
	} % end of my-eqn-pic %%%%%%%%%%%%%%%%%%%%%%%%%%%%%%%%
	\Biggr ] 
	\, 
	\Biggl [ 
	\sup_{w_{2}}
	\myeqnpic{(80,100)}{(-5, 0)}{ %begin my-eqn-pic %%%%
	    \thicklines
		\drawline(0,80)(0,20)(30,20)
		\drawline(10,80)(50,80)
		\put(  0,80){\picpivone}
     		\put( 40,90){$w_{2}$}
   		\put( 30, 0){$0$}
	} % end of my-eqn-pic %%%%%%%%%%%%%%%%%%%%%%%%%%%%%%%%
	\Biggr ] 
	\, 
	\Biggl [ 
	\sup_{w_{1}} 
	\myeqnpic{(90,100)}{(30, 0)}{ %begin my-eqn-pic %%%%
	    \thicklines
		\put( 50,20){\pictrim}
		\put( 65,20){\picpivtwo}
    		\drawline( 50,80)(100,80)
    		\drawline( 75,20)(100,20)
    		\put( 90, 90){$w_{1}$}
    		\put(100,  0){$0$}
	} % end of my-eqn-pic %%%%%%%%%%%%%%%%%%%%%%%%%%%%%%%%
	\Biggr ] \, \, 
	\Biggl [ 
	\myeqnpic{(50,100)}{(-10, 0)}{ %begin my-eqn-pic %%%%%
	    \thicklines
		\put(  0,20){\pictrir}
    		\put(  0,-5){$0$}
	} % end of my-eqn-pic %%%%%%%%%%%%%%%%%%%%%%%%%%%%%%%%
	\Biggr ] .
\enalign 
 
By Lemma~\ref{lem-triangles}, the factors on the right side obey 
\eqalign
	\myeqnpic{(50,100)}{(-10, 0)}{ %begin my-eqn-pic %%%%%
	\begin{picture}(50,100)(-10,0)
	    \thicklines
		\put(  0,20){\pictrir}
    		\put( 10, 0){$0$}
	\end{picture}
	} % end of raisebox, picture 
	%%%%%%%%%%%%%%%%%%%%%%%%%%%%%%%%%%%%%%%%%%%%%%%%%%%%%%%%	
	& 
	\leq 
	1+ \pol{3}_{0,p_{c}}(0)  = 1 + O(\oneOd), 
	\lbeq{bd-triangle}
	\\
	\myeqnpic{(80,100)}{(-5, 0)}{ %begin my-eqn-pic %%%%%%%%%
	    \thicklines
		\drawline(0,80)(0,20)(30,20)
		\drawline(10,80)(50,80)
		\put(  0,80){\picpivone}
     		\put( 40,90){$w_{2}$}
   		\put( 30, 0){$0$}
	} % end of my-eqn-pic %%%%%%%%%%%%%%%%%%%%%%%%%%%%%%%%%
	& 
	= 
	p \sum_{(0,v)} 
	\myeqnpic{(80,100)}{(-5, 0)}{ %begin my-eqn-pic %%%%%%%%
	    \thicklines
		\drawline(50,80)(0,80)(0,20)(30,20)
     		\put( 30,90){$w_{2}$}
   		\put( 30, 0){$v$}
	} % end of my-eqn-pic %%%%%%%%%%%%%%%%%%%%%%%%%%%%%%%%%%
	\leq 
	p \sum_{(0,v)} [ \pol{3}_{0,p_{c}}(w_{2} -v) + \delta_{w_{2}, v} ]
	\leq p \Omega O(\oneOd) + p 
	= O(\oneOd) , 
	\lbeq{bd-optriangle}
	\\
	\myeqnpic{(50,100)}{(-30, 0)}{ %begin my-eqn-pic %%%%%%%
	    \thicklines
		\drawline(0,20)(-30,50)(0, 80)
     		\put(-10,90){$w_{3}$}
   		\put(  0, 0){$0$}
	} % end of my-eqn-pic %%%%%%%%%%%%%%%%%%%%%%%%%%%%%%%%%%%
	& 
	\leq 
	\delta_{w_{3}, 0} + \pol{2}_{0,p_{c}}(w_{3}) 
	\leq  1 + O(\oneOd) ,
	\lbeq{bd-opbubble}
	\\
	\myeqnpic{(90,100)}{(30, 0)}{ %begin my-eqn-pic %%%%%%%
	    \thicklines
		\put( 50,20){\pictrim}
		\put( 65,20){\picpivtwo}
    		\drawline( 50,80)(100,80)
    		\drawline( 75,20)(100,20)
    		\put( 90, 90){$w_{1}$}
    		\put(100,  0){$0$}
	} % end of my-eqn-pic %%%%%%%%%%%%%%%%%%%%%%%%%%%%%%%%
	& 
	\leq 
	\Biggl [ 
	\myeqnpic{(50,100)}{(-20, 0)}{ %begin my-eqn-pic %%%%
	    \thicklines
    		\drawline(-15,0)(0,30)(15,0)(-15,0)
    		\put(20,  0){$0$}
	} % end of my-eqn-pic %%%%%%%%%%%%%%%%%%%%%%%%%%%%%%%%
	\Biggr ] \, 
	\Biggl [ 
	\sup_{w} 
	\myeqnpic{(80,100)}{(-5, 0)}{ %begin my-eqn-pic %%%%%%%
	    \thicklines
		\put( 0,20){\picpivtwo}
    		\drawline( 50,80)(0,80)(0,20)
		\drawline(10,20)(50,20)
    		\put( 40, 90){$w$}
    		\put( 50,  0){$0$}
	} % end of my-eqn-pic %%%%%%%%%%%%%%%%%%%%%%%%%%%%%%%%
	\Biggr ] 
	= [ 1 + O(\oneOd) ] \, O(\oneOd) 
	= O(\oneOd) .
\enalign 
Thus the first diagram of \refeq{f2bd.res} is bounded 
by $[ 1 + O(\oneOd) ]^{2} O(\oneOd)^{2}  = O(\oneOd^{2}) $. 
Similarly, the second diagram of \refeq{f2bd.res} is bounded by 
\eq
	\Biggl [ 
	\sup_{w_{4}}
	\myeqnpic{(50,100)}{(-30, 0)}{ %begin my-eqn-pic %%%%%%%
	    \thicklines
		\drawline(0,20)(-30,50)(0, 80)
     		\put(-10,90){$w_{4}$}
	} % end of my-eqn-pic %%%%%%%%%%%%%%%%%%%%%%%%%%%%%%%%%%
	\Biggr ] 
	\, 
	\Biggl [ 
	\sup_{w_{3}}
	\myeqnpic{(80,100)}{(-5,0)}{ %begin my-eqn-pic %%%%%%%
	    \thicklines
		\drawline(50,80)(10,80)
		\drawline(0,80)(0,20)(50,20)
		\put(  0,80){\picpivone}
     		\put( 40,90){$w_{3}$}
   		\put( 50, 0){$0$}
	} % end of my-eqn-pic %%%%%%%%%%%%%%%%%%%%%%%%%%%%%%%%%
	\Biggr ] 
	\, 
	\Biggl [ 
	\sup_{w_{2}}
	\myeqnpic{(80,100)}{(-5,0)}{ %begin my-eqn-pic %%%%%%%%%
	    \thicklines
		\drawline(50,80)(0,80)(0,20)(50,20)
     		\put( 40,90){$w_{2}$}
   		\put( 50, 0){$0$}
	} % end of my-eqn-pic %%%%%%%%%%%%%%%%%%%%%%%%%%%%%%%%%
	\Biggr ] 
	\, 
	\Biggl [ 
	\sup_{w_{1}} 
	\myeqnpic{(80,100)}{(-5,0)}{ %begin my-eqn-pic %%%%%%%%%
	    \thicklines
		\put(0,20){\picpivtwo}
    		\drawline(50,80)(0,80)(0,20)
		\drawline(10,20)(50,20)
    		\put(40, 90){$w_{1}$}
    		\put(50,  0){$0$}
	} % end of my-eqn-pic %%%%%%%%%%%%%%%%%%%%%%%%%%%%%%%%%%%
	\Biggr ] \, \, 
	\Biggl [ 
	\myeqnpic{(50,100)}{(-10, 0)}{ %begin my-eqn-pic %%%%%%%
	    \thicklines
		\put(  0,20){\pictrir}
    		\put(  5, 0){$0$}
	} % end of my-eqn-pic %%%%%%%%%%%%%%%%%%%%%%%%%%%%%%%%%%
	\Biggr ] 
	= O(\oneOd^{2}) 
	.
\en

A similar analysis can be carried out for other values of $n$,
leading to \refeq{fhatbd.2}. 
 
Finally, we consider the bound on
$| \hat{\phi}^{(n)}_{h,p}(0) - \hat{\phi}^{(n)}_{h,p}(k) |$.  
For this, we write 
$\hat{\phi}^{(n)}_{h,p}(0) - \hat{\phi}^{(n)}_{h,p}(k) 
= \sum_{x} \phi^{(n)}_{h,p} (0,x) [ 1 - \cos (k\cdot x) ]$ 
and bound $\phi^{(n)}_{h,p} (0,x)$ as above. Now in step 2, we use
both the triangle and the weighted bubble diagrams at
$p_c$, together with the bound $1-\cos(k \cdot x) \leq k^2x^2/2d$,
with the result
\eq 
	| \hat{\phi}^{(n)}_{h,p}(0) - \hat{\phi}^{(n)}_{h,p}(k) | 
	\leq O(\oneOd^{n}) e^{-h(n+1)}\, [1-\Dhat(k)] , \quad (n \geq 1) . 
\en  
This completes the proof of \refeq{fnbd}.

%%%%%%%%%%%%%%%%%%%%%%%%%%%%%%%%%%%%%%%%%%%%%%%%%%%%%%%%%%%%%%%%%%%%%%%%%%%%%%%
\subsubsection{Bounds involving $\hat{\Phi}^{(n)}$ }

The bounds  \refeq{Phinbd}
on $\Phihat$ can be obtained in the same way.  The only difference
between $\hat{\phi}^{(n)}$ and $\hat{\Phi}^{(n)}$ is in the level-$n$ 
expectation, which involves $F_1'$ for $\hat{\phi}^{(n)}$ and 
$F_1''= \{ F_{1}' (v_{n-1}, u_{n}; A) \ON \tilde{C}_{n}  \}$ for 
$\hat{\Phi}^{(n)}$.  Since  $F_{1}''$ is a subset of the event \refeq{F1pnbd.1}, 
the bounds for $\hat{\phi}^{(n)}$ also apply for $\hat{\Phi}^{(n)}$,
apart from a factor $p \Omega \leq 1+O(\lambda)$ due to the sum over $u_{n+1}$.  

We turn now to the remaining bound \refeq{Phinbd.2}, which involves
the extraction of a factor $M_{h,p}$.  
By definition,
\eq
\lbeq{Phi010h}
	\hat{\Phi}^{(01)}_{0,p}(0) - \hat{\Phi}^{(01)}_{h,p}(0)
	= p \sum_{(u_0, v_0): u_0 \neq 0} 
	\expec{I[E_{0,b}''(0, u_0, v_0)] \, 
	[1 - e^{-h|\tilde{C}(0)|} ]}_b .
\en
This is bounded above by $p\sum_{(u_0, v_0): u_0 \neq 0} 
\prob{0 \dbc u_0 \AND 0 \conn G}$.  But the event in this expression is
contained in the event that there is a $w \in \Zd$ such that
$\{0 \conn u_0\} \circ \{0 \conn w \} \circ \{ w \conn u_0\} 
\circ \{w \conn G\}$, and hence, as required, \refeq{Phi010h} is bounded by
\eq
	p \sum_{(u_0, v_0): u_0 \neq 0} \sum_w 
	\tau_0(0, w) \tau_0(w, u_0) \tau_0(u_0, 0) M_{h,p} 
	\leq p \Omega O(\oneOd) M_{h,p} .  
\en

For $n \geq 1$, we can proceed in a similar fashion.  For simplicity, 
we illustrate the argument for $n=2$, for which 
\eq
\lbeq{Phiz(2)}
	\Phi_{h,p}^{(2)}(0, u_2, v_2) = 
	\expec{I[E_{0}''] \expec{Y_1'' \expec{Y_2''}_2 }_1 }_0 .
\en 
We begin by writing the difference 
$\Phi_{0,p}^{(2)}(0, u_2, v_2)- \Phi_{h,p}^{(2)}(0, u_2, v_2)$ 
as a telescoping sum.
To abbreviate the notation, we denote the nested expectation \refeq{Phiz(2)}
by $(0'' 1'' 2'')_h$.  Then  
\eqarray
        (0'' 1'' 2'') _{h=0} - (0'' 1'' 2'') _{h} 
        & = & \bigl [ (0'' 1'' 2'')_{h=0} - (0'')_{h} (1'' 2'')_{h=0} \bigr ]
        + \bigl [ (0'')_{h} (1'' 2'')_{h=0} - (0'' 1'')_{h} (2'')_{h=0} \bigr ]
        \nonumber \\
        & & + \bigl [ (0'' 1'')_{h} (2'')_{h=0} - (0'' 1'' 2'')_{h} \bigr ] .
\enarray 
The three terms on the right side are treated similarly. 
For example, the second term is given by 
\eq 
	(0'')_{h} (1'' 2'')_{h=0} - (0'' 1'')_{h} (2'')_{h=0} 
	= 
	\expec{I[E_{0, b}''] e^{-h|\tilde{C}_0(0)|} \, 
	\bigl \langle Y_{1, b}'' (1 - e^{-h|\tilde{C}_1(v_0)|} ) \, 
	\langle Y_{2, b}''  
	\rangle_{2, b} 
	\bigr \rangle_{1, b} 
	}_{0, b} . 
\en 
The innermost expectation can be bounded, as in \refeq{F1pbd.2}, by
\eq
	\langle Y_{2, b}''  
	\rangle_{2, b} \leq 
	\sum_{w_{1} \in \tilde{C}_{1}} 
	p \sum_{(u_{2}, v_{2})}
	\biggl [ 
	\myeqnpic{(140,100)}{(-35, 0)}{ %begin my-eqn-pic %%%%%%%
	    \thicklines
    		\drawline( 0,80)(50,80)
		\put( 50, 20){\pictrir} 
    		\put(-30,90){$v_{1}$}
    		\put( 20, 5){$w_{1}$}
    		\put( 85,45){$u_{2}$}
	} % end of my-eqn-pic %%%%%%%%%%%%%%%%%%%%%%%%%%%%%%%%%%%
	\biggr ] .  		
\en 
In the middle expectation, 
the factor $1 - e^{-h|\tilde{C}_1(v_0)|}$ can be 
interpreted as a requirement that
$\tilde{C}_1(v_0)$ should be connected to $G$, so that 
\eq
	\bigl \langle Y_{1, b}'' \, 
	(1 - e^{-h|\tilde{C}_1(v_0)|} ) \, 
	I[w_{1} \in \tilde{C}_{1}] \, 
	\bigr \rangle_{1} 
	\leq 
	\bigl \langle Y_{1,b}'' \, 
	I[v_{0} \conn G \AND w_{1} \in \tilde{C}_{1}] \, 
	\bigr \rangle_1 . 
\en 
Using the bound of \refeq{F1ppbd.event} 
for $Y_{1,b}'' I[w_{1} \in \tilde{C}_{1}]$, this is bounded above by
\eq
	\sum_{w_{0} \in \tilde{C}_{0}}
	\Biggl \langle 
	\biggl \{
	I \biggl [ 
	\myeqnpic{(150,100)}{(-75,0)}{ %begin my-eqn-pic %%%%%%
    		\drawline(-50,80)(50,80)
		\put(  0, 20){\pictrim}
    		\put(-70, 90){$v_{0}$}
    		\put( 40, 90){$w_{1}$}
    		\put(-50,  0){$w_{0}$}
    		\put( 30,  0){$u_{1}$}
	} % end of my-eqn-pic  %%%%%%%%%%%%%%%%%%%%%%%%%%%%%%%%%
	\biggr ] 
	\; 
	+ 
	\; 	I 
	\biggl [ 
	\myeqnpic{(180,100)}{(-25, 0)}{ %begin my-eqn-pic %%%%%
    		\drawline(  0,80)(150,80)
   		\drawline( 50,20)(100,20)
    		\drawline( 50,20)( 50,80) 
    		\drawline(100,20)(100,80) 
    		\put(-20,90){$v_{0}$}
    		\put(130,90){$w_{1}$}
    		\put( 10, 0){$w_{0}$}
    		\put(100, 0){$u_{1}$}
	} % end of my-eqn-pic  %%%%%%%%%%%%%%%%%%%%%%%%%%%%%%%%
	\biggr ] 
	\biggr \} \, 
	I [ v_{0} \conn G ]
	\Biggr \rangle .
	\lbeq{F1ppdiff.bd1}
\en 

\begin{figure}
\begin{center}
\includegraphics[scale = 0.5]{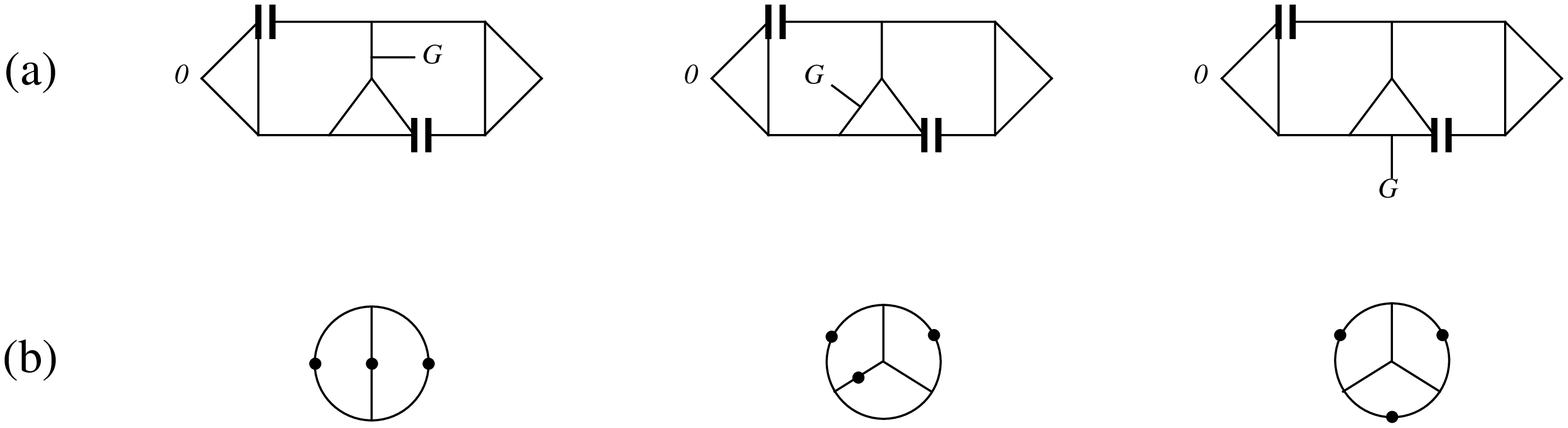} 
\end{center}
\caption{
(a) Examples of diagrams arising in bounding \refeq{F1ppdiff.bd1}.
(b) Feynman diagrams arising in bounding these diagrams.}
\label{fig-Phi-sec3}
\end{figure}

Compared with \refeq{F1ppbd.2}, there is now an extra condition 
$v_{0} \conn G$.  
This connection to $G$ corresponds diagrammatically to the addition
of a vertex from which a connection to $G$ emerges.  We proceed as
in the previous diagrammatic bounds, using the BK
inequality. A factor $M_{h,p}$ arises from the connection to $G$.
This factor is multiplied by a sum of diagrams.
Explicitly, the diagrams are those obtained by adding an extra vertex
to any one of the fourteen lines in each of the two diagrams appearing
on the right side of \refeq{f2bd.res}.  These diagrams can then
be bounded in terms of the triangle diagram, apart from a few  
cases where the triangle alone is insufficient
to estimate the diagrams.  Three such
cases are depicted in Figure~\ref{fig-Phi-sec3}, 
together with resulting Feynman diagrams that cannot be reduced
to triangles.  These irreducible diagrams
can be bounded using the square diagram for the 
nearest-neighbour model in sufficiently high dimensions.
For the spread-out model, we illustrate the argument for the leftmost
diagram in Figure~\ref{fig-Phi-sec3}.  This diagram results from
construction~2 of Section~\ref{sub-ind.1} applied to the triangle,
and is therefore finite for $d>6$ by \refeq{cor-Gind} and 
Theorem~\ref{thm-R1}.  Moreover, it converges to 1 as $L \to \infty$,
by an application of the dominated convergence theorem as in
\cite[Lemma~5.9]{HS90a}.  However, the contribution leading to the limiting
value 1 arises from the case where the lines in the Feynman diagram
all contract to a point, and this contribution was not present originally
and need not be included in the bound.  Thus the diagram can be bounded
by $O(\oneOd^{2})$, where we increase $\oneOd$ if necessary to achieve this.
The overall result is 
\eq
	| \Phihat_{0,p}^{(2)}(0) - \Phihat_{h,p}^{(2)}(0) | \leq 
	O(\oneOd^{2})M_{h,p}. 
\en 

%%%%%FIGFIGFIGFIGFIGFIGFIGFIGFIGFIGFIGFIGFIGFIGFIGFIGFIGFIGFIGFIG
%%%%%FIGFIGFIGFIGFIGFIGFIGFIGFIGFIGFIGFIGFIGFIGFIGFIGFIGFIGFIGFIG
%%%%FIGFIGFIGFIGFIGFIGFIGFIGFIGFIGFIGFIGFIGFIGFIGFIGFIGFIGFIGFIG
%%%%FIGFIGFIGFIGFIGFIGFIGFIGFIGFIGFIGFIGFIGFIGFIGFIGFIGFIGFIGFIG

Similar bounds can be obtained for general $n \geq 1$, yielding
the bound
\eq 
        | \Phihat_{0,p}^{(n)}(0) - \Phihat_{h,p}^{(n)}(0) | \leq 
        O( \oneOd^{n} ) M_{h}    
\en   
of \refeq{Phinbd.2}. 
This completes the proof of Lemma~\ref{lem-Phinbd}.
\qed 

The method of proof of Lemma~\ref{lem-Phinbd} also gives the bound
\eq
\lbeq{x2Phin}
	\sum_x |x|^2 |\Phi^{(\bar{n})}_{h,p}(0,x)| \leq O(\oneOd^{\bar{n}}).
\en
Arguing as in the proof of Lemma~\ref{lem-Pi00bd}, we also have
\eq
\lbeq{x2Phi00}
	\sum_x |x|^2 |\Phi^{(00)}_{h,p}(0,x)| 
	= p\Omega \left(-\nabla^2_k \hat{D}(0) \right) 
		(1-M_{h,p}) + O(\oneOd).
\en
Therefore, for $p \leq p_c$,
\eq
\lbeq{nablaPhibd}
	-\nabla_k^2 \sum_{n=0}^\infty \hat{\Phi}_{0,p}^{(n)}(0) 
	= p\Omega \left(-\nabla^2_k \hat{D}(0) \right) + O(\oneOd).
\en

%%%%%%%%%%%%%%%%%%%%%%%%%%%%%%%%%%%%%%%%%%%%%%%%%%%%%%%%%%%%%%%%%%%%%%%%%%%%%%%
%%%%%%%%%%%%%%%%%%%%%%%%%%%%%%%%%%%%%%%%%%%%%%%%%%%%%%%%%%%%%%%%%%%%%%%%%%%%%%%
\subsection{The cut-the-tail lemma and bounds on the remainder} 
\label{sub-bd.rem}

The following lemma will be used to bound the remainder term
$\hat{\rho}^{(n)}_{h,p}(k)$ of \refeq{rhodef}.  
It will be used again in Sections~\ref{sec-2M.exp.err} and \ref{sec-Psidef}.  
The lemma is called the ``cut-the-tail'' lemma,
because it is used to cut off a $G$-free connection between two points,
at a pivotal bond.  Its proof is deferred to Section~\ref{sec-cut-the-tail}.

\begin{lemma}
\label{lem-tail.tau1}
Let $x$ be a site, $\{u,v\}$ a bond, and $E$ an increasing event. 
Then for a set of sites $A$ with $A \ni u$, 
and for $p \leq p_{c}$, $h \geq 0$ (assuming no infinite cluster
when $(h,p)=(0,p_c)$), 
\eq 
	\expec { \Ind [ E \ON \tilde{C}^{\{u,v\}}(A) ]  \,  
	\tau_{h,p}^{\tilde{C}^{\{u,v\}}(A)} (v, x) }
	\leq \frac{1}{1 - p M_{h,p}} \, \prob{E} \, 
	\tau_{h,p}(v,x) .
\en 
\end{lemma}

The remainder of this section will be devoted to the proof of the
following lemma.  The method of proof combines the cut-the-tail lemma with
standard diagrammatic estimates.
 
\begin{lemma}
\label{lem-Rnbd}
For $n \geq 1$, and for $h \geq 0, p < p_c$ or  $h>0, p = p_c$, 
\eq
\lbeq{lemRnbd.1}
	|\hat{R}_{h,p}^{(j)}(k) |
	\leq O(\oneOd^j) e^{-hj} \, (\chi_{h,p} +1) M_{h,p},
	\quad
	| \hat{r}_{h,p}^{(n)}(k) |
	\leq O(\oneOd^n) e^{-hn} \,  \chi_{h,p} ,	
\en 
and hence
\eq
	| \hat{\rho}^{(n)}_{h,p}(k) | 
	\leq  O(\oneOd) e^{-h}\, (\chi_{h,p} +1) M_{h,p} .  
\en 
\end{lemma}

\Proof
By definition,  $\rho_{h,p}^{(n)} (0,x)
	=  \sum_{j=1}^n  R_{h,p}^{(j)}(0,x) 
	+  r_{h,p}^{(n)}(0,x)$,
so it suffices to prove \refeq{lemRnbd.1}.  
By definition,
\eqarray
	 R_{h,p}^{(j)}(0,x) 
	 & = &   (-1)^{j-1} 
	\Ebold_0I[E_0''] \Ebold_1 Y_1''  \cdots 
	\Ebold_{j-1} Y_{j-1}'' \Ebold_j I[F_2] ,
	 \\   
	r_{h,p}^{(n)}(0,x) & = & (-1)^n 
	\Ebold_0 I[E_0''] \Ebold_1 Y_1''  \cdots 
	\Ebold_{n-1} Y_{n-1}'' \Ebold_n Y_n  ,
\enarray

The term $r_{h,p}^{(n)}$ differs from $\phi_{h,p}^{(n)}$ only 
in the level-$n$ expectation, which is 
\eq
	\expec{Y_n}_n =
	P\bigl ( F_1(v_{n-1}, x; \tilde{C}_{n-1})\bigr )
	= P\bigl ( v_{n-1} \ct{\tilde{C}_{n-1}} x \AND v_{n-1} \nc G \bigr ) .  
\en 
Combining \refeq{F1exp} and \refeq{Cut1} gives
\eq
	\expec{Y_n}_n
	= 
	\expec{I[F_1'(v_{n-1}, x; \tilde{C}_{n-1})]}_{n} 
	+ 
	p \sum_{(u_{n}, v_{n})} 
	\expec{I[F_1''(v_{n-1}, u_{n}, v_{n}; \tilde{C}_{n-1})] 
	\tau_{h,p}^{\tilde{C}_n^{\{u_{n}, v_{n} \}} } (v_{n}, x)  
	}_{n} .
	\lbeq{Rn.3}
\en 
We have already derived a bound on the first term, namely $e^{-h}$
times \refeq{F1pbd.2}.

For the second term of \refeq{Rn.3},
we wish to employ Lemma~\ref{lem-tail.tau1}.  Because $F_1'$
is not increasing, due to its $G$-free condition, we first note 
that 
\eqsplit
\lbeq{F1baron}
	F_1''(v_{n-1}, u_{n}, v_{n}; \tilde{C}_{n-1})) 
	& =
	\bigl \{ F_1'(v_{n-1}, u_{n}, v_{n};\tilde{C}_{n-1}) ) 
	\textrm{ occurs on } \tilde{C}_n \bigr \}
	\\
	&  
	\mbox{\large $\subset$} 
	\bigcup_{
		\substack{w_{n-1} \in \tilde{C}_{n-1}\\ v_{n}' \in \Zd} 
		}
	\bigl \{ \bar{F}_{1,b}'(v_{n-1}, u_{n}, w_{n-1}, v_{n}') 
	\textrm{ occurs on } \tilde{C}_n
	\bigr \}.
\ensplit 
The event $\bar{F}_{1,b}'(v_{n-1}, u_{n}, w_{n-1}, v_{n}')$ 
defined in \refeq{F1bar-def} is an increasing event, 
and we can apply the cut-the-tail lemma to obtain
\eqalign
\lbeq{F1tail}
	& \expec{I[F_1''(v_{n-1}, u_{n}, v_{n}; \tilde{C}_{n-1})] \, 
	\tau_{h,p}^{\tilde{C}^{\{u_{n}, v_{n} \}} } (v_{n}, x)  
	} 
	\nnb 
	& 
	\qquad \qquad 
	\leq 
	\frac{1}{1 - p M_{h,p}} 
	\sum_{ \substack{w_{n-1} \in \tilde{C}_{n-1}\\ v_{n}' \in \Zd}  }
	\prob{\bar{F}_{1,b}'(v_{n-1}, u_{n}, w_{n-1}, v_{n}'} \, 
	\tau_{h,p}  (v_{n}, x)  .
\enalign 
As a result,  
\eq
	\expec{Y_n}_n \leq 
	e^{-h} \sum_{w_{n-1} \in \tilde{C}_{n-1}}  
	\Biggl [ 
	\myeqnpic{(130,100)}{(-35,0)}{ %begin my-eqn-pic %%%%%%
	    \thicklines
    		\drawline( 0,80)(50,80)
		\put( 50,20){\pictrir} 
    		\put(-30,90){$v_{n-1}$}
    		\put( 25, 0){$w_{n-1}$}
    		\put( 85,45){$x$}
	} % end of my-eqn-pic  %%%%%%%%%%%%%%%%%%%%%%%%%%%%%%%%
	\Biggr ] 
	+ 
	\frac{p}{1 - p M_{h,p}} 
	\sum_{(u_{n}, v_{n})} 
	\sum_{ w_{n-1} \in \tilde{C}_{n-1} }
	\Biggl [ 
	\myeqnpic{(150,100)}{(-35,0)}{ %begin my-eqn-pic %%%%%%
	    \thicklines
    		\drawline( 0,80)(50,80)
		\put( 50,20){\pictrir}
    		\put(-30,90){$v_{n-1}$}
    		\put( 25, 0){$w_{n-1}$}
    		\put( 90,45){$u_{n}$}
	} % end of my-eqn-pic  %%%%%%%%%%%%%%%%%%%%%%%%%%%%%%%%
	\Biggr ] 
	\tau_{h,p}  (v_{n}, x) . 
	\lbeq{F1bd.2}
\en 
Note that $p(1-pM_{h,p})^{-1} \leq p_c(1-p_c)^{-1} = O(\oneOd)$.
We use this and obtain a bound for $\hat{r}^{(n)}_{h,p}(k)$
in terms of nested expectations.  
The resulting nested expectation can be bounded as has been done for 
$\hat{\phi}^{(n)}(k)$ in Section~\ref{sub-bound.basic}, and 
the resulting diagrams are the same 
apart from a factor of $\chi_{h,p}$ arising from
the factor $\tau_{h,p}(v_n,x)$ in \refeq{F1bd.2}.  
Thus we obtain 
\eq
\lbeq{Rz0bd}
	| \hat{r}_{h,p}^{(n)}(k) |
	\leq O(\oneOd^n) e^{-h(n+1)} + O(\oneOd^n) e^{-hn} \chi_{h,p} 
	=  O(\oneOd^n ) e^{-hn} \,  \chi_{h,p} . 
\en

The analysis is similar for $\hat{R}_{h,p}^{(j)}$, $j \geq 1$.   Here
the level-$j$ expectation is the probability of the event 
$F_2(v_{j}, x; \tilde{C}_{j-1}) = \bigl \{
v_{j} \conn x \IN \Zd\backslash \tilde{C}_{j-1} 
\AND v_{j} \ct{\tilde{C}_{j-1}} G 
\bigr \}$.  By Lemmas~\ref{lem-F2dcmp} and
\ref{lem-F345cut}, 
\eqalign 
	\expec{I[F_{2}(v, x; A)]} 
	& 
	\leq 
	\expec{I[F_{3}(v, x; A)]}
	+ \expec{I[F_{4}(v, x; A)]}
	\nnb 
	& 
	= \sum_{i=3,4} 
	\biggl [ 
	\expec{I[F_i'(v, x; A)]} 
	+ 
	p \sum_{(u', v')} 
	\expec{I[F_i''(v, u', v' ; A)] \,  
	\tau_{h,p}^{  \tilde{C}^{\{ u', v' \}}(v)  }
	(v', x) 
	} 
	\biggr ] . 
	\lbeq{Rj.3}
\enalign
In order to bound the above terms, we introduce 
an auxiliary increasing event 
\eq
	\bar{F}_2'(v, x, w) = 
	\bigl \{
	(v \conn x) \circ (x \conn w) \circ (w \conn G) 
	\bigr \} 
	= 
	\biggl \{
	\myeqnpic{(140,100)}{(-15,0)}{ %begin my-eqn-pic %%%%%
	    \thinlines
		\drawline(0,80)(50,80)(50,20)(100,20)
    		\put(-10,90){$v$}
    		\put( 50,90){$x$}
    		\put( 35, 0){$w$}
		\put(105, 0){$G$}
	} % end of my-eqn-pic  %%%%%%%%%%%%%%%%%%%%%%%%%%%%%%%% 
	\biggr \}
\en 
and note that  
\eq
	F_3'(v, x; A) \cupd F_4'(v, x; A) \subset 
	\bigcup_{w \in A} 
	\bar{F}_2'(v, x, w)    . 
	\lbeq{F4bdd.3}
\en 
Thus the first term of \refeq{Rj.3} can be bounded by
\eq
	M_{h,p} \, \sum_{w \in A} 
	\biggl \{
	\myeqnpic{(70,100)}{(-15,0)}{ %begin my-eqn-pic %%%%%
	    \thinlines
		\drawline(0,80)(50,80)(50,20)
    		\put(-10,90){$v$}
    		\put( 50,90){$x$}
    		\put( 35, 0){$w$}
	} % end of my-eqn-pic  %%%%%%%%%%%%%%%%%%%%%%%%%%%%%%%% 
	\biggr \}
	.  
\en 
For the second term, using an analogue of \refeq{F1baron}
to apply the cut-the-tail lemma, we bound the expectation in the second
term of \refeq{Rj.3} by  
\eq  
	\frac{1}{1 - p M_{h,p}} \sum_{w \in A} 
	\prob{\bar{F}_2'(v, u', w) }  \, 
	\tau_{h,p}  (v', x)  .
\en
As a result, we have a bound 
\eq
\lbeq{F2bd.2}
	\langle I[F_{2}] \rangle_{j} \leq 
	M_{h,p} \, \sum_{w \in \tilde{C}_{j-1}} 
	\biggl \{
	\myeqnpic{(100,100)}{(-35,0)}{ %begin my-eqn-pic %%%%%
	    \thinlines
		\drawline(0,80)(50,80)(50,20)
    		\put(-30,90){$v_{j-1}$}
    		\put( 50,90){$x$}
    		\put( 35, 0){$w$}
	} % end of my-eqn-pic  %%%%%%%%%%%%%%%%%%%%%%%%%%%%%%%% 
	\biggr \}
	+ 
	\frac{M_{h,p}}{1 - p M_{h,p}} \sum_{w \in \tilde{C}_{j-1}} 
	\biggl \{
	\myeqnpic{(140,100)}{(-35,0)}{ %begin my-eqn-pic %%%%%
	    \thinlines
		\drawline(0,80)(50,80)(50,20)
    		\put(-30,90){$v_{j-1}$}
    		\put( 50,90){$u_{j}$}
    		\put( 35, 0){$w$}
	} % end of my-eqn-pic  %%%%%%%%%%%%%%%%%%%%%%%%%%%%%%%% 
	\biggr \} \, 
	\tau_{h,p}(v_{j}, x) .
\en 
The rest of the work is routine.  We have nested expectations with the 
rightmost expectation bounded as above.  
We estimate the nested expectation from 
right to left as usual.  
The other expectations of $Y_{1}''$ 
are dealt with in the standard manner by using \refeq{F1ppbd.2}, 
and we extract
a factor $e^{-h}$ from each expectation except for the rightmost 
one.  Since one new vertex $w$ has been added to the diagrams,
the resulting diagrams can be bounded in terms of the triangle 
diagram, to give 
\eq
	|\hat{R}_{h,p}^{(j)}(k) |
	\leq O(\oneOd^j\, e^{-hj}) M_{h,p} 
	+ O(\oneOd^j ) e^{-hj} \chi_{h,p} M_{h,p} 
	= O(\oneOd^j) e^{-hj} \, (\chi_{h,p} +1) M_{h,p}. 
\en 
\qed

%%%%%%%%%%%%%%%%%%%%%%%%%%%%%%%%%%%%%%%%%%%%%%%%%%%%%%%%%%%%%%%%%%%%%%%%%%%%%%%
%%%%%%%%%%%%%%%%%%%%%%%%%%%%%%%%%%%%%%%%%%%%%%%%%%%%%%%%%%%%%%%%%%%%%%%%%%%%%%%
\subsection{Proof of Proposition~\protect\ref{prop-taubd} completed} 
\label{sub-taubd}
% previously \label{sec-taubd}

In this section, we prove Proposition~\ref{prop-taubd}.
We fix $p=p_c$ throughout the section, and usually drop the corresponding
subscript from the notation.  We consider $h>0$, and continue to treat
the nearest-neighbour and spread-out models simultaneously.

In view of Lemmas~\ref{lem-Phinbd} and \ref{lem-Rnbd}, we can take
the limit $N \to \infty$ in the expansion
\refeq{oneF4.tauhat1} to obtain
\eq
\lbeq{tauNinfty}
        \tauhat_{h,p_c}(k) = 
        \frac{\hat{\phi}_{h}(k) + \Rhat_{h}(k) }
	{ 1 - \Phihat_{h}(k)} ,
\en
where 
\eq
	\hat{\phi}_{h}(k) = \sum_{n=0}^\infty \hat{\phi}_{h}^{(n)}(k), \quad
	\Rhat_{h}(k) = \sum_{j=1}^\infty  \Rhat_{h}^{(j)}(k) , \quad
	\Phihat_{h}(k) = \sum_{n=0}^\infty \Phihat_{h}^{(n)}(k).
\en

Note that the event $F_2(v_{n-1},x;A)$ is empty when $h=0$, and therefore
$\Rhat_{0,p}^{(j)}(k) = 0$ for all $p$ and $j$.  Hence, since 
$\chi_{0,p_c}= \infty$, setting $h=0$ and $k=0$ in
\refeq{tauNinfty} gives
\eq
	\infty = \frac{\hat{\phi}_{0}(0)}{ 1 - \Phihat_{0}(0)}.
\en
Since $\hat{\phi}_{0}(0)$ and $\Phihat_{0}(0)$ have been proven to be
finite, we conclude that 
\eq
\lbeq{Phi00is1}
	\Phihat_{0}(0) =1.
\en

The proof of \refeq{tauwant} proceeds by obtaining upper and lower
bounds for each of the numerator and denominator of \refeq{tauNinfty}.
The following lemma provides a first step in this direction.

\begin{lemma}
\label{lem-tau.numden.bd}
For $p = p_c$, $h>0$, and $k \in [-\pi,\pi]^d$,
\eqalign  
	\hat{\phi}_{h}(k) + \Rhat_{h}(k) 
	& = 1 - M_h + O(\oneOd) e^{-h} (\chi_h M_h +1) ,
	\lbeq{tau.num.bd}
	\\
	{}1 - \Phihat_{h}(k)  
	& = p_c \Omega \left [ 
	\{ 1 + O(\oneOd) \} M_h 
	+ \{ 1 - M_h + O(\oneOd) (M_h + e^{-2h}) \}  [ 1- \Dhat(k) ] 
	\right ] . 
	\lbeq{tau.den.bd}
\enalign 
\end{lemma}

\Proof
We first prove \refeq{tau.num.bd}.  By \refeq{f00bd.0} and
\refeq{fnbd}, 
\eq
	\sum_{n=0}^\infty \hat{\phi}_{h}^{(n)}(k) = \hat{\phi}_{h}^{(00)}(k)
	+ \hat{\phi}_{h}^{(01)}(k) + \sum_{n=1}^\infty \hat{\phi}_{h}^{(n)}(k)
	= 1 - M_h + O(\oneOd) e^{-2h}.
\en
By Lemma~\ref{lem-Rnbd}, $| \sum_{j=1}^\infty \Rhat_{h}^{(j)} (k)| 
\leq O(\oneOd) e^{-h}  ( \chi_h + 1) M_h$. 
Combining these gives \refeq{tau.num.bd}. 

By \refeq{Phi00is1},
\eq 
	1 -  \Phihat_h (k) 
	=  \left [ \Phihat_0 (0) - \Phihat_h (0) \right ] 
	+ \left [ \Phihat_h (0) - \Phihat_h (k) \right ] .
	\lbeq{tau.den.dcmp.1}
\en 
By \refeq{Phi00bd.0} and \refeq{Phinbd.2},
\eq
\lbeq{54.1}
	\Phihat_0 (0) - \Phihat_h (0)
	= p_c\Omega  M_{h} (1 + O(\oneOd)) .
\en 
By \refeq{Phi00bd.1} and \refeq{Phinbd}, 
\eq
\lbeq{54.2}
	\Phihat_h(0) - \Phihat_h(k)
	= p_c \Omega \left [  1 - M_{h}   
	+ O(\oneOd ) ( M_{h} + e^{-2h} )    
	\right ] [1 - \Dhat(k)] .
\en 
Combining \refeq{54.1} and \refeq{54.2} then gives  \refeq{tau.den.bd}.
\qed

We handle the term in \refeq{tau.num.bd} involving the product $\chi_h M_h$ 
using the following lemma.
\begin{lemma}
\label{lem-chiM.bd}
For $p= p_c$, $h>0$,
\eq
	\chi_h M_h = [ 1 + O(\oneOd) ] ( 1 - M_h) 
	+ O( \oneOd ) 
	\leq 1 + O(\oneOd) . 
	\lbeq{chiM.bd1}
\en 
\end{lemma}

\Proof 
Putting $k=0$ in  Lemma~\ref{lem-tau.numden.bd}, and using \refeq{pc}, gives 
\eq
	\chi_h = \tauhat_h (0) 
	= \frac{1 - M_h + O(\oneOd ) ( \chi_h M_h + 1)} 
	{M_h  \{ 1 + O(\oneOd) \} } .
\en 
We multiply both sides by $M_h$ and solve for $\chi_h M_h$, obtaining
\eq
	\chi_h M_h 
	= [ 1 + O(\oneOd) ] ( 1 - M_h) + O( \oneOd ) ,
\en 
as required.
\qed

Using Lemma~\ref{lem-chiM.bd}, we can now obtain good bounds on the
magnetization $M_{h}$. 

\begin{lemma}
\label{lem-Mzbd}
For $p = p_c$ and $h>0$, 
\eq
	\frac{e^{-h}}{2e} \leq 1 - M_h \leq e^{-h} 
	\lbeq{Mz.bd.apriori}
\en 
and 
\eq
	\sqrt { K_3 (1-e^{-h})} \leq M_h  \leq  \sqrt { K_4 (1-e^{-h})} ,
	\lbeq{Mz.bd.comp}
\en 
with $K_3$ and $K_4$ independent of $\oneOd$.  
\end{lemma}

\Proof 
For the upper bound of \refeq{Mz.bd.apriori}, we simply note that
$1 - M_h = \prob{0 \nc G} \leq \prob{0 \nin G } = e^{-h}$.  
The lower bound follows by first bounding $1-M_h$
below by the probability that $0 \nin G$ and all bonds emanating from
$0$ are vacant.  This gives  $1-M_h \geq  e^{-h} (1-p_c)^{\Omega} 
\geq \frac{e^{-h}}{2e}$, using \refeq{pc} in the last step.

The second bound requires more work.  
We first consider $h$ such that $e^{-h} \leq \frac{1}{2}$.  
In this case, it follows from
the upper bound of \refeq{Mz.bd.apriori} that $\frac{1}{2} \leq M_h \leq 1$,
and \refeq{Mz.bd.comp} follows trivially from that.  We therefore 
restrict attention in what follows, without further mention,
to $h$ such that $e^{-h} \in (\frac{1}{2},1)$.

By \refeq{chiM.bd1},  
\eq
\lbeq{M2de}
        \frac{d M^2_h}{dh}
	=  2 M_{h} \chi_{h} 
	= 2 (1 + O(\oneOd)) (1 - M_h)  + O( \oneOd ) .
\en  
This gives the differential inequalities 
\eq 
	 c_1 -  c_2 M_z \leq \frac{d M^2_h}{dh}  \leq  c_3
	\lbeq{Mbd.diff.1}
\en 
where $c_1, c_2, c_3$ are constants of the form $2 + O(\oneOd)$. 

We first integrate the upper bound, and find that 
\eq
	M_h^2 \leq   c_3 h .
\en 
Using this in the lower bound of \refeq{Mbd.diff.1}, we obtain
\eq
	 c_1 -  c_2 \sqrt{c_3 h}  \leq  \frac{d M^{2}_h } {d h}.
\en 
Integration then gives 
\eq
	M_h^2 \geq c_1 h - \frac{2}{3} c_2 \sqrt{c_3} h^{3/2} \geq ch, 
\en 
for some $c>0$.  The desired bounds then follow from the fact that
$h$ is bounded above and below by multiples of $1-e^{-h}$, for the
range of $h$ under consideration.
\qed

We are now in a position to prove \refeq{tauwant}, by applying
Lemmas~\ref{lem-chiM.bd} and \ref{lem-Mzbd} to the
estimates on the numerator and denominator of
\refeq{tauNinfty} given in Lemma~\ref{lem-tau.numden.bd}.  
For the numerator, using 
\refeq{Mz.bd.apriori} and the uniform bound \refeq{chiM.bd1} on $\chi_h M_h$,  
we obtain 
\eq 
\lbeq{ulnum}
	\left [ (2e)^{-1} + O(\oneOd) \right ] e^{-h}
	\leq \hat{\phi}_h(k) + \hat{R}_h(k) 
	\leq  
	[1 + O(\oneOd)] e^{-h} .
\en  
This is sufficient for our needs.

Next, we derive an upper bound
for the denominator, starting from \refeq{tau.den.bd}.  Using \refeq{pc}, 
\refeq{Mz.bd.apriori} and \refeq{Mz.bd.comp}, we have 
\eqarray
	1-\hat{\Phi}_h(k) 
	& \leq  &
	[ 1 + O(\oneOd) ] 
	\left [
	 M_h + \{e^{-h} + O(\oneOd) M_h \} [ 1 - \Dhat(k)] 
	\right ] .
	\nonumber \\ 
\lbeq{udenom}	
	& \leq &
	[ 1 + O(\oneOd) ](\sqrt{K_4(1-e^{-h})} + [1-\hat{D}(k)]) .
\enarray
For the lower bound, it follows from  \refeq{tau.den.bd}, \refeq{pc} and
\refeq{Mz.bd.apriori} that
\eq
\lbeq{2eehbd}
	1-\hat{\Phi}_h(k) 
	\geq 
	\{1 + O(\oneOd)\} 
	\left [ 
	M_h + \{(2e)^{-1}e^{-h} + O(\oneOd) M_h \} [ 1 - \Dhat(k)] 
	\right ] . 
\en 
This implies
\eq
\lbeq{ldenom}
	1-\hat{\Phi}_h(k) 
	\geq 
	\mbox{const.} 
	\left [ 
	[ 1 - \Dhat(k)] + \sqrt{1-e^{-h}}
	\right ]  
\en
with the constant independent of $\lambda$, as follows.
When $e^{-h} \in (\frac{1}{2},1)$, \refeq{ldenom} follows from
the lower bound of \refeq{Mz.bd.comp}.  When $e^{-h} \in [0,\frac{1}{2}]$,
\refeq{2eehbd} is bounded below by a constant since $M_h \geq \frac{1}{2}$,
and \refeq{ldenom} then follows.

Combining \refeq{ulnum}, \refeq{udenom} and \refeq{ldenom} then gives
\refeq{tauwant}.

%%%%%%%%%%%%%%%%%%%%%%%%%%%%%%%%%%%%%%%%%%%%%%%%%%%%%%%%%%%%%%%%%%%%%%%%%%%%%%%
%%%%%%%%%%%%%%%%%%%%%%%%%%%%%%%%%%%%%%%%%%%%%%%%%%%%%%%%%%%%%%%%%%%%%%%%%%%%%%%
\subsection{Proof of the cut-the-tail lemma}
\label{sec-cut-the-tail}

In this section, we prove Lemma~\ref{lem-tail.tau1}.  The proof makes
use of the following result.

\begin{lemma}
\label{lem-CCtilcompare}
Let $p \in [0, p_c]$ and $h \geq 0$ (assuming no infinite cluster
for $(h,p)=(0,p_c)$).
For an increasing event $F$,  
\eq 
\lbeq{cutwarmup}
	\prob{(F \AND v \nc G) \ON \tilde{C}^{\{u,v\}}(v) } 
	\leq \frac{1}{1 - p M_{h,p} } 
	\prob{F \ON C(v) \AND v \nc G} .
\en  
\end{lemma}

\Proof 
To abbreviate the notation, we write
$\tilde{F}$ for $\{F \ON \tilde{C}^{\{u,v\}}(v)\}$
and $\tilde{C}$ for $\tilde{C}^{\{u,v\}}$. 
We wish to bound the left side of \refeq{cutwarmup}
by replacing $\Ctilde(v)$  by $C(v)$. 
To begin, we recall example~(3) below Definition~\ref{def-event-on} and
write the left side of \refeq{cutwarmup} as  
\eq
\lbeq{cutwarm1}
	\prob{ \tilde{F} \AND \tilde{C}(v) \cap G = \emptyset }
	= \prob{ \tilde{F} \AND {C}(v) \cap G = \emptyset }
	+ \prob{ \tilde{F} \AND \tilde{C}(v) \cap G = \emptyset \AND
	{C}(v) \cap G \neq \emptyset}.
\en
Since $\tilde{F} \subset \{ F \ON C(v) \}$ for $F$ increasing, 
\eq
\lbeq{CCtilcmp.az}
	\prob{ \tilde{F} \AND {C}(v) \cap G = \emptyset }
	\leq \prob{ F \ON C(v) \AND {C}(v) \cap G = \emptyset }.
\en

In the second term on the right side of \refeq{cutwarm1}, the event
$\{C(v) \cap G \neq \emptyset\}$ can be replaced by the event
$\{\{u,v\}\mbox{ is occupied} \AND \{u \conn G \INSIDE 
\Zd\backslash\tilde{C}(v) \}\}$.
Hence we may apply Lemma~\ref{lem-cond.0} to this term.
After doing so, we bound 
$P(u \conn G \INSIDE \Zd\backslash\tilde{C}(v))$ by $M_{h,p}$, to obtain
\eq
\lbeq{CCtilcmp.second.2z}
	\prob{\tilde{F} \AND \tilde{C}(v) \cap G = \emptyset 
        \AND C(v) \cap G \neq \emptyset} 
        \leq
	pM_{h,p} 
	\prob{\tilde{F}  \AND \tilde{C}(v) \cap G = \emptyset} .
\en

Combining \refeq{cutwarm1}--\refeq{CCtilcmp.second.2z}, we have 
\eq 
\lbeq{CCtilcmp.bz}
        \prob{\tilde{F} \AND \tilde{C}(v) \cap G = \emptyset  }  
        \leq 
        \prob{F \ON C(v) \AND C(v) \cap G = \emptyset } 
	+ p \, M_{h,p} \, 
        \prob{ \tilde{F} \AND \tilde{C}(v) \cap G = \emptyset } . 
\en 
Solving \refeq{CCtilcmp.bz} for 
$\prob{\tilde{F} \AND \tilde{C}(v) \cap G = \emptyset }$ then gives 
the desired result. 
\qed

We are now able to prove the cut-the-tail lemma, which asserts that
for increasing $E$,
\eq 
\lbeq{ctt.want}
	\expec { \Ind [ E \ON \tilde{C}^{\{u,v\}}(A) ]  \,  
	\tau_{h,p}^{\tilde{C}^{\{u,v\}}(A)} (v, x) }
	\leq \frac{1}{1 - p M_{h,p}} \, P(E) \, 
	\tau_{h,p}(v,x) .
\en 

\bigskip
\noindent 
{\bf Proof of Lemma~\ref{lem-tail.tau1}.} 
We first note that by Lemma~\ref{lem-cond.0}, the left side
of \refeq{ctt.want} can be written as
\eq
\lbeq{lem-tail.tau.3}
	\prob{ E \ON \tilde{C}(A) \AND 
	(v \conn x \AND v \nc G) 
	\INSIDE \Zd \backslash \tilde{C}(A)} .
\en
When 
$v \conn x \INSIDE \Zd \backslash \tilde{C}(A)$, 
this in particular means that 
$\tilde{C}(A) \not\ni v$, and thus 
$\tilde{C}(A) \cap \tilde{C}(v) = \emptyset$. 
We can then rewrite \refeq{lem-tail.tau.3} as 
\eq 
	\prob {E \ON \tilde{C}(A) \AND 
	\left ( v \conn x \AND v \nc G \right ) 
	\INSIDE \Zd \backslash \tilde{C}(A) 
	\AND \tilde{C}(v) \cap \tilde{C}(A) = \emptyset
	} .
\en
Because $\{v \conn x\}$ and $\{v \nc G\}$ depend only on bonds/sites 
connected to $v$, and because $\tilde{C}(v)
\subset \Zd \backslash \tilde{C}(A)$ when 
$\tilde{C}(A) \cap \tilde{C}(v) = \emptyset$, the above is equal to	
\eq
	\prob {
	E \ON \tilde{C}(A) \AND 
	\left ( v \conn x \AND v \nc G \right ) \INSIDE  
	\tilde{C}(v)
	\AND \tilde{C}(v) \cap \tilde{C}(A) = \emptyset
	} . 
	\lbeq{lemtail1.1}
\en

Since $E$ is increasing, 
and recalling Definition~\ref{def-event-on}(c), we have 
\eq
	\{ E \ON \tilde{C}(A) \AND \tilde{C}(v) \cap \tilde{C}(A) = \emptyset 
	\} 
	\subset \{ E \ON \Zd \backslash \tilde{C}(v)\}
	= \{ E \INSIDE \Zd \backslash \tilde{C}(v) \}.
\en
Recalling that ``occurs in'' and ``occurs on'' are the same 
for $\tilde{C}$, 
\refeq{lemtail1.1} is therefore bounded above by 
\eqalign 
\lbeq{cutlem101hara}
	\prob{ \left ( v \conn x \AND v \nc G \right ) \ON  
	\tilde{C}(v) \AND
	E \INSIDE \Zd \backslash \tilde{C}(v) } . 
\enalign 
Now by Lemma~\ref{lem-cond.0}, the above quantity is equal to 
\eq
	\bigl \langle I[
	\left ( v \conn x \AND v \nc G \right ) \ON  \tilde{C}(v)
	]
	\, 
	\langle I[ E \INSIDE \Zd \backslash \tilde{C}(v)] 
	\rangle \bigr \rangle . 
\en 

Finally, since $E$ is increasing, this is bounded above by
\eq
	P(E)\prob{(v \conn x \AND v \nc G) \ON \tilde{C}(v) }.
\en
The proof is completed by applying Lemma~\ref{lem-CCtilcompare} to
estimate the final factor on the right side, noting that ``occurs on
$C(v)$''
can be replaced by ``occurs in $C(v)$'' after applying the lemma.
\qed

%%%%%%%%%%%%%%%%%%%%%%%%%%%%%%%%%%%%%%%%%%%%%%%%%%%%%%%%%%%%%%%%%%%%%%%%%%%%%%%
%%%%%%%%%%%%%%%%%%%%%%%%%%%%%%%%%%%%%%%%%%%%%%%%%%%%%%%%%%%%%%%%%%%%%%%%%%%%%%%
%%%%%%%%%%%%%%%%%%%%%%%%%%%%%%%%%%%%%%%%%%%%%%%%%%%%%%%%%%%%%%%%%%%%%%%%%%%%%%%
%%%%%%%%%%%%%%%%%%%%%%%%%%%%%%%%%%%%%%%%%%%%%%%%%%%%%%%%%%%%%%%%%%%%%%%%%%%%%%%
\section{Refined $k$-dependence using the two-$M$ scheme}
\label{sec-2M.exp.err}
\setcounter{equation}{0}

In this section, we go part way to improving the bounds of
Proposition~\ref{prop-taubd} to the asymptotic statement of
Theorem~\ref{thm-tauasymp}, using the two-$M$ scheme for the 
expansion.  In Section~\ref{sub-2Mbd.infinite},
we show that we can take $M,N \to \infty$ in
\refeq{taugPiUk}, and prove existence of the limit 
$\lim_{h \downarrow 0}\hat{\tau}_{h,p_c}(k)$ of Theorem~\ref{thm-tauasymp}.  
In Section~\ref{sub-2Mbd.num}, 
the numerator resulting from the limit $M,N \to \infty$ in \refeq{taugPiUk}
will be shown to be equal
to $\hat{\phi}_{0,p_c}(0) + \ohone + O(k^2)$.
In Section~\ref{sub-2Mbd.k-den}, we will extract the leading
$k^2$-dependence of the limiting denominator of \refeq{taugPiUk}.
This will prove \refeq{tauirpc}.
Extraction of the leading $h$-dependence of the denominator will be postponed
to Section~\ref{sec-Psidef}.

We begin by presenting some
new methods for bounding diagrams, which will be required in both
Sections~\ref{sec-2M.exp.err} and \ref{sec-Psidef}.

%%%%%%%%%%%%%%%%%%%%%%%%%%%%%%%%%%%%%%%%%%%%%%%%%%%%%%%%%%%%%%%%%%%%%%%%%%%%%%%
%%%%%%%%%%%%%%%%%%%%%%%%%%%%%%%%%%%%%%%%%%%%%%%%%%%%%%%%%%%%%%%%%%%%%%%%%%%%%%%
\subsection{Diagrammatic methods}
\label{sec-diagmethods}

In this section, we describe two methods for estimating diagrams.

The first method involves an application of the dominated convergence
theorem, in a manner that will be used repeatedly.  We illustrate
this method in the simplest example where it is useful.

\begin{example}
\label{ex-dct}
For $p \leq p_c$, consider the sum
\eq
	\sum_{x,y \in \Zd} P_p( (0 \conn x) \circ 
	(x \conn y) \circ (y \conn 0) \AND 0 \conn G) .
\en
Diagrammatically, the above event corresponds to a square with vertices
at $0,x,y$, and a fourth vertex from which a connection to $G$ emerges.  
A naive estimate, which we do not want to use, would be to use BK 
to bound the above sum by the square
diagram $1+\pol{4}_{h,p}(0)$ 
times the magnetization.  This is a useless bound when $p=p_c$
and $d \leq 8$,
because the square diagram then diverges.  Instead, we use the dominated
covergence theorem, as follows.  First, the above probability
is bounded above by 
$\tau_{0,p_c}(0,x) \tau_{0,p_c}(x,y) \tau_{0,p_c}(y,0)$, which
is summable since the triangle diagram is finite in sufficiently high
dimensions for the nearest-neighbour model and for sufficiently spread-out
models for $d>6$.  
On the other hand, the above probability
is also bounded by $M_{h,p}$, which goes to zero
as $h \to 0$.  It therefore follows from the dominated
convergence theorem that
\eq
	\lim_{h \to 0} \sum_{x,y \in \Zd} P_p( (0 \conn x) \circ 
	(x \conn y) \circ (y \conn 0) \AND 0 \conn G)  =0.
\en
\end{example}

\medskip

As was just 
pointed out in Example~\ref{ex-dct}, the square diagram is infinite at 
the critical point, for $d \leq 8$.  
However, there is a method for employing a square diagram for $d>6$
when $h>0$, if at least
one of the lines comprising the square corresponds to $\tau_{h,p_c}(0,x)$.
In this case, 
the square is finite for all $d>6$, with a controlled rate of
divergence, for $d \leq 8$, 
as $h \to 0$.  The remainder of this section describes
this observation in more detail, and sets the stage for its use
in our later diagrammatic estimates.

We begin with an elementary estimate for the integrals defined by
\eq
        I_{m,n}^{(d)}(h) = \int_{[-\pi, \pi]^{d}} d^{d}k 
        \frac{1}{(k^{2}+ \sqrt{h})^{m} (k^{2})^{n} } ,
\en  
with $m,n \geq 0$ (not necessarily integers) and $h \geq 0$.

\begin{lemma}
\label{lem-int}
Let $m,n \geq 0$.
If $d \leq 2n$, then $I_{m,n}^{(d)}(h)= \infty$ for all $h \geq 0$.
As $h \to 0$, 
\eq
        I_{m,n}^{(d)}(h) \sim \left \{ 
        \begin{array}{ll}
        const. h^{\frac{d - 2(m+n)}{4}} 
        & 2n < d < 2 (n+m)  \\
	const. |\log h| 
        & d = 2 (n+m), \,  m>0  \\
        const. & d > 2(n+m) .
        \end{array}
        \right . 
\en  
\end{lemma}

\Proof 
For $d \leq 2n$, 
$I_{m,n}^{(d)}(h) \geq (\pi^2 d + \sqrt{h})^{-m} 
\int_{[-\pi, \pi]^{d}} d^{d}k \, k^{-2n} = \infty$.

For $d > 2(n+m)$, $I_{m,n}^{(d)}(h) \leq I_{m,n}^{(d)}(0) =
\int_{[-\pi, \pi]^{d}} d^{d}k \, k^{-2(n+m)} < \infty$, and by 
the monotone convergence theorem, 
$\lim_{h \to 0} I_{m,n}^{(d)}(h) = I_{m,n}^{(d)}(0)$.

For $2n < d < 2(n+m)$, or for $d=2(n+m)$ with $m>0$, 
the integral diverges as $h \to0$ and its asymptotic
behaviour is given by that of the integral over $|k|\leq 1$.
Switching to polar coordinates, and 
writing $\omega_{d}$ for the solid angle in $d$-dimensions, this gives
\eq
        I_{m,n}^{(d)}(h) \sim \omega_{d} \int_{0}^{1} dk 
        \frac{k^{d-1}}{(k^{2} + \sqrt{h})^{m} k^{2n}} 
        \sim \frac{\omega_{d}}{2} h^{\frac{d - 2(m+n)}{4}} 
        \int_{0}^{h^{-1/2}} dr 
        \frac{r^{(d-2)/2}}{(1+r)^{m} r^n} 
\en  
where we made the change of variables $r = k^{2}h^{-1/2}$. 
The integral is finite as $h \to 0$ 
if $2n < d < 2(n+m)$, and it diverges logarithmically if 
$d=2(n+m)$ with $m>0$.  This completes the proof. 
\qed
\medskip

We define the square diagram containing one $G$-free line, at $p=p_c$, to be
\eq
	{\sf S}_h  = \sum_{w,x,y \in \Zd} \tau_{h,p_c}(0,w) \tau_{0,p_c}(w,x)
	\tau_{0,p_c}(x,y) \tau_{0,p_c}(y,0).
\en
By the monotone convergence theorem, the Parseval relation, the upper
bound of Proposition~\ref{prop-taubd} and the infra-red bound \refeq{irbd},
\eqarray
	{\sf S}_h & = & \lim_{p \to p_c}
	\sum_{w,x,y \in \Zd} \tau_{h,p_c}(0,w) \tau_{0,p}(w,x)
	\tau_{0,p}(x,y) \tau_{0,p}(y,0)
	\nnb
	& = & \lim_{p \to p_c}
	\int_{[-\pi,\pi]^d} \frac{d^dk}{(2\pi)^d} \hat{\tau}_{h,p_c}(k)
	\hat{\tau}_{0,p}(k)^3 \leq \textrm{const.} I_{1,3}^{(d)}(h).
\enarray
It then follows immediately from Lemma~\ref{lem-int} that
${\sf S}_h \leq O(h^{(d-8)/4})$ for $6<d<8$, that ${\sf S}_h \leq O(|\log h|)$
for $d=8$, and that ${\sf S}_h =O(1)$ for $d>8$.

If we replace the basic quantity of Example~\ref{ex-dct} by
\eq
\lbeq{ShMhex}
	\sum_{x,y \in \Zd} P_{p_c}( 0 \conn x \AND 0 \nc G) P_{p_c}( 
	(x \conn y) \circ (y \conn 0) \AND 0 \conn G) ,
\en
then the naive estimate rejected in Example~\ref{ex-dct} can be used to
bound \refeq{ShMhex} 
above by ${\sf S}_h M_{h,p_c}$.  Using the bounds mentioned above
for ${\sf S}_h$ and the upper bound on the magnetization of 
Lemma~\ref{lem-Mzbd}, we have
\eq
\lbeq{deltaddef}
	{\sf S}_h M_h \leq O(h^{\delta(d)}), \quad \mbox{where } 
	h^{\delta(d)} = \left \{
	\begin{array}{ll}
	h^{(d-6)/4} & (6<d<8) \\
	h^{1/2} |\log h	| & (d=8) \\
	h^{1/2}  & (d>8).
	\end{array}
	\right.
\en

We will obtain upper bounds similar to \refeq{ShMhex}
by bounding a pair of nested expectations.  The probability
involving the connection to $G$ will come from one expectation,
and the probability involving the $G$-free connection will
come from a second expectation.  
To produce a bound in terms of a probability of a $G$-free 
connection, we will use the 
generalization of the BK inequality given in the following lemma.

Let $E$ be an event specifying that 
finitely many pairs of sites are connected, possibly disjointly.
In particular, $E$ is increasing.
We say that $E$ occurs and is $G$-free if $E$ occurs
and the clusters of all the sites
for which connections are specified in its definition 
do not intersect the random set $G$ of green sites.  The following lemma is
a BK inequality for $G$-free connections, in which the upper bound
retains a $G$-free condition on one part of the event only.

\begin{lemma}
\label{lem-onehline}
Let $E_{1},E_2$ be events of the above type.  Then 
\eq
\lbeq{E1E2oneh}
        \prob{(E_{1} \circ E_{2}) 
        \textnormal{ occurs and is $G$-free} } 
        \leq \prob{E_{1} 
	\textnormal{ occurs and is $G$-free} } \, \prob{E_{2}} .
\en  
\end{lemma}

\Proof 
Given an event $F$,
we denote by $[F]_n$ the event that $F \INSIDE [-n,n]^d$.  It suffices to
show that
\eq
\lbeq{E1E2onehn}
        \prob{[(E_{1} \circ E_{2}) 
        \textnormal{ occurs and is $G$-free}]_n } 
        \leq \prob{[E_{1} 
	\textnormal{ occurs and is $G$-free}]_n } \, \prob{[E_{2}]_n} ,
\en 
since \refeq{E1E2oneh} then follows by letting $n \to \infty$.
This finite volume argument is used to deal with the fact that the
usual BK inequality  \cite[Theorem 2.15]{Grim89}
applies initially to events depending on only
finitely many bonds.

Given a bond-site configuration, we define $C(G)_n$ to be the set of sites 
in $[-n,n]^d$
which are connected to the green set $G$.  
Conditioning on $C(G)_n$, we have 
\eq 
        \prob{ [(E_{1} \circ E_{2}) 
	\textnormal{ occurs and is $G$-free}]_n } 
        = \sum_{\gamma} \prob{
        C(G)_n= \gamma \AND [(E_{1} \circ E_{2})  
        \textnormal{ occurs and is $G$-free}]_n } ,
\en 
where the sum is taken over all subsets $\gamma$ of sites 
in $[-n,n]^d$.  
%We denote by $\gamma^s$ the set of sites in $\gamma$.
When $C(G)_n = \gamma$, bonds touching but not in $\gamma$ are vacant and 
we can replace the event
$[(E_{1} \circ E_{2})$ occurs and is $G$-free$]_n$ by 
$[(E_{1} \circ E_{2})$ occurs in $\Zd \backslash \gamma]_n$.  
Thus we have 
\eq
        \prob{ [(E_{1} \circ E_{2}) \mbox{ occurs and is G-free}]_n } 
        = \sum_{\gamma} \prob{
        C(G)_n=\gamma \AND [(E_{1} \circ E_{2}) 
        \INSIDE \Zd \backslash \gamma ]_n
        } .
\en

Since the event $[E_{1} \circ E_{2} \INSIDE \Zd \backslash \gamma ]_n$
depends only on bonds and sites in $[-n,n]^d$
which do not touch $\gamma$, while the event
$C(G)_n=\gamma$ depends only on bonds and sites which do touch $\gamma$,  
the probability factors to give
\eq     
        \sum_{\gamma} \prob{C(G)_n=\gamma} \, 
        \prob{[(E_{1} \circ E_{2}) \INSIDE \Zd \backslash \gamma]_n} .
\en 
Now we can apply the usual BK inequality (in the reduced lattice consisting
of bonds and sites in $[-n,n]^d $ which do not touch $\gamma$) 
to the latter probability, to obtain an upper bound 
\eq 
        \sum_{\gamma} \prob{C(G)_n=\gamma} \, 
        \prob{[E_{1} \INSIDE \Zd \backslash \gamma]_n} \, 
        \prob{ [E_{2} \INSIDE \Zd \backslash \gamma]_n} .
\en 
Since $E_{2}$ is increasing, this is bounded above, as required, by
\eq
        \sum_{\gamma} \prob{C(G)_n=\gamma} \, 
        \prob{[E_{1} \INSIDE \Zd \backslash \gamma]_n} \, \prob{[ E_{2}]_n } 
	= \prob{[E_{1} \mbox{ occurs and is $G$-free}]_n} \, \prob{[E_{2}]_n}.
\en 
\qed

The two methods exemplified by Example~\ref{ex-dct} and by use of ${\sf S}_h$
will be prominent
in the diagrammatic estimates used in the remainder of this 
paper.  The latter method gives error estimates and is therefore stronger
than the
former, which does not.  However, when it does not affect our final
result, we will sometimes
use the dominated convergence method when stronger bounds in terms 
of ${\sf S}_h$ could also be obtained.
We now illustrate the methods with two examples,
in which the quantity
\eq
\lbeq{R1}
	A_h(k) 
	= - \sum_x e^{ik\cdot x} p_c\sum_{(u_0,v_0)}
	p_c \sum_{(u_1,v_1)} 
	\langle I[E_0''(0,u_0,v_0)] \langle
	I[F_4''(v_0,u_1,v_1;\tilde{C}_0)]
	\tau^{\tilde{C}_1}_{h,p_c}(v_1,x)
	\rangle_1 \rangle_0
\en 
will be bounded at $p=p_c$ first
using dominated convergence and then using ${\sf S}_h$.   The term
$A_h(k)$ is a contribution to the Fourier transform $\hat{S}_{h,p_c}^{(1)}(k)$
of the $n=1$ case of \refeq{Rtilndef}, via Lemma~\ref{lem-F345cut}.  We drop
the subscripts $p_c$ in the two examples.

\begin{example}
\label{ex-R1dct}
We now illustrate the use of dominated convergence to conclude that
$A_h(k) = \ohone$.  As a start,
we take absolute values inside the sum over $x$ to obtain a $k$-independent
upper bound.

\noindent {\em Step 1.}  The event $F_4''(v_0,u_1,v_1;\tilde{C}_0)$
is a subset of  
\eq
	\bar{F}_4''(v_0,u_1,v_1;\tilde{C}_0)
	= \bar{F}_4 (v_0,u_1,v_1;\tilde{C}_0) \ON \tilde{C}_1, 
\en 
where
$\bar{F}_4$ is the increasing
event that there exist $w_0 \in \tilde{C}_0$ and
$w_1 \in \tilde{C}_1$ such that there are disjoint connections
$v_0 \conn w_1$, $w_1 \conn G$, $w_1 \conn u_1$, $u_1 \conn w_0$,
$w_0 \conn G$.  Then we apply the cut-the-tail Lemma~\ref{lem-tail.tau1}
to bound the inner expectation in \refeq{R1} by 
$(1-p_c M_h)^{-1}P(\bar{F}_4) \tau_h(v_1,x)$.

\noindent {\em Step 2.}
Next, we bound $P(\bar{F}_4)$ by the probability that there exists 
$w_0 \in \tilde{C}_0$ such that there are disjoint connections
$v_0 \conn u_1$, $u_1 \conn w_0$, $w_0 \conn G$.  Applying BK to
bound this, and also to bound the outer expectation, this leads to
an upper bound for $|A_h(k)|$ by 
\eq
	[1+O(\oneOd)] M_{h} \, \chi_{h} \, \sum_{v_0,u_1}
	%\biggl [
	\myeqnpic{(130,100)}{(-50, 0)}{ %begin my-eqn-pic %%%%
	    \thicklines
		\put(  0,20){\pictril} 
		\put(  0,80){\picpivone}
    		\drawline(  0,20)(50,20)
    		\drawline( 10,80)(50,80)
    		\drawline( 50,20)(50,80)
    		\put(-50,40){$0$}
    		\put( 50,90){$u_{1}$}
    		\put(-10,95){$v_{0}$}
    		%\put( 50, 0){$w_{1}$}
	} % end of my-eqn-pic %%%%%%%%%%%%%%%%%%%%%%%%%%%%%%%% 
	%\biggr ] 
	\leq 
	[1+O(\oneOd)] \, 	
	\biggl [
	\myeqnpic{(60,100)}{(-50, 0)}{ %begin my-eqn-pic %%%%
	    \thicklines
		\put(  0,20){\pictril} 
    		\put(-50,40){$0$}
	} % end of my-eqn-pic %%%%%%%%%%%%%%%%%%%%%%%%%%%%%%%% 
	\biggr ] \, 
	\biggl [
	\sup_{x}
	\myeqnpic{(70,100)}{(-20, 0)}{ %begin my-eqn-pic %%%%
	    \thicklines
		\put(  0,80){\picpivone}
    		\drawline(  0,20)(50,20)
    		\drawline( 10,80)(50,80)
    		\drawline( 50,20)(50,80)
    		\put(-20, 0){$0$}
    		\put(-20,95){$x$}
	} % end of my-eqn-pic %%%%%%%%%%%%%%%%%%%%%%%%%%%%%%%% 
	\biggr ] 	
	\leq 
	O(\oneOd)  . 
\en
In the above, the factor $O(\oneOd)$ arises as 
in \refeq{bd-optriangle}, and we used Lemma~\ref{lem-chiM.bd} to bound
$M_h \chi_h$.

\noindent {\em Step 3.}  
Since the summand of \refeq{R1} is bounded above by 
$(1-p_c M_h)^{-1}P(\bar{F}_4)  \leq O(M_h^2 ) = O(h)$, 
it goes to zero pointwise as $h \to 0$.

\noindent {\em Step 4.}
By Steps~2 and 3 and the dominated convergence theorem, \refeq{R1} is $\ohone$.
\end{example}

\begin{example}
\label{ex-R1Sh}
We now illustrate the use of Lemma~\ref{lem-onehline} to conclude that
$A_h(k) = O(h^{\delta(d)})$.

\noindent{\em Step 1.}
We first apply the cut-the-tail lemma as in Step~1 of Example~\ref{ex-R1dct}.

\noindent{\em Step 2.}
We wish to extract a $G$-free line from the connections required
by $E_0''$, but there is a subtlety associated with the fact that
$C_0$ is only required to be $G$-free on $\tilde{C}_0$.  The following
device will allow this to be handled.
Let $\expec{\, \cdot \,}^{\tilde{}}$ denote the 
\emph{conditional}\/ expectation, under the 
condition that $\{u_0, v_0\}$ is vacant.
By the definition of ``occurs on $\tilde{C}$'' in
Definition~\ref{def-event-on}(c), we can rewrite the nested expectation
appearing in \refeq{R1} as
\eq
	\langle I[E_0'(0,u_0,v_0)] \langle
	I[F_4''(v_0,u_1,v_1; C_0)]
	\tau^{\tilde{C}_1}_{h,p_c}(v_1,x)
	\rangle_1 \rangle_0^{\tilde{}}.
\en
Here, in particular, we have used the fact that
$\tilde{C}_0^{\{u_{0}, v_{0}\}}(0)=C_0(0)$ when $\{u,v\}$ is vacant.  
We apply the BK inequality 
to estimate $P(\bar{F}_4)$, this time extracting all the 
disjoint connections. As a result, $|A_h(k)|$ can now be bounded by 
\eq
	\frac{1}{1-p_cM_h} 
	\sum_{x,w_0, w_{1}} p_c \sum_{(u_0,v_0)} p_c \sum_{(u_1,v_1)} 
	\langle I[E_0'(0,u_0,v_0)] I[w_0 \in C_0] \rangle^{\tilde{}}
	\tau_0(v_0,w_1) \tau_0(w_1, u_1) \tau_0(u_1, w_0)  M_h^2
	\tau_h(v_1,x).
\en

\noindent{\em Step 3.}
The remaining conditional expectation involves disjoint connections 
$0 \conn u_0$, $0 \conn w$, $w \conn u_0$,
$w \conn w_0$, all $G$-free, for some $w$.  
Lemma~\ref{lem-onehline} can be used to bound this conditional
expectation by $\tau_0(0,u_0) \tau_0(0,w) \tau_0(w,u_0)$ times 
$\langle I[w \conn w_0 , w \nc G] \rangle^{\tilde{}}$, using  a
slight generalization of Lemma~\ref{lem-onehline} to conditional probabilities,
and the fact
that $\langle I[a \conn b] \rangle^{\tilde{}} \leq \tau_0(a,b)$.
Now, for any event $E$, 
\eq 
	\prob{E} 
	\geq 
	\prob{E \AND \{u, v\} \textrm{ is vacant}} 
	= (1-p_c) \expec{I[E]}^{\tilde{}} . 
\en 
This leads to an upper bound for \refeq{R1} by $O(\chi_h M_h^2)$ times
the diagram
\eq
	\myeqnpic{(130,100)}{(-50, 0)}{ %begin my-eqn-pic %%%%
	    \thicklines
		\put(  0,20){\pictril} 
		\put(  0,80){\picpivone}
    		\drawline( 10,80)(100,80)
    		\drawline(100,20)(100,80)
		\dottedline{6}(0,20)(100,20)
    		\put(-50,40){$0$}
		\put( 50,80){\circle*{8}}
%     		\put(100,90){$u_{1}$}
%     		\put(-10,95){$v_{0}$}
%     		\put(100, 0){$w_{1}$}
	} % end of my-eqn-pic %%%%%%%%%%%%%%%%%%%%%%%%%%%%%%%% 
\en
where thick lines represent $\tau_0$ and the dotted line 
represents $\tau_h$.  This diagram
is bounded above by the triangle times ${\sf S}_h$, leading to an overall
bound $O(\chi_h M_h^2{\sf S}_h) = O(h^{\delta(d)})$, 
which is stronger
than the bound obtained in Example~\ref{ex-R1dct}.
\end{example}

%%%%%%%%%%%%%%%%%%%%%%%%%%%%%%%%%%%%%%%%%%%%%%%%%%%%%%%%%%%%%%%%%%%%%%%%%%%%%%%
%%%%%%%%%%%%%%%%%%%%%%%%%%%%%%%%%%%%%%%%%%%%%%%%%%%%%%%%%%%%%%%%%%%%%%%%%%%%%%%
\subsection{The two-$M$ scheme to infinite order}
\label{sub-2Mbd.infinite}

In this section, we fix $p=p_c$ and drop subscripts $p_c$ from the notation.
The bounds of Lemma~\ref{lem-xiXiURbds}
below, together with the estimates for
$\hat{\phi}_h^{(n)} (k)$ and $\Phihat_h^{(n)} (k)$ obtained in 
Section~\ref{sec-part.exp.bd}, imply that 
we can take the limit
$M,N \to \infty$ in \refeq{taugPiUk}, obtaining
\eq
\lbeq{taugPiUkinf}
	\hat{\tau}_h(k) = 
	\frac{\hat{\phi}_h(k)  
	+ \hat{\xi}_h(k) 
	+ \hat{U}_h(k) 
	+ \hat{S}_h(k) 
	}
	{1- \hat{\Phi}_h(k)
	- \hat{\Xi}_h(k) 
	}.
\en
On the right side, we introduced 
$\hat{\phi}_h(k)= \sum_{n=0}^\infty \hat{\phi}_h^{(n)}(k)$,
$\hat{\xi}_h(k) =\sum_{n=1}^\infty \sum_{m=0}^\infty \hat{\xi}^{(n,m)}_h(k) $,
$\hat{U}_h(k) = \sum_{n=1}^\infty \sum_{m=1}^{\infty} \hat{U}^{(n,m)}_h(k) $,
$\hat{S}_h(k) =\sum_{n=1}^{\infty} \hat{S}^{(n)}_h(k)$,
$\hat{\Phi}_h(k) = \sum_{n=0}^\infty \hat{\Phi}_h^{(n)}(k)$,
$\hat{\Xi}_h(k) = \sum_{n=1}^\infty \sum_{m=0}^\infty  \hat{\Xi}^{(n,m)}_h(k)$.
Absolute convergence of the sums is guaranteed by Lemma~\ref{lem-xiXiURbds}.

Moreover, it follows from Lemma~\ref{lem-xiXiURbds} that
$\hat{\xi}_h(k)$, $\hat{U}_h(k)$, $\hat{S}_h(k)$ and $\hat{\Xi}_h(k)$
vanish in the limit $h \downarrow 0$.  Since 
$\lim_{h \downarrow 0} \hat{\phi}_h(k) = \hat{\phi}_0(k)$
and
$\lim_{h \downarrow 0} \hat{\Phi}_h(k) = \hat{\Phi}_0(k)$ by
\refeq{f0-bond}--\refeq{Phin-bond}, Lemma~\ref{lem-Phinbd} and the dominated
convergence theorem, it follows that
\eq
\lbeq{tauhat0limex}
	\lim_{h \downarrow 0}\hat{\tau}_{h,p_c}(k) 
	= \frac{\hat{\phi}_{0,p_c}(k)}{1-\hat{\Phi}_{0,p_c}(k)}.
\en
This proves existence of the limit stated in Theorem~\ref{thm-tauasymp}.

In addition, Lemma~\ref{lem-xiXiURbds}
will provide some of the estimates to be used in
the asymptotic analysis of the numerator and denominator of
\refeq{taugPiUkinf}. 

\begin{lemma}
\label{lem-xiXiURbds}
For $h \geq 0$, $p=p_c$, $k \in [-\pi,\pi]^d$, and for all $n,m \geq 0$,
\eqalign
	\lbeq{2-M.xi.bd}
	\sum_x | {\xi}_h^{(n,m)}(0,x) | , \; 
	\sum_x \left | {\Xi}_h^{(n,m)} (0,x) \right |
	& \leq O(\oneOd^{n+m} )h^{1/2},
	\\
	\lbeq{2-M.u.bd}
	\sum_x \left | {u}_h^{(n,m)}(0,x) \right | 
	& \leq O(\oneOd^{n+m} )
	\\ 
	\lbeq{2-M.S.bd}
	\sum_x \left | S^{(n)}(0,x) \right | 
	& \leq O(\oneOd^{n} ) \, O( h^{\delta(d)}), 
	\\
	\lbeq{2-M.U.bd}
	\sum_x \sum_{n, m} \left | {U}_h^{(n,m)} (0,x) \right |
	& \leq \ohone . 
\enalign
\end{lemma}

\Proof
Each of the the above quantities is given in 
\refeq{xin0def}--\refeq{UNdef} by a nested expectation
in which one of the expectations involves the factor $W'$ or $W''$ defined in
\refeq{alndef}--\refeq{alnppdef}, or a factor of $F_{4}$.  
In bounding them, we take absolute values and bound
the difference in $W'$ or $W''$ using the triangle inequality.  The absolute
values are taken inside the sum over $x$ defining the Fourier transform,
using $|e^{ik \cdot x}| \leq 1$.  This gives bounds uniform in $k$.

Our general strategy is the same as that in 
Section~\ref{sec-part.exp.bd}, which is to bound nested expectations 
from right to left, and then decompose the resulting diagram 
having lines  
$\tau_{0,p_{c}}$ or $\tau_{h, p_{c}}$ into triangles and squares.  
In the following, we comment on the special features relevant for each 
quantity.

\smallskip\noindent
\textbf{ Bounds on $\hat{\xi}^{(n,m)}_h$ and $\hat{\Xi}^{(n,m)}_h$.}
These two terms are almost the same, and we discuss only $\hat{\xi}^{(n,m)}_h$.
We begin with $\xi^{(n,0)}_h$, which is bounded by 
\eq
	\xi^{(n,0)}_h(0,x) \leq 
	\Ebold_0 I[E_{0}''] \Ebold_1 Y_{1}''  \cdots 
	\Ebold_{n-1} Y_{n-1}'' \Ebold_n (I[(F_{3}')_{n}] + I[(F_{5}')_{n}]) , 
\en 
where $(F_{j}')_{n}$ denotes the event $F_{j}'$ on 
level-$n$.  The rightmost expectation can be bounded using BK by 
\eqalign
	\langle I[(F_{3}')_{n}] + I[(F_{5}')_{n}] \rangle_{n}
	&
	\leq 
	\sum_{w_{n-1} \in \tilde{C}_{n-1}} 
	\biggl \langle 
	I \biggl [ 
	\myeqnpic{(170,100)}{(-35,0)}{ %begin my-eqn-pic %%%%%
	    \thinlines
		\drawline(0,80)(50,80)(50,20)(100,20)
    		\put(-30,90){$v_{n-1}$}
    		\put( 50,90){$x$}
    		\put( 20, 0){$w_{n-1}$}
		\put(110, 0){$G$} 
	} % end of my-eqn-pic  %%%%%%%%%%%%%%%%%%%%%%%%%%%%%%%% 
	\biggr] 
	+ 
	I \biggl [ 
	\myeqnpic{(150,100)}{(-45,0)}{ %begin my-eqn-pic %%%%%
	    \thinlines
		\drawline(0,80)(50,80)
		\drawline(0,20)(25,80)
		\put( 50,20){\pictrir}
    		\put(-40,90){$v_{n-1}$}
    		\put( 85,45){$x$}
    		\put( 30, 0){$w_{n-1}$}
    		\put(-20,-10){$G$}
		\put( 25,80){\circle*{8}}
	} % end of my-eqn-pic  %%%%%%%%%%%%%%%%%%%%%%%%%%%%%%%% 	
	\biggr] 
	\biggr \rangle
	\nnb 
	& 
	\leq 
	M_{h} \sum_{w_{n-1} \in \tilde{C}_{n-1}} 
	\biggl \{ 
	\myeqnpic{(100,100)}{(-35,0)}{ %begin my-eqn-pic %%%%%
	    \thicklines
		\drawline(0,80)(50,80)(50,20)
    		\put(-30,90){$v_{n-1}$}
    		\put( 50,90){$x$}
    		\put( 30, 0){$w_{n-1}$}
	} % end of my-eqn-pic  %%%%%%%%%%%%%%%%%%%%%%%%%%%%%%%% 
	+ 
	\myeqnpic{(120,100)}{(-35,0)}{ %begin my-eqn-pic %%%%%
	    \thicklines
		\drawline(0,80)(50,80)
		\put( 50,20){\pictrir}
    		\put(-30,90){$v_{n-1}$}
    		\put( 85,45){$x$}
    		\put( 30, 0){$w_{n-1}$}
		\put( 25,80){\circle*{8}}
	} % end of my-eqn-pic  %%%%%%%%%%%%%%%%%%%%%%%%%%%%%%%% 	
	\biggr \} . 
\enalign
The other expectations involve $Y_{j}''$ and have already been 
bounded by \refeq{F1ppbd.2}.  The overall result is diagrams which are 
either given by the diagrams that bound $\hat{\phi}_h(k)$ in
Section~\ref{sub-bound.basic}, but
with an additional vertex added in the penultimate loop, or by
the diagrams that arose in bounding $\hat{R}_h(k)$ in Section~\ref{sub-bd.rem}.
These diagrams all have critical dimension $6$ (see \refeq{dcdef})
and are $O(\oneOd^n)$.

The bounds on $\xi^{(n,m)}_h$ for $m \geq 1$ are similar.  
We estimate expectations from right to left, as usual.
The estimate of the expectation at level-$(n+1)$ introduces
a factor $I[w_{n} \in \tilde{C}_{n} ]$, and we wish to bound
\eq
	\langle (W'')_{n} I[w_{n} \in \tilde{C}_{n} ] 
	\rangle_{n} 
	\leq 
	\langle 
	I[F_{3}' \ON \tilde{C}_{n} \AND w_{n} \in \tilde{C}_{n}] 
	+ I[F_{5}' \ON \tilde{C}_{n} \AND w_{n} \in \tilde{C}_{n}] 
	\rangle_{n} .	
\en 
This is bounded by  
\eqalign
	& 
	\sum_{w_{n-1} \in \tilde{C}_{n-1}} 
	\biggl \langle I \biggl [ 
	\myeqnpic{(210,100)}{(-35,0)}{ %begin my-eqn-pic %%%%%
		\drawline(0,80)(50,80)(50,20)(150,20)
		\drawline(100,20)(100,80)
    		\put(-30,90){$v_{n-1}$}
    		\put( 40,90){$u_{n}$}
    		\put( 20, 0){$w_{n-1}$}
		\put(160,10){$G$} 
		%\put(100,20){\circle*{8}}
		\put(90,85){$w_{n}$}
	} % end of my-eqn-pic  %%%%%%%%%%%%%%%%%%%%%%%%%%%%%%%% 
	\biggr ] 
	+ I \biggl [ 
	\myeqnpic{(180,100)}{(-45,0)}{ %begin my-eqn-pic %%%%%
		\drawline(0,80)(50,80)(50,20)(100,20)
		\drawline(50,50)(100,50)
    		\put(-30,90){$v_{n-1}$}
    		\put( 40,90){$u_{n}$}
    		\put( 10, 0){$w_{n-1}$}
 		\put(110,10){$G$} 
		\put(100,60){$w_{n}$}
		%\put( 50,50){\circle*{8}}
	} % end of my-eqn-pic  %%%%%%%%%%%%%%%%%%%%%%%%%%%%%%%% 
	\biggr ] 
	+ I \biggl [ 
	\myeqnpic{(180,100)}{(-45,0)}{ %begin my-eqn-pic %%%%%
		\drawline(-10,80)(50,80)(50,20)(100,20)
		\drawline(20,80)(-10,20)
    		\put(-40,90){$v_{n-1}$}
    		\put( 50,90){$u_{n}$}
    		\put( 20, 0){$w_{n-1}$}
		\put(-40, 0){$w_{n}$} 
    		\put(110,10){$G$}
		%\put(  0,20){\circle*{8}}
	} % end of my-eqn-pic  %%%%%%%%%%%%%%%%%%%%%%%%%%%%%%%% 
	\biggr ] 
	\nnb 
	& \quad
	+ I \biggl [ 
	\myeqnpic{(220,100)}{(-45,0)}{ %begin my-eqn-pic %%%%%
		\drawline(0,80)(50,80)(50,20)(100,20)(100,80)(50,80)
		\drawline(100,80)(150,80)
		\drawline(0,55)(25,80)
		\put(-30,35){$G$}
    		\put(-40,90){$v_{n-1}$}
    		\put(140,90){$w_{n}$}
    		\put( 20, 0){$w_{n-1}$}
    		\put(100, 0){$u_{n}$}
		\put( 25,80){\circle*{8}}
	} % end of my-eqn-pic  %%%%%%%%%%%%%%%%%%%%%%%%%%%%%%%% 
	\biggr ] 
	+ I \biggl [ 
	\myeqnpic{(160,100)}{(-40,0)}{ %begin my-eqn-pic %%%%%
		\drawline(0,80)(100,80)
		\put( 50,20){\pictrim}
		\drawline(0,55)(25,80)
		\put(-30,35){$G$}
    		\put(-40,90){$v_{n-1}$}
    		\put( 90,90){$w_{n}$}
    		\put(-10, 0){$w_{n-1}$}
    		\put( 70, 0){$u_{n}$}
		\put( 25,80){\circle*{8}}
	} % end of my-eqn-pic  %%%%%%%%%%%%%%%%%%%%%%%%%%%%%%%% 
	\biggr ] 
	+ I \biggl [ 
	\myeqnpic{(160,100)}{(-40,0)}{ %begin my-eqn-pic %%%%%
		\drawline(0,80)(100,80)
		\put( 50,20){\pictrim}
		\drawline(0,50)(50,65)
		\put(-30,35){$G$}
    		\put(-40,90){$v_{n-1}$}
    		\put( 90,90){$w_{n}$}
    		\put(-10, 0){$w_{n-1}$}
    		\put( 70, 0){$u_{n}$}
		\put( 50,65){\circle*{8}}
	} % end of my-eqn-pic  %%%%%%%%%%%%%%%%%%%%%%%%%%%%%%%% 
	\biggr ] 
	+ I \biggl [ 
	\myeqnpic{(160,100)}{(-35,0)}{ %begin my-eqn-pic %%%%%
		\drawline(0,80)(100,80)
		\drawline(100,55)(75,80)
		\put(100,35){$G$}
		\put( 50,20){\pictrim}
    		\put(-35,90){$v_{n-1}$}
    		\put( 90,90){$w_{n}$}
    		\put(-10, 0){$w_{n-1}$}
    		\put( 70, 0){$u_{n}$}
		\put( 75,80){\circle*{8}}
	} % end of my-eqn-pic  %%%%%%%%%%%%%%%%%%%%%%%%%%%%%%%% 	
	\biggr ] \biggr \rangle , 
	\lbeq{Wppw-event.2}
	\\ 
\intertext{which is bounded using BK by} 
	& 
	M_{h} \sum_{w_{n-1} \in \tilde{C}_{n-1}} 
	\biggl \{ 
	\myeqnpic{(190,100)}{(-35,0)}{ %begin my-eqn-pic %%%%%
	    \thicklines
		\drawline(0,80)(50,80)(50,20)(150,20)
    		\put(-30,90){$v_{n-1}$}
    		\put( 40,90){$u_{n}$}
    		\put( 20, 0){$w_{n-1}$}
		\put(140, 0){$w_{n}$} 
		\put(100,20){\circle*{8}}
	} % end of my-eqn-pic  %%%%%%%%%%%%%%%%%%%%%%%%%%%%%%%% 
	+ 
	\myeqnpic{(140,100)}{(-35,0)}{ %begin my-eqn-pic %%%%%
	    \thicklines
		\drawline(0,80)(50,80)(50,20)
		\drawline(0,20)(100,20)
    		\put(-30,90){$v_{n-1}$}
    		\put( 40,90){$u_{n}$}
    		\put(-30, 0){$w_{n-1}$}
		\put( 90, 0){$w_{n}$} 
	} % end of my-eqn-pic  %%%%%%%%%%%%%%%%%%%%%%%%%%%%%%%% 
	+ 
	\myeqnpic{(140,100)}{(-35,0)}{ %begin my-eqn-pic %%%%%
	    \thicklines
		\drawline(0,20)(50,20)(50,80)
		\drawline(0,80)(100,80)
    		\put(-30,90){$v_{n-1}$}
    		\put( 50, 0){$u_{n}$}
    		\put(-30, 0){$w_{n-1}$}
		\put( 90,90){$w_{n}$} 
	} % end of my-eqn-pic  %%%%%%%%%%%%%%%%%%%%%%%%%%%%%%%% 
	\nnb 
	& \hspace{20mm}
	+ 
	\myeqnpic{(200,100)}{(-35,0)}{ %begin my-eqn-pic %%%%%
	    \thicklines
		\drawline(0,80)(50,80)(50,20)(100,20)(100,80)(50,80)
		\drawline(100,80)(150,80)
    		\put(-40,90){$v_{n-1}$}
    		\put(140,90){$w_{n}$}
    		\put( 20, 0){$w_{n-1}$}
    		\put(100, 0){$u_{n}$}
		\put( 25,80){\circle*{8}}
	} % end of my-eqn-pic  %%%%%%%%%%%%%%%%%%%%%%%%%%%%%%%% 	
	+ 
	\myeqnpic{(160,100)}{(-40,0)}{ %begin my-eqn-pic %%%%%
	    \thicklines
		\drawline(0,80)(100,80)
		\put( 50,20){\pictrim}
    		\put(-40,90){$v_{n-1}$}
    		\put( 90,90){$w_{n}$}
    		\put(-10, 0){$w_{n-1}$}
    		\put( 70, 0){$u_{n}$}
		\put( 25,80){\circle*{8}}
	} % end of my-eqn-pic  %%%%%%%%%%%%%%%%%%%%%%%%%%%%%%%% 	
	+ 
	\myeqnpic{(160,100)}{(-40,0)}{ %begin my-eqn-pic %%%%%
	    \thicklines
		\drawline(0,80)(100,80)
		\put( 50,20){\pictrim}
    		\put(-40,90){$v_{n-1}$}
    		\put( 90,90){$w_{n}$}
    		\put(-10, 0){$w_{n-1}$}
    		\put( 70, 0){$u_{n}$}
		\put( 50,65){\circle*{8}}
	} % end of my-eqn-pic  %%%%%%%%%%%%%%%%%%%%%%%%%%%%%%%% 	
	+ 
	\myeqnpic{(160,100)}{(-35,0)}{ %begin my-eqn-pic %%%%%
	    \thicklines
		\drawline(0,80)(100,80)
		\put( 50,20){\pictrim}
    		\put(-35,90){$v_{n-1}$}
    		\put( 90,90){$w_{n}$}
    		\put(-10, 0){$w_{n-1}$}
    		\put( 70, 0){$u_{n}$}
		\put( 75,80){\circle*{8}}
	} % end of my-eqn-pic  %%%%%%%%%%%%%%%%%%%%%%%%%%%%%%%% 	
	\biggr \} . 
	\lbeq{Wppbd.2}
\enalign 
In the above, we used the fact that $v_{n-1}' \dbc u_{n}$ 
\emph{through} $\tilde{C}_{n-1}$, so that we can choose $w_{n-1}$ on 
either side of the two disjoint paths connecting $v_{n-1}'$ and $u_{n}$. 
Combined with the bound \refeq{F1ppbd.2} on $Y_{j}''$,  
the result can be seen to be bounded by diagrams with critical dimension
$6$.  Bounding these diagrams in terms of triangles yields
\eq
	\bigl | \hat{\xi}_h^{(n,m)}(k) \bigr | 	
	\leq O(\oneOd^{n+m} ) M_h . 
\en

%%%%%%%%%%%%%%%%%%%%%%%%%%%%%%%%%%%%%%%%%%%%%%%%%%%%%%%%%%%%%%%%%%%%%%%%%%%%%%%
%\subsubsection{Bound on $\hat{u}^{(n,m)}_h$}
%\label{subsub-2Mbd.U}
\smallskip\noindent
\textbf{Bound on $\hat{u}^{(n,m)}_h$.}
The term $\hat{u}^{(n,m)}_h$ is almost the same as 
the term $r_{h}^{(n)}$ bounded in Section~\ref{sub-bd.rem}, 
except that $\hat{u}^{(n,m)}_h$ contains $W''$ rather than $Y''$
in one internal expectation.
The bounds proceed exactly as in the proof of Lemma~\ref{lem-Rnbd},
except that \refeq{Wppbd.2} is used for the level-$n$ expectation.
The bound on the level-$n$ expectation introduces an additional
vertex, compared to the bound on $r_{h}^{(n)}$, which raises
the critical dimension to $6$.  This leads to
\eq
	\left | \hat{u}_h^{(n,m)}(k) \right | 	
	\leq O(\oneOd^{n+m} ) M_h \chi_h = O(\oneOd^{n+m} ) , 
\en 
where in the last step we used the bound $\chi_h M_h = O(1)$
of Lemma~\ref{lem-chiM.bd}.

%%%%%%%%%%%%%%%%%%%%%%%%%%%%%%%%%%%%%%%%%%%%%%%%%%%%%%%%%%%%%%%%%%%%%%%%%%%%%%%
%\subsubsection{Bound on $\protect\Hat{\protect\Tilde{R}}^{(n)}$}
\smallskip\noindent
\textbf{Bound on $\protect\hat{S}^{(n)}_h$.}
We bound $\hat{S}^{(n)}_h$ in
the manner illustrated for the case $n=1$ in
Example~\ref{ex-R1Sh}.  In this method, a $G$-free line is extracted from the
level-$(n-1)$ expectation.  The result is
\eq
	\bigl | \hat{S}^{(n)}_h(k) \bigr | 	
	\leq 
	O(\oneOd^{n} ) \,O(h^{\delta(d)}) .
\en

Since Example~\ref{ex-R1Sh} involved the extraction of a $G$-free
line from the level-$0$ expectation, we now describe in more
detail how the corresponding step is performed for the level-$(n-1)$
expectation, when $n>1$. 
When $n>1$, after bounding expectations at levels $n$ and higher,
the level-$(n-1)$ expectation is given by 
\eqalign 
	& 
	\bigl \langle 
	Y_{n-1}'' \, I[w_{n-1} \in \tilde{C}_{n-1} ] 
	\bigr \rangle_{n-1}
	= 
	\bigl \langle 
	Y_{n-1}' \, I[w_{n-1} \in C_{n-1} ] 
	\bigr \rangle_{n-1}^{\tilde{}}
	\nnb 
	& 
	= 
	\bigl \langle 
	 I[F_{1, b}' \AND w_{n-1} \in C_{n-1} 
	 \AND C_{n-1} \cap G = \emptyset] 
	\bigr \rangle_{n-1}^{\tilde{}}
	\nnb 
	& 
	\leq 
	\Biggl \langle 
	I \biggl [ 
	\thinlines\picFoneppwNMONEa 
	\bigcup \thinlines\picFoneppwNMONEb
	\biggr ] \, 
	I [ C_{n-1} \cap G = \emptyset ] I[w_{n-1} \in C_{n-1} ]
	\Biggr \rangle_{n-1}^{\tilde{}} ,
\enalign
using the conditional expectation introduced in Example~\ref{ex-R1Sh}.
Using Lemma~\ref{lem-onehline} 
to bound the above by corresponding diagrams, this gives
\eq 
\lbeq{goodY1bd}
	\bigl \langle 
	Y_{n-1}'' \, I[w_{n-1} \in \tilde{C}_{n-1} ] 
	\bigr \rangle_{n-1}
	\leq 
	\biggl [ 
	\picFoneppwGfreeNMONEa 
	+ \thicklines\picFoneppwGfreeNMONEb
	\biggr ] I[w_{n-1} \in C_{n-1} ], 
\en 
where the thick solid lines represent $\tau_{p_{c}}$, and the dotted lines 
represent $\tau_{h, p_{c}}$, both in the \emph{conditional}\/ expectation 
with $\{u_{n-1}, v_{n-1}\}$ vacant.   The conditional expectation
can then be handled as in Example~\ref{ex-R1Sh}.  An example 
of a typical diagram arising in bounding $\hat{S}^{(4)}$ is
\eq	
	\myeqnpic{(300,140)}{(-35,0)}{ %begin my-eqn-pic %%%%%
	    \thicklines
		\drawline( 10,80)( 85,80)
		\drawline(125,80)(200,80)
		\drawline(  0,20)( 35,20)
		\drawline( 75,20)(100,20)
		\dottedline{6}(100,20)(200,20)
		\drawline(150,80)(150,120)
		\drawline(230,50)(270,50)
		\drawline(215,65)(255,85)
		\put(  0,80){\picpivone}
		\put( 65,20){\picpivtwo}
		\put(115,80){\picpivone}
		\put(  0,20){\pictril}
		\put(200,20){\pictrir}
		\put( 50,20){\pictrim}
		\put(100,20){\pictrimu}
    		\put(-50,40){$0$}
		\put(155,110){$G$}
		\put(260,80){$G$}
	} % end of my-eqn-pic  %%%%%%%%%%%%%%%%%%%%%%%%%%%%%%%% 	
	\qquad . 
\en
The resulting bound is
\eq
	\bigl | \hat{S}^{(n)}_h(k) \bigr | 	
	\leq \chi_h M_h^2 {\sf S}_h O(\oneOd^n) =
	O(\oneOd^{n} ) \,O(h^{\delta(d)}) .
\en

\smallskip\noindent
\textbf{Bound on $\hat{U}^{(n,m)}_h$.}
This bound is the most involved one.  By definition, 
$U^{(n,m)}_h$ contains one $W''$ at level-$n$ and one $F_2$ at level-$(n+m)$.  
The bounds \refeq{Wppbd.2} on $W''$ and \refeq{F2bd.2} on $F_{2}$ 
each introduce an additional vertex, and when combined, give
rise to a diagram with critical dimension 8.
For some of the diagrams, we can use the method of Example~\ref{ex-R1Sh},
but for others we are unable to extract a $G$-free line to compensate
for a subdiagram with critical dimension 8 and we must resort instead
to the dominated convergence method of Example~\ref{ex-R1dct}.

We first consider the case
$m \geq 2$, for which there is at least one expectation of $Y_{j}''$ occuring 
between the level-$n$ expectation of $W''$ and the level-$(n+m)$
expectation of $F_{2}$.  This allows us to use the
bound \refeq{goodY1bd} on $Y_{j}''$ involving one $G$-free line, 
for one of the expectations between levels-$n$ and 
$(n+m)$.  The resulting diagrams can be bounded in terms of ${\sf S}_h$
and $n+m-1$ triangles, with each triangle contributing $O(\oneOd)$.  
A typical diagram contributing in this case is
\eq
        \myeqnpic{(300,140)}{(-35,-40)}{ %begin my-eqn-pic %%%%%
            \thicklines
                \drawline(10,80)( 50,80)
                \drawline(60,80)(100,80)
                \dottedline{6}(100,80)(150,80)
                \drawline(150,80)(200,80)
                \drawline(210,80)(300,80)
                \drawline(  0,20)( 85,20)
                \drawline(125,20)(250,20)
                \drawline( 50,20)( 50,80)
                \drawline(150,20)(150,80)
                \drawline(200,20)(200,80)
                \drawline(250,-20)(250,80)
                \drawline( 70,20)( 70,-20)
                \put(  0,80){\picpivone}
                \put( 50,80){\picpivone}
                \put(200,80){\picpivone}
                \put(115,20){\picpivtwo}
                \put(  0,20){\pictril}
                \put(100,20){\pictrim}
                \put(-50,40){$0$}
                \put( 75,-35){$G$}
                \put(255,-35){$G$}
        } % end of my-eqn-pic  %%%%%%%%%%%%%%%%%%%%%%%%%%%%%%%%         
	\qquad .
\en
This gives the bound
\eq
\lbeq{Rmn.mgeq2}
	\left| \hat{U}_h^{(n,m)}(k) \right|
	\leq O(\oneOd^{n+m-1}) {\sf S}_h M_h^2 \chi_h
	\leq O(\oneOd^{n+m-1})O(h^{\delta(d)}) \quad \quad (m \geq 2). 
\en

Next, we consider the case $m = 1$, in which
$W''$ and $F_{2}$ appear in the two innermost expectations. 
For the innermost expectation of $F_{2}$, we use \refeq{F2bd.2}
(with the shift $j \to n+1$ in indices).
To bound $W''$ on level-$n$, we use \refeq{Wppw-event.2},
and give two separate arguments, one for the first and sixth terms
on the right side of \refeq{Wppw-event.2}, and one for the second
through fifth terms.

The contributions from the second through fifth terms of 
\refeq{Wppw-event.2} can be handled using the bound
\refeq{goodY1bd} to estimate $Y_{n-1}''$.
The resulting diagrams can be bounded above by ${\sf S}_h$ and
$n$ triangles, yielding an overall bound by 
$O(\oneOd^n {\sf S}_h M_h^2 \chi_h) = O(\oneOd^n)O( h^{\delta(d)})$.
However, this method does not apply to the first and sixth
terms of \refeq{Wppw-event.2}, because it leads to diagrams in
which a square subdiagram arises from the innermost expectation
in such a way that we cannot extract a $G$-free line to produce
${\sf S}_h$.   An example is the diagram
\eq
        \myeqnpic{(250,140)}{(-35,-40)}{ %begin my-eqn-pic %%%%%
            \thicklines
                \drawline(10,80)( 85,80)
                \drawline(125,80)(250,80)
                \drawline(  0,20)( 35,20)
                \drawline( 75,20)(200,20)
                \drawline(150,-20)(150,20)
                \drawline(200,-20)(200,80)
                \put(  0,80){\picpivone}
                \put(115,80){\picpivone}
                \put( 65,20){\picpivtwo}
                \put(  0,20){\pictril}
                \put(100,20){\pictrimu}
                \put( 50,20){\pictrim}
                \put(-50,40){$0$}
                \put(125,-30){$G$}
                \put(205,-30){$G$}
        } % end of my-eqn-pic  %%%%%%%%%%%%%%%%%%%%%%%%%%%%%%%%         
	\qquad . 
\en

For these remaining two cases, we use the dominated convergence theorem.
As an upper bound, we neglect the connection to $G$ required by $W''$,
to obtain
\eq
	(\textrm{first and sixth terms of 
	\refeq{Wppw-event.2}}) 
	\leq 
	\sum_{w_{n-1} \in \tilde{C}_{n-1}} 
	\biggl \{ 
	\myeqnpic{(140,100)}{(-35,0)}{ %begin my-eqn-pic %%%%%
	    \thicklines
		\drawline(0,80)(50,80)(50,20)(100,20)
    		\put(-30,90){$v_{n-1}$}
    		\put( 40,90){$u_{n}$}
    		\put( 20, 0){$w_{n-1}$}
		\put( 90, 0){$w_{n}$} 
	} % end of my-eqn-pic  %%%%%%%%%%%%%%%%%%%%%%%%%%%%%%%% 
	+ 
	\myeqnpic{(160,100)}{(-35,0)}{ %begin my-eqn-pic %%%%%
	    \thicklines
		\drawline(0,80)(100,80)
		\put( 50,20){\pictrim}
    		\put(-35,90){$v_{n-1}$}
    		\put( 90,90){$w_{n}$}
    		\put(-10, 0){$w_{n-1}$}
    		\put( 70, 0){$u_{n}$}
	} % end of my-eqn-pic  %%%%%%%%%%%%%%%%%%%%%%%%%%%%%%%% 	
	\biggr \} . 
	\lbeq{Wppbd.smp}
\en 
This leads to diagrams with critical dimension 6, which are $O(\oneOd^{n})$.
However, each term with fixed $x$ is a huge sum over various 
vertices and pivotal bonds.  Having fixed all of them, 
the summand, being a nested expectation of an indicator function, 
and having a connection to $G$, is bounded above by $M_h$. 
In particular, it  goes to zero pointwise as $M_h \to 0$.
Hence, by the dominated convergence theorem, the sum over $n$ of
these contributions
is bounded by $\chi_hM_h \oh = \oh$.

Combining the above yields the bound 
\eq
	\sum_{m,n} \left | \hat{U}_h^{(n,m)} (k) \right | \leq \oh .
\en 

This completes the proof of Lemma~\ref{lem-xiXiURbds}.
\qed

%%%%%%%%%%%%%%%%%%%%%%%%%%%%%%%%%%%%%%%%%%%%%%%%%%%%%%%%%%%%%%%%%%%%%%%%%%%%%%%
%%%%%%%%%%%%%%%%%%%%%%%%%%%%%%%%%%%%%%%%%%%%%%%%%%%%%%%%%%%%%%%%%%%%%%%%%%%%%%%
\subsection{Asymptotic behaviour of the numerator}
\label{sub-2Mbd.num}

Fix $p=p_c$.
We now apply the bounds of Lemma~\ref{lem-xiXiURbds} to prove
that the numerator of \refeq{taugPiUkinf} is given by
\eq
\lbeq{numasy}
	\hat{\phi}_h(k)  
	+ \hat{\xi}_h(k) 
	+ \hat{U}_h(k) 
	+ \hat{S}_h(k) 
	= \hat{\phi}_0(0) + \ohone + O(k^2) .
\en
The constant $\hat{\phi}_0(0)$ is equal to
$1+O(\oneOd)$, by \refeq{f00bd.0} and Lemma~\ref{lem-Phinbd}.

Summation of \refeq{2-M.xi.bd}--\refeq{2-M.S.bd} over $m,n$, together with 
\refeq{2-M.U.bd}, gives
$| \hat{\xi}_h(k) + \hat{U}_h(k) + \hat{S}(k)|  \leq \oh $. 
To prove \refeq{numasy}, it therefore suffices to show that
\eq
	\hat{\phi}_h(k) 
	= \hat{\phi}_0(0) + O(h^{1/2}) + O(k^2) .
	\lbeq{2-M.R.bd0}
\en
For this, we make the decomposition 
\eq
\lbeq{fhksplit}
	\hat{\phi}_h(k) = \hat{\phi}_0(0) 
	- \left [ \hat{\phi}_0(0) - \hat{\phi}_h(0) \right ] 
	- \left [ \hat{\phi}_h(0) - \hat{\phi}_h(k) \right ] . 
\en 
The second term of \refeq{fhksplit} is $O(M_h)$, by \refeq{f00bd.0}
and the analogue of \refeq{Phinbd.2} for $\hat{\phi}_h(0)$.
The third term is $O(k^2)$, by \refeq{f00bd.0} and \refeq{fnbd}.
This proves \refeq{numasy}.

%%%%%%%%%%%%%%%%%%%%%%%%%%%%%%%%%%%%%%%%%%%%%%%%%%%%%%%%%%%%%%%%%%%%%%%%%%%%%%%
%%%%%%%%%%%%%%%%%%%%%%%%%%%%%%%%%%%%%%%%%%%%%%%%%%%%%%%%%%%%%%%%%%%%%%%%%%%%%%%
\subsection{$k$-dependence of the denominator}
\label{sub-2Mbd.k-den}

Fix $p=p_c$.
Since $\hat{\Phi}_0(0)=1$ by \refeq{Phi00is1}, 
the denominator of \refeq{taugPiUkinf} 
can be written as
\eq
\lbeq{denom.decomp}
	1 - \hat{\Phi}_h(k) - \hat{\Xi}_h(k) =
	\left [ \Phihat_h(0) - \Phihat_h(k) \right ]
	+ \left [ \hat{\Xi}_h (0) - \hat{\Xi}_h(k)  \right ]
	+ \left [ \Phihat_0(0) - \Phihat_h(0) - \hat{\Xi}_h (0) \right ].
\en
The last term is 
independent of $k$ and will be treated in Section~\ref{sec-Psidef}
using the second expansion.  
In this section, we prove that
\eqarray
	\hat{\Xi}_h(0) - \hat{\Xi}_h(k) & = & \ok h^{1/2} ,
\lbeq{Xi-k-dep}
	\\
	\Phihat_h(0) - \Phihat_h(k) & = & 	
	 - \nabla_k^2 \Phihat_0(0)  \,  \frac{k^2}{2d} 
	+ \ok k^2 
	+ \oh k^2 .
\lbeq{Phi-k-dep}
\enarray
This shows that the second term in \refeq{denom.decomp} is an 
error term, and extracts the leading $k^2$-dependence of the first term.
The constant $-\nabla_k^2 \Phihat_0(0)$ of \refeq{Phi-k-dep} was seen to
be finite and positive in \refeq{nablaPhibd}.  

\smallskip \noindent
{\bf Proof of \refeq{Xi-k-dep}.}
By the triangle inequality,
\eq
	h^{-1/2} |\hat{\Xi}_h(0) - \hat{\Xi}_h(k) |
	\leq \sum_{n=1}^\infty \sum_{m=0}^\infty \sum_x 
	h^{-1/2} |\Xi_h^{(n,m)} (x)| \,  |1 - \cos (k \cdot x) | . 
\en
It was shown in Lemma~\ref{lem-xiXiURbds} that
$h^{-1/2}\sum_x  |\Xi_h^{(n,m)} (x)| \leq O(\oneOd^{n+m} )$.
Thus the right side is summable, uniformly in $h$ and $k$.  On the other hand,
the summand on the right side goes to zero as $k \to 0$.  The 
dominated convergence theorem then gives \refeq{Xi-k-dep}.
\qed

\smallskip \noindent
{\bf Proof of \refeq{Phi-k-dep}.}
We begin with the decomposition
\eq
\lbeq{Phi-k-decomp}
	\Phihat_h(0) - \Phihat_h(k) 
	= \left [ \Phihat_0(0) - \Phihat_0(k) \right ]
	- \left [  [ \Phihat_0(0) - \Phihat_0(k) ]
	- [ \Phihat_h(0) - \Phihat_h(k) ] 
	\right ].
\en
The first term on the right side can be written as
\eq 
\lbeq{Phimain0k}
	\Phihat_0(0) - \Phihat_0(k) 
	= - \nabla_k^2 \Phihat_0(0) \frac{k^2}{2d}
	+ k^2 \sum_x \Phi_0(0,x) 
	k^{-2} \left ( 1 - \cos (k \cdot x) - \frac{(k \cdot x)^2}{2} 
	\right ) ,
\en
since the first term and the contribution to the second term
from $- (k \cdot x)^2 /2$ cancel by symmetry. 
The summand of the second term is bounded uniformly in $k$ by a summable
function of $x$, since
$k^{-2}|1 - \cos (k \cdot x) - \frac{1}{2}(k \cdot x)^2| \leq O(x^2)$
and $\sum_x x^2 |\Phi(0,x)| < \infty$ by \refeq{x2Phin}--\refeq{x2Phi00}.
Since $k^{-2} [1 - \cos (k \cdot x) - \frac{(k \cdot x)^2}{2}] \to 0$ 
pointwise in $x$ as $k \to 0$, the second term of \refeq{Phimain0k}
is $\ok k^2$ by the dominated convergence theorem.
Therefore
\eq 
\lbeq{2ndexp.den.kk}
	\Phihat_0(0) - \Phihat_0(k) 
	= - \nabla_k^2 \Phihat_0(0) \frac{k^2}{2d}
	+ \ok k^2 .
\en

The absolute value of the second term on the right side 
of \refeq{Phi-k-decomp} is bounded above by 
\eq
	\sum_x \left | \left \{ \Phi_0(0,x) - \Phi_h (0,x) \right \} 
	( 1 - \cos (k \cdot x) ) \right | 
	\leq \frac{k^2}{2d} \sum_x \sum_{n=0}^\infty  
	|x|^2 
	\left | \Phi_0^{(n)}(0,x) - \Phi_h^{(n)} (0,x) \right | .
\en 
By \refeq{Phi0-bond}--\refeq{Phin-bond}, the
summand on the right side goes to zero pointwise as $h \to 0$, and it 
is bounded by $|x|^2 | \Phi_0^{(n)}(0,x)|$, which is summable in $x,n$ 
by \refeq{x2Phin}--\refeq{x2Phi00}.  It therefore follows from the dominated
convergence theorem that

\eq
	\left | [ \Phihat_0(0) - \Phihat_0(k) ]
	- [ \Phihat_h(0) - \Phihat_h(k)  ] \right | 
	\leq \oh \frac{k^2}{2d} .
	\lbeq{2ndexp.den.cross}
\en 
Equation~\refeq{Phi-k-dep} then follows from 
\refeq{2ndexp.den.cross} and \refeq{2ndexp.den.kk}.
\qed

Equation~\refeq{tauirpc} of 
Theorem~\ref{thm-tauasymp} is now an immediate consequence of
\refeq{tauhat0limex}, \refeq{2-M.R.bd0} and \refeq{Phi-k-dep},
with $CD^{-2}$ equal to 
$\hat{\phi}_0(0)[-\frac{1}{2d}\nabla_k^2\hat{\Phi}_0(0)]^{-1}$.

%%%%%%%%%%%%%%%%%%%%%%%%%%%%%%%%%%%%%%%%%%%%%%%%%%%%%%%%%%%%%%%%%%%%%%%%%%%%%%%
%%%%%%%%%%%%%%%%%%%%%%%%%%%%%%%%%%%%%%%%%%%%%%%%%%%%%%%%%%%%%%%%%%%%%%%%%%%%%%%
%%%%%%%%%%%%%%%%%%%%%%%%%%%%%%%%%%%%%%%%%%%%%%%%%%%%%%%%%%%%%%%%%%%%%%%%%%%%%%%
%%%%%%%%%%%%%%%%%%%%%%%%%%%%%%%%%%%%%%%%%%%%%%%%%%%%%%%%%%%%%%%%%%%%%%%%%%%%%%%
\section{The second expansion and refined $h$-dependence}
\label{sec-Psidef}

We will now complete the proof of Theorem~\ref{thm-tauasymp},
by establishing \refeq{tauasy}.
We fix $p=p_c$, and drop subscripts $p_c$ from the notation.
We assume without further mention that $d \gg 6$ for the
nearest-neighbour model, and that $d>6$ and $L \gg 1$ for the
spread-out model.

%%%%%%%%%%%%%%%%%%%%%%%%%%%%%%%%%%%%%%%%%%%%%%%%%%%%%%%%%%%%%%%%%%%%%%%%%%%%%%%
%%%%%%%%%%%%%%%%%%%%%%%%%%%%%%%%%%%%%%%%%%%%%%%%%%%%%%%%%%%%%%%%%%%%%%%%%%%%%%%
\subsection{The refined $h$-dependence}
\label{sub-Psi.overview}

It already follows from \refeq{numasy} and 
\refeq{denom.decomp}--\refeq{Phi-k-dep} that
\eq
	\hat{\tau}_h(k) = \frac{\hat{\phi}_0(0) + \oh + O(k^2)}
	{-\frac{1}{2d} \nabla_k^2 \hat{\Phi}_0(0) k^2 
	[1 + \okone + \oh] + \okone h^{1/2} +
	[\hat{\Phi}_0(0)- \hat{\Phi}_h(0) - \hat{\Xi}_h(0)]}
	.
\en
It therefore suffices to show that
\eq
\lbeq{propcons}
	\hat{\Phi}_0(0)- \hat{\Phi}_h(0) - \hat{\Xi}_h(0)
	=h^{1/2}[K + \ohone ], 
\en
for some positive constant $K$.
Equation~\refeq{tauasy} then follows, with
\eq
\lbeq{CD2}
	C = 2^{3/2} K^{-1} \hat{\phi}_0(0), \quad
	D^2 = -\frac{1}{2d} \nabla_k^2 \hat{\Phi}_0(0)2^{3/2} K^{-1}.
\en
These are positive constants, since $\hat{\phi}_0(0) = 1+O(\oneOd)$ as 
explained below \refeq{2-M.R.bd0}, and 
$-\nabla_k^2 \hat{\Phi}_0(0)$ is positive by \refeq{nablaPhibd}.
These values for $C$ and $D^{-2}$ are 
consistent with the identification of $CD^{-2}$ at the end of 
Section~\ref{sub-2Mbd.k-den}.

Equation~\refeq{propcons}
will be a consequence of the following two propositions.

\begin{prop}
\label{prop-Phi}
There is a positive constant $K_1$, with $K_1 = 1+O(\oneOd)$, 
such that for $h >0$,
\eq
	- \frac{d}{dh} \Phihat_h(0) = \left[ K_1 + \ohone \right] \chi_h .
\en
\end{prop} 

\begin{prop}
\label{prop-Xi}
There is a constant $K_2$, with $|K_2| \leq O(\oneOd)$, 
such that for $h >0$,
\eq
	- \hat{\Xi}_h(0) = \left[ K_2 + \ohone \right] M_h .
\en 
\end{prop}

\medskip
\noindent 
\textbf{Proof of \refeq{propcons} assuming Propositions~\ref{prop-Phi} 
and \ref{prop-Xi}.}
The two propositions imply that
\eq
\lbeq{denom.M}
	\hat{\Phi}_0(0)- \hat{\Phi}_h(0) - \hat{\Xi}_h(0) 
	= \int_0^h \left[ K_1 + \ou \right] \chi_u \, du 
	+ \left[ K_2 + \ohone \right] M_h
	= \left[ K_1 + K_2 + \oh \right] M_h.
\en
To prove \refeq{propcons}, it
therefore suffices to show that 
$M_h = \left[ \mbox{const.} + \oh \right] h^{1/2}$.
To see this, we note that by \refeq{denom.M}, 
\refeq{numasy} and \refeq{denom.decomp},
\eq
\lbeq{K1K2Mh}
	\chi_h = \tauhat_h(0) 
	= \frac{\hat{\phi}_0(0) + \oh } { [K_1 + K_2 + \oh] M_h  } .
\en 
Therefore 
\eq
	\frac{d M_h^2}{dh}  = 2 M_h \, \chi_h 
	= 2\hat{\phi}_0(0)(K_1 + K_2)^{-1} + \oh ,
\en 
and hence, by integrating and then taking the square root, we have
the desired result
\eq
	M_h =   [ (2\hat{\phi}_0(0))^{1/2}(K_1 + K_2)^{-1/2} + \oh] h^{1/2}.
\en 
The constant $K$ of \refeq{propcons} is therefore given by 
\eq
\lbeq{K2formula}
	K^2 = 2\hat{\phi}_0(0)(K_1 + K_2) = 2 +O(\oneOd),
\en
and \refeq{propcons} is proved. 
\qed

The proofs of Propositions~\ref{prop-Phi} and \ref{prop-Xi} are
similar, and both involve the use of a second expansion.
Before discussing the second expansion
for the derivative of $\Phihat_h(0)$ in detail, we begin by
considering the leading contribution.

%%%%%%%%%%%%%%%%%%%%%%%%%%%%%%%%%%%%%%%%%%%%%%%%%%%%%%%%%%%%%%%%%%%%%%%%%%%%%%%
%%%%%%%%%%%%%%%%%%%%%%%%%%%%%%%%%%%%%%%%%%%%%%%%%%%%%%%%%%%%%%%%%%%%%%%%%%%%%%%
\subsection{Leading behaviour of the $h$-derivative of $\hat{\Phi}_h(0)$}
\label{sub-Psi.leading}

By definition, 
$\Phi_{h}(0,x)= \sum_{j=0}^\infty\Phi_{h}^{(j)}(0,x)$,
with the $j^{\rm th}$ 
term in the sum given by \refeq{Phindef}.  The leading behaviour 
of $\Phi_h(0,x)$ is given by the $j=0$ term
\eq
	\Phi_h^{(0)}(0,v_0) = p_c \sum_{u_0 : v_0-u_0 \in \Omega}
	\langle I[E_0''(0,u_0,v_0)] \rangle .
\en
By definition, 
\eq
\lbeq{IE0dif}
	- \frac{d}{dh} \langle I[E_0''(0,u_0,v_0)] \rangle
	= \langle |C(0)| I[0 \dbc u_0]
	e^{-h|C(0)|} \rangle^{\tilde{}}
	= \sum_y  
	\langle I[0 \conn y \AND 0 \dbc u_0 \AND 0 \nc G]  \rangle^{\tilde{}},
\en
where we are using the notation introduced in Example~\ref{ex-R1Sh} in which
$\langle \cdot \rangle^{\tilde{}}$ denotes
expectation conditional on $\{u_0,v_0\}$ being vacant.  
We may now proceed to derive an expansion
for the connection $0 \conn y$, as in the argument leading up
to \refeq{tauhHR}.

For this, we introduce
\eqarray
	L_0'(0,u_0,a_0) & = & E_0'(0,u_0) \cap \{ 0 \dbc a_0 \} \\
	L_0''(0,u_0,a_0,b_0) & = & L_0'(0,u_0,a_0) \ON 
	\tilde{C}^{\{a_0,b_0\}}(0).
\enarray
As in \refeq{expan0}, the right side of \refeq{IE0dif} can be seen to
be given by
\eq
	\sum_y \langle I[L_0'(0,u_0,y)]  \rangle^{\tilde{}}
	+ \sum_y p_c\sum_{(a_0,b_0)}
	\langle I[L_0''(0,u_0,a_0,b_0)]  
	\tau_h^{\tilde{C}^{\{a_0,b_0\}}(0)}(b_0,y)
	\rangle^{\tilde{}}.
\en
Initially, the restricted two-point function in the second term should
be with respect to the conditional expectation 
$\langle \cdot \rangle^{\tilde{}}$ rather than the usual expectation.  However, 
there is no difference between the two expectations here,
since $u_0 \in \tilde{C}^{\{a_0,b_0\}}(0)$ when $L_0''(0,u_0,a_0,b_0)$
occurs.  Now we proceed as in the derivation of the one-$M$ scheme of the
expansion.  The result is
\eq
\lbeq{2ndexp-0}
	- \frac{d}{dh} \Phi_h^{(0)}(0,v_0)
	= \Ucal^{(0)}_h(0,v_0) + \Vcal^{(0)}_h(0,v_0) \chi_h
	+ \Ecal^{(0)}_h(0,v_0),
\en
where 
\eqalign
	\Ucal_h^{(0)}(0,v_0) 
	& = 
	\sum_y p_c \sum_{u_0  \in v_0 - \Omega }
	\langle I[L_0'(0,u_0,y)]  \rangle^{\tilde{}}
	+ \sum_y
	\sum_{\ell=1}^\infty (-1)^\ell 
	\tilde{\Ebold}_{0} I[L_0''] \, 
	{\Ebold}_{1}  Z_1'' \, 
	\cdots 
	{\Ebold}_{\ell-1}  Z_{\ell-1}'' \, 
	{\Ebold}_{\ell}  Z_{\ell}'  , 
	\\
\lbeq{Vcal0def}
	\Vcal_h^{(0)}(0,v_0) 
	& = 
	\sum_y p_c \sum_{u_0  \in v_0 - \Omega}
	\langle I[L_0'(0,u_0,y)]  \rangle^{\tilde{}}
	+ \sum_y \sum_{\ell=1}^\infty (-1)^\ell
	\tilde{\Ebold}_{0}  I[L_0''] \, 
	{\Ebold}_{1}  Z_1'' \, 
	\cdots 
	{\Ebold}_{\ell-1}  Z_{\ell-1}'' \, 
	{\Ebold}_{\ell}  Z_{\ell}''  , 
	\\
	\Ecal^{(0)}(0,v_0)  
	& =  \sum_y \sum_{\ell=1}^\infty (-1)^{\ell-1} 
	\tilde{\Ebold}_{0}  I[L_0'']  \, 
	{\Ebold}_{1}   Z_1'' \, 
	\cdots 
	{\Ebold}_{\ell-1} Z_{\ell-1}'' \, 
	{\Ebold}_{\ell}  I[F_2 (b_{\ell-1}, y; \tilde{C}_{\ell-1})] ,
	\hspace{1cm}
\enalign
with
\eq
	Z_\ell' = I[F_1'(b_{\ell-1},y;\tilde{C}_{\ell-1})],
	\quad
	Z_\ell'' = I[F_1''(b_{\ell -1}, a_\ell ,b_\ell; \tilde{C}_{\ell-1})]
\en
and sums with factors $p_c$ tacitly understood as in the first expansion.
The bounds on these terms are the same as for the
first expansion, except that now there is an additional summed vertex $u_0$
on the diagrammatic loop corresponding to the leftmost expectation.
The diagrams in the first expansion, occuring for $\phi$ or $\Phi$,
have critical dimension 6 {\em after}\/ adding an additional vertex.
As we will discuss more generally in Section~\ref{sub-Phi.diff.bounds}, 
the methods of Sections~\ref{sec-part.exp.bd} 
and \ref{sec-diagmethods} then justify our having
already taken the expansion to infinite order in \refeq{2ndexp-0}, and imply
that $\hat{\Ucal}_h^{(0)}(0)$ and $\hat{\Vcal}_h^{(0)}(0)$ are $O(1)$ and
that the error term $\hat{\Ecal}_h^{(0)}(0)$ is lower order than $\chi_h$.
Therefore
\eq
	- \frac{d}{dh} \Phihat_h^{(0)}(0)
	=  [\hat{\Vcal}_h^{(0)}(0)+ \oh ] \chi_h  .
\en

%%%%%%%%%%%%%%%%%%%%%%%%%%%%%%%%%%%%%%%%%%%%%%%%%%%%%%%%%%%%%%%%%%%%%%%%%%%%%%%
%%%%%%%%%%%%%%%%%%%%%%%%%%%%%%%%%%%%%%%%%%%%%%%%%%%%%%%%%%%%%%%%%%%%%%%%%%%%%%%
\subsection{Differentiation of $\hat{\Phi}_h(0)$}
\label{sub-Phidiff}

In general,  
when $\Phihat_h^{(N)}(0)$ is given by a $(N+1)$-fold nested expectation,
the result of the differentiation will not be as simple as it was for
the case $N=0$.
By the product rule, the differentiation of a nested expectation
will give rise to a sum of terms with
one of the nested expectations differentiated in each term.
Typically, we are differentiating the $n^{{\rm th}}$ expectation in
an expression of the form 
\eq
\lbeq{PhihatN.form1}
	\Phihat^{(N)}_h(0) = (-1)^{N} 
	\Ebold_{0} I[E_0''] \, 
	\Ebold_{1} Y_1'' 
	\cdots \Ebold_{n-1} Y_{n-1}'' \, \Ebold_{n} Y_{n}'' 
	\Ebold_{n+1} Y_{n+1}'' \,
        \Ebold_{n+2} Y_{n+2}'' \,
  	\ldots 
	\Ebold_{N} Y_{N}''
	.
\en
Writing out the $h$-dependence of the $n^{\rm th}$ expectation
explicitly, and, for later convenience, switching to the
conditional expectation of Example~\ref{ex-R1Sh} for the $n^{\rm th}$
and $(n \pm 1)^{\rm st}$ expectations, gives
\eq
\lbeq{PhihatN.form2}
	\Phihat^{(N)}_h(0) = (-1)^{N}   
	{\Ebold}_{0} I[E_0''] \, 
	{\Ebold}_{1} Y_1'' \, 
	\cdots 
	\tilde{\Ebold}_{n-1} Y_{n-1}' \, 
	\tilde{\Ebold}_{n} Y_{n,b}' e^{-h |C_{n}|} \, 
	\tilde{\Ebold}_{n+1} Y_{n+1}' \, 
        {\Ebold}_{n+2} Y_{n+2}'' \, 
        \ldots 
	{\Ebold}_{N} Y_{N}'' . 
\en

The effect of applying $-\frac{d}{dh}$ to the factor $e^{-h |C_{n}|}$
is simply to multiply by $|C_n|$, and this new factor can be represented
by $\sum_{y \in \Zd} I[y \in C_n]$.  The expression is otherwise unchanged.
It is just as easy to work with
$y$ fixed, rather than summed, and so we postpone the summation over $y$
until the last step.  Diagrammatically, the connection to $y$ presents
the usual diagrams for $\hat{\Phi}^{(N)}_h(0)$, with a new line joining the
diagram to $y$.  If that line were independent of the rest of the diagram,
we could factor the expectation to obtain a diagram of $\hat{\Phi}^{(N)}_h(0)$
with an additional vertex $a'$, multiplied by $\tau_h(a',y)$, and summed over
$a'$.  Since the diagrams with additional vertex have critical dimension 6,
summing over $y$ would yield the desired result $\mbox{const.}\chi_h$ for
the derivative, with the constant $O(\oneOd^N)$.  
Of course, this presumed independence
is not actually present, and we must perform an expansion in order
to factor out the two-point function.  This is the role of the second 
expansion.  We will perform the second expansion using the one-$M$ scheme
of Section~\ref{sec-exp.rem}.

The result will be of the form
\eq
\lbeq{PhiNdif}
	-\frac{d}{dh} \Phihat_h^{(N)}(0) = 
	\hat{\Ucal}_h^{(N)}(0) + \hat{\Vcal}_h^{(N)}(0) \chi_h 
	+ \hat{\Ecal}_h^{(N)}(0).
\en
Each of the three terms on the right side will turn out to
involve a factor $O(\lambda^N)$, to allow the sum over $N$ to
be performed.  It will also turn out that
$\hat{\Ucal}_h^{(N)}(0)$ and $\hat{\Vcal}_h^{(N)}(0)$ are bounded
uniformly in $h$, and that $\hat{\Ecal}_h^{(N)}(0)$ diverges 
more slowly than $\chi_h$.

%%%%%FIGFIGFIGFIGFIGFIGFIGFIGFIGFIGFIGFIGFIGFIGFIGFIGFIGFIGFIGFIG
%%%%%FIGFIGFIGFIGFIGFIGFIGFIGFIGFIGFIGFIGFIGFIGFIGFIGFIGFIGFIGFIG
\begin{figure}
\begin{center}
\includegraphics[scale = 0.5]{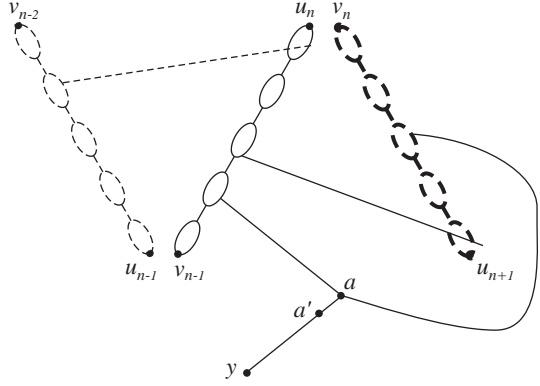} 
\end{center}
\caption{
Schematic diagram of the choice of cutting
bond.  Solid lines represent 
$C_{n}$, dashed lines $C_{n-1}$, and bold dashed lines represent 
$\Bcal_{n+1}$. 
}
\label{fig-Phi-cut}
\end{figure}
%%%%%FIGFIGFIGFIGFIGFIGFIGFIGFIGFIGFIGFIGFIGFIGFIGFIGFIGFIGFIGFIG
%%%%%FIGFIGFIGFIGFIGFIGFIGFIGFIGFIGFIGFIGFIGFIGFIGFIGFIGFIGFIGFIG

The first step in the second expansion is the identification
of a suitable pivotal bond at which to sever the connection to $y$.
We begin by applying Fubini's theorem to interchange the $n^{{\rm th}}$ and 
$(n+1)^{{\rm st}}$ (bond/site) expectations, 
and regard the clusters $C_{n \pm 1}$
as being fixed.  The occurrence of the events $(F_1')_l$ on levels 
$l=n-1,n,n+1$ enforces
a compatibility between $C_n$ and $C_{n \pm 1}$, in the sense that 
certain connections are required to occur for $C_n$.  
These connections
are depicted schematically in Figure~\ref{fig-Phi-cut}.
We omit any discussion of the easier special cases where it is
the expectation at level-0 or $N$ that is differentiated.

We recall the definition of the \emph{backbone} of a 
cluster $C_{n+1}$ connecting $v_{n}$ to $u_{n+1}$, namely the set of
all sites $x$ for which there are disjoint connections 
$x \conn v_{n}$ and $x \conn u_{n+1}$.
We denote this backbone as $\Bcal(C_{n+1}) \equiv \Bcal_{n+1}$.
We also recall the existence of an ordering of the pivotal bonds for 
the connection from a site to a set of sites, as defined in 
Definition~\ref{def-percterms}(d).
The \emph{cutting bond} is defined to be 
the last pivotal bond $(a',a)$ for the connection 
$y \to \{v_{n-1}, u_{n}\} \cup \Bcal_{n+1}$.   
It is possible that no such pivotal bond exists, and in 
that case, no expansion is required.

In choosing the cutting bond, 
we require it to be pivotal for $\{v_{n-1}, u_{n}\}$ 
to preserve (on the $a$ side of the cluster $C_{n}$) 
the backbone structure of the cluster $C_{n}$ which is 
required by $F_1'(v_{n-1}, u_{n}; C_{n-1})_n$.   
Also, we require the cutting bond to be pivotal for
$\Bcal_{n+1}$ to ensure that we do not 
cut off as a tail something which may be needed 
to ensure that, as required by 
$F_{1}'(v_{n}, u_{n+1}; C_{n})_{n+1}$, the last sausage
of $C_{n+1}$ is connected through $C_{n}$.   

Having chosen the cutting bond, we now begin to set the scene
for the expansion.  This requires an examination of the
overall conditions present in the level-$n$ expectation.
In addition to $F_{1}'(v_{n-1}, u_{n}; C_{n-1})_{n}$ itself,
there are conditions arising from 
$F_{1}'(v_{n}, u_{n+1}; C_{n})_{n+1}$.  The latter event can
be decomposed as
\eq
\lbeq{F1F0Hcut}
	F_{1}'(v_{n}, u_{n+1}; C_{n})_{n+1} 
	= F_{0}(v_{n}, u_{n+1})_{n+1} \cap H_{\rm cut}(\Bcal_{n+1})_{n}, 
\en 
where
\begin{align}
   	F_{0}(v_{n}, u_{n+1})_{n+1} 
	= &
	\event{v_{n} \conn u_{n+1} \AND v_{n} \nc G}
	\\
	H_{\rm cut}(\Bcal_{n+1})_{n} 
	= &
	\Bigl \{ 
	C_{n} \textrm{ intersects } \Bcal_{n+1} 
	\textrm{ such that the level-$(n+1)$ 
	connections satisfy} 
	\nonumber \\
\lbeq{Hcutdef}
	& 
	(v_{n+1}' \dbc u_{n+1} \textrm{ through } C_n )
	\AND 
	(v_n \conn u_{n+1}' \IN \Zd \backslash C_n)
	\Bigr \} 
	, 
\end{align}
with $(u_{n+1}',v_{n+1}')$ the last pivotal bond for the level-$(n+1)$
connnection from $v_n$ to $u_{n+1}$ required
by $(F_1)_{n+1}$.
(If there is no such pivotal bond, the requirements in \refeq{Hcutdef}
are replaced by $v_n \dbc u_{n+1}$ through $C_n$).
Our task now is to rewrite the overall level-$n$ condition
$F_{1}'(v_{n-1}, u_{n}; C_{n-1})_{n} \cap 
H_{\rm cut}(\Bcal_{n+1})_{n} \cap \event{y \in C_{n}}_{n}$
into a form suitable for generating the expansion. 

Recall the definition of $\tilde{C}^{(a,a')}(A)$ given in 
Definition~\ref{def-percterms}.
We define several events, as in Section~\ref{sec-exp.rem}.  
These events depend on $y,v_{n-1}, u_{n}, C_{n-1}, \Bcal_{n+1}$,
but to simplify the notation we make only the $y$-dependence explicit
in the notation.  Let
\begin{align}
\lbeq{H1def}
	H_{1}(y)_{n}
	= \,  & \event{y \conn v_{n-1}} \cap 
	F_{1}'(v_{n-1}, u_{n}; C_{n-1})_{n} \cap 
	H_{\rm cut}(\Bcal_{n+1})_{n} ,
	\\
\lbeq{H1pdef}
	H_{1}'(y)_{n}
	= \,  & H_{1}(y)_{n} \cap 
	\event{y \dbc \{v_{n-1}, u_{n}\} \cup \Bcal_{n+1} } ,
	\\ 
\lbeq{H1ppdef}
	H_{1}''(a, a')_{n} 
	= \, & 
	\{ H_{1}'(a)_n \ON  
	\tilde{C}_{n}^{\{a,a'\}}(\{v_{n-1}, u_{n}\} \cup \Bcal_{n+1}) 
	\} ,
	\\
	H_{1}(a, a', y)_{n}
	= \, & 
	H_{1}(y)_{n} \cap 
	\event {(a', a) \textrm{ is the last occupied pivotal bond for }
	y \to \{v_{n-1}, u_{n}\} \cup \Bcal_{n+1}} .
\end{align}
Then the overall level-$n$ event $H_{1}(y)_{n}$ is the disjoint union 
\eq
\lbeq{H1decomp}
	H_{1}(y)_{n} = 
	H_{1}'(y)_{n} \bigcup^{\cdot}
	\biggl ( \bigcup_{(a,a')}^{\cdot} 
	H_{1}(a, a', y)_n \biggr ) . 
\en  
In \refeq{H1decomp}, configurations in $H_1(y)_n$ have been classified
according to the last pivotal bond $(a', a)$. The appearance of
$H_{1}'$ corresponds to the possibility that there is no such pivotal 
bond, and in this case, no expansion will be required.   

For the configurations in which there is a pivotal bond, we will use
the following important lemma. 
\begin{lemma}
\label{lem-F1cut}
The events $H_{1}(a, a', y)_n$ and $H_{1}''(a, a')_{n}$ obey 
\begin{align}
	H_{1}(a, a', y)_n
	= \, & H_{1}''(a, a')_{n}
	\cap 
	\event{ ( y \conn a' \AND y \nc G) \INSIDE \Zd \backslash 
         \tilde{C}_n^{\{a,a'\}}(\{v_{n-1}, u_{n} \} \cup \Bcal_{n+1}) } 
	 \nnb 
	 & 
	 \cap \event{\{a, a'\} \textnormal{ is occupied}}.
\lbeq{11cuteq}
\end{align}
\end{lemma}

Before proving the lemma, we note that together with \refeq{H1decomp} and
Lemma~\ref{lem-cond.0} it implies the identity
\eq
	\expec{I[H_{1}(y)_{n}]}_{n}^{\tilde{}} = 
	\expec{I[H_{1}'(y)_{n}]}_{n}^{\tilde{}} +
	p_c \sum_{(a,a')}  
	\langle I[ H_{1}''(a, a')_{n}] \, 
	\tau_{h}^{\tilde{C}_n^{\{a,a'\}}
	(\{v_{n-1}, u_{n} \} \cup \Bcal_{n+1})}(a', y) \rangle_{n}^{\tilde{}} . 
	\lbeq{H1-cut.2}
\en 
This will be the point of departure for the second expansion.
Initially, the restricted two-point function appearing in the above equation
should be with respect to the conditional, rather than the usual expectation.
However, there is no difference between the two.  To see this,
note that the event that $a' \conn y$ in  
$\Zd \backslash \tilde{C}_n^{\{a,a'\}}(\{v_{n-1}, u_{n} \} \cup \Bcal_{n+1})$
is independent of the bond $\{u_{n}, v_{n}\}$, since this bond
touches the set $\tilde{C}_n^{\{a,a'\}}
(\{v_{n-1}, u_{n} \} \cup \Bcal_{n+1})$.  Therefore either expectation
can be used for the restricted two-point function, 
and for simplicity, we will use the ordinary unconditional expectation.

\bigskip
\noindent 
\textbf{Proof of Lemma~\ref{lem-F1cut}.} 
To abbreviate the notation, we write  
$A = \{v_{n-1}, u_{n} \}  \cup \Bcal_{n+1}$, and define
\eq
	F_{\rm piv} = \{(a', a) \textrm{ is pivotal for }y \to A \}.
\en
By definition of $H_{1}(a, a', y)_n$, 
\eq
\lbeq{H1is}
	H_{1}(a, a', y)_n 
	= \event{ \{a,a'\} \textrm{ is occupied }}
	\cap H_{1}(y)_n \cap 
	\event{a \dbc A} 
	\cap F_{\rm piv} . 
\en 

We introduce the events
\begin{align}
    	\Fcal_1 
	& = \event{a \conn v_{n-1}}_{n} \cap 
	\event{a \dbc A}_{n} \cap 
	F_{1}'(v_{n-1}, u_{n}; C_{n-1})_{n} \cap 
	H_{\rm cut}(\Bcal_{n+1})_{n} ,
	\\
	\Fcal_2 
	& = \event{y \conn a' \AND y \nc G}_{n} , 
\end{align} 
and claim that
\eqsplit 
\lbeq{Hclaim}
	H_{1}(a, a', y)_n 
	& = 
	\{\Fcal_1 \ON \tilde{C}^{\{a,a'\}}(A)\} 
	\cap \{\Fcal_2 \INSIDE \Zd\backslash \tilde{C}^{\{a,a'\}}(A)\} 
	\\
	& \quad \cap \event{ \{a,a'\} \textrm{ is occupied }} 
	\cap F_{\rm piv} . 
\ensplit
Assuming the claim, 
Lemma~\ref{lem-pivotal2} can then be employed to 
rewrite the last event in the above, to give 
\eqalign
	H_{1}(a, a', y)_n 
	& = 
	\bigl \{ (\Fcal_{1} \AND a \conn A) \ON \tilde{C}^{\{a, a'\}}(A) 
	\bigr \} \cap \event{ \{a,a'\} \textrm{ is occupied }}
	\nnb 
	& \quad \cap
        \bigl \{ 
        (\Fcal_{2} \AND y \conn a') \INSIDE 
        \Zd\backslash \tilde{C}^{\{a, a'\}}(A) 
	\bigr \}    . 
\enalign
In view of the definitions of $\Fcal_1$ and $\Fcal_2$, this implies
the desired identity \refeq{11cuteq}. 

It remains to prove \refeq{Hclaim}.
Combining \refeq{H1is} and \refeq{H1def}, we have
\eqalign
	H_{1}(a, a', y)_n 
	& =   \event{ \{a,a'\} \textrm{ is occupied }}
	\cap
	\event{y \conn v_{n-1}} \cap 
	F_{1}'(v_{n-1}, u_{n}; C_{n-1})_{n} 
	\nnb 
	& \quad 
	\cap 
	H_{\rm cut}(\Bcal_{n+1})_{n}
	\cap 
	\event{a \dbc A} 
	\cap F_{\rm piv} .
\enalign
To see that this can be written in the form \refeq{Hclaim}, we will
analyze the various events in the above expression.

We begin with $\event{y \conn v_{n-1}}$, and note that
\eqalign
\lbeq{yvpf.0}
	&
	\event{y \conn v_{n-1}} 
	\cap F_{\rm piv}
	\\ 
	& \nonumber \hspace{1cm}
	= \event{y \conn a'}  \cap
	\event{ \{a,a'\} \textrm{ is occupied }} 
	\cap
	\{ a \conn v_{n-1} \ON \tilde{C}^{\{a, a'\}}(A) \}
	\cap F_{\rm piv} .
\enalign
In fact, the right side is clearly contained in the left side.  Conversely,
for a configuration on the left side, since $v_{n-1} \in A$,
the bond $(a',a)$ must also
be pivotal for $y \conn v_{n-1}$, and this implies that
$\event{y \conn a'}$, that $\{a',a\}$ is occupied, and that
$a \conn v \ON \tilde{C}^{\{a, a'\}}(v_{n-1})$.  Since
$\tilde{C}^{\{a, a'\}}(v_{n-1}) \subset \tilde{C}^{\{a, a'\}}(A)$,
this implies 
$\{ a \conn v_{n-1} \ON \tilde{C}^{\{a, a'\}}(A) \}$.
This proves \refeq{yvpf.0}.
Now, by Lemma~\ref{lem-pivotal2}, $F_{\rm piv}$
is the intersection of the events
$\{ y \conn a' \INSIDE \Zd\backslash \tilde{C}^{\{a, a'\}}(A) \}$ and
$\{ a \conn A \ON \tilde{C}^{\{a, a'\}}(A) \}$, and hence
it follows from \refeq{yvpf.0} that
\eqalign
\lbeq{yvpf}
	\event{y \conn v_{n-1}} 
	\cap F_{\rm piv}
	& =  
	\event{ \{a,a'\} \textrm{ is occupied }}
	\cap
	\{ a \conn v_{n-1} \ON \tilde{C}^{\{a, a'\}}(A) \}
	\nnb 
	& \quad 
        \cap
	\{ y \conn a' \IN \Zd\backslash \tilde{C}^{\{a, a'\}}(A) 
	\} .
\enalign

Next we prove that
\eqalign
\lbeq{F1pf.1}
	&
	F_{1}'(v_{n-1}, u_{n}; C_{n-1})_{n}
	\cap F_{\rm piv}
	\cap \event{y \conn v_{n-1}}
	\cap 
	\event{\{a, a'\} \textnormal{ is occupied }}
	\nnb 
	& \hspace{1cm}
	=  \left\{ F_{1}'(v_{n-1}, u_{n}; C_{n-1}) 
        \ON  \tilde{C}_{n}^{\{a,a'\}}(A) \right\} 
        \cap 
	\left\{ y \nc G \IN 
	\Zd\backslash \tilde{C}_{n}^{\{a,a'\}}(A) \right\} 
	\nnb 
	& \hspace{1.5cm}
        \cap
	\{ y \conn a' \textnormal{ in } 
	\Zd\backslash \tilde{C}^{\{a, a'\}}(A) 
	\}
	\cap
	\{ a \conn v_{n-1} \ON  \tilde{C}^{\{a, a'\}}(A) \} 
	\nnb 
	& \hspace{1.5cm} 
	\cap 
	\event{\{a, a'\} \textnormal{ is occupied }}.
\enalign  
As a first observation, we note that by \refeq{yvpf}, the second last line
in the above can be replaced by $\{ y \conn v_{n-1}\} \cap F_{\rm piv}$,
and we will interpret the right side in this way.  To prove \refeq{F1pf.1},
we begin by supposing we have a configuration in the left side.  To show
that it is in the right side, it suffices to 
show that the first two events on the right side must then occur.
By the $G$-free condition in $F_1'$, together with the fact that 
$y \in C(v_{n-1})$, it follows that $\{ v_{n-1} \nc G \ON 
\tilde{C}^{\{a, a'\}}(A)\}$ and that $\{ y \nc G \IN 
\Zd\backslash \tilde{C}_{n}^{\{a,a'\}}(A) \}$.  As for the bond
connections required by $F_1'$, these are conditions on the backbone 
$\Bcal_n$, which are independent of bonds not touching 
$\tilde{C}^{\{a, a'\}}(A)$ since $F_{\rm piv}$ occurs.  Thus these connections
must occur on $\tilde{C}^{\{a, a'\}}(A)$, and we have shown that the left
side of \refeq{F1pf.1} is contained in the right side.
Conversely, given a configuration on the right side, we need to show
that $F_1'$ occurs.  The necessary bond connections again occur since 
$F_{\rm piv}$ occurs.  To see that, in addition, $v_{n-1} \nc G$, 
we note that when $\{ y \conn v_{n-1}\} \cap F_{\rm piv}$ occurs,
$C_{n} (v_{n-1}) = \tilde{C}_{n}^{\{a,a'\}}(a) \cupd 
\tilde{C}_{n}^{\{a,a'\}}(a')$, where the union is disjoint.  
But for a configuration on the right side of \refeq{F1pf.1}, 
the two clusters forming
this disjoint union must be $G$-free.  This completes the proof of
\refeq{F1pf.1}.

Turning now to $H_{\rm cut}(\Bcal_{n+1})_{n}$, we claim that 
\eq
\lbeq{Hcutpf}
	H_{\rm cut}(\Bcal_{n+1})_{n}  \cap F_{\rm piv}
% 	\\ && \nonumber \hspace{1cm}
	=  \{ H_{\rm cut}(\Bcal_{n+1})_{n} \ON \tilde{C}^{\{a, a'\}}(A) \}
         \cap F_{\rm piv} . 
\en
In fact, if the left side occurs, then the right side occurs because
$F_{\rm piv}$ requires $(a',a)$ to be pivotal for $y$'s connection
to $\Bcal_{n+1}$ and hence all $C_n$'s connections to $\Bcal_{n+1}$
are independent of bonds not touching $\tilde{C}^{\{a, a'\}}(A)$.
Conversely, the right side is contained in the left side for the same reason.

Finally, we claim that
\eq
\lbeq{adbcApf}
	\event{a \dbc A}  \cap F_{\rm piv}
	=  \{ a \dbc A \ON \tilde{C}^{\{a, a'\}}(A) \} \cap F_{\rm piv}. 
\en
In fact, 
because $\event{a \dbc A}$ is increasing, the right side is contained in
the left side.  Conversely, for a configuration on the left side,
it must be the case that $\{ a \dbc A \ON \tilde{C}^{\{a, a'\}}(a) \}$,
and since $\tilde{C}^{\{a, a'\}}(a) \subset \tilde{C}^{\{a, a'\}}(A)$,
the right side occurs.

The event $H_1(a,a',y)_n$ is the intersection of the events occurring
on the left sides of \refeq{F1pf.1}, \refeq{Hcutpf}
and \refeq{adbcApf}.  Therefore it
is the intersection of the events occurring 
on the right sides of these equations.  A rearrangement of these right side
events then gives \refeq{Hclaim} and completes the proof. 
\qed

\medskip

We are now in a position to obtain the identity \refeq{PhiNdif},
starting from \refeq{H1-cut.2} and using the one-$M$ scheme. 
With $N$ and $n$ fixed, the first step is to 
rewrite $\tau^{\tilde{C}}_h$ in \refeq{H1-cut.2} using \refeq{tAF12}.  
The result is 
\eqalign
\lbeq{H1F12}
	\expec{I[H_{1}(y)_{n}]}_{n}^{\tilde{}}
	= & 
	\expec{I[H_{1}'(y)_{n}]}_{n}^{\tilde{}}
	+
	p_c \sum_{(a_0,b_0)}  
	\expec{I[ H_{1}''(a_0,b_0)_{n}] }_{n}^{\tilde{}} \, \tau_{h}(b_0, y) 
	\nnb 
	& - 
	p_c \sum_{(a_0,b_0)}  
	\bigl \langle I[ H_{1}''(a_0,b_0)_{n}] 
	\langle I[F_{1}(b_0, y; \tilde{C}_n^{\{a_0,b_0\}}
	(\{v_{n-1}, u_{n} \} \cup \Bcal_{n+1})) 
	\rangle_{(n,1)} 
	\bigr \rangle_{n}^{\tilde{}}
	\nnb 
	& +
	p_c \sum_{(a_0,b_0)}  
	\bigl \langle I[ H_{1}''(a_0,b_0)_{n}] 
	\langle I[F_{2}(b_0, y; \tilde{C}_n^{\{a_0,b_0\}}
	(\{v_{n-1}, u_{n} \} \cup \Bcal_{n+1}))
	\rangle_{(n,1)} 
	\bigr \rangle_{n}^{\tilde{}} . 
\enalign 
We further expand the term containing $F_{1}$ using 
\refeq{Yniterate},  leaving the term containing $F_{2}$ as it is. 
Let $\tilde{C}_{(n,0)}= \tilde{C}_n^{\{a_0,b_0\}}
(\{v_{n-1}, u_{n} \} \cup \Bcal_{n+1})$.
For $j \geq 1$, let 
$\tilde{C}_{(n,j)}= \tilde{C}_{(n,j)}^{\{a_j,b_j\}}(b_{j-1})$,
$Z_{(n,j)}' = I[F_{1}'(b_{j-1}, y; \tilde{C}_{(n,j-1)}]$, and
$Z_{(n,j)}' = I[F_{1}''(b_{j-1}, a_j, b_j; \tilde{C}_{(n,j-1)}]$.

The first iteration can be written schematically as
\eqalign
\lbeq{secexp-Phi.Hnexp}
	\expec{(H_{1})_{n}}_{n}^{\tilde{}} 
	= & 
	\expec{(H_{1}')_{n}}_{n}^{\tilde{}} 
	+
	\expec{(H_{1}'')_{n} }_{n}^{\tilde{}} \, \tau
	- 
	\expec{(H_{1}'')_{n}  
	\langle (F_1)_{(n,1)} \rangle_{(n,1)} }_{n}^{\tilde{}}
	+ 
	\expec{(H_{1}'')_{n}
	\langle (F_{2})_{(n,1)} \rangle_{(n,1)} }_{n}^{\tilde{}} 
	\nnb 
	= & 
	\expec{(H_{1}')_{n}}_{n}^{\tilde{}} 
	+
	\expec{(H_{1}'')_{n} }_{n}^{\tilde{}} \, \tau_{h}
	- 
	\expec{(H_{1}'')_{n} 
	\langle Z_{(n,1)}' \rangle_{(n,1)} }_{n}^{\tilde{}}
	- 
	\expec{(H_{1}'')_{n} 
	\langle Z_{(n,1)}'' \rangle_{(n,1)} }_{n}^{\tilde{}} \tau 
	\nnb 
	& 
	+ 
	\expec{(H_{1}'')_{n}  
	\expec{Z''_{(n,1)} 
	\langle (F_1)_{(n,2)} \rangle_{(n,2)} 
	}_{(n,1)} }_{n}^{\tilde{}} 
	\nnb 
	&
	+ 
	\expec{(H_{1}'')_{n}
	\langle (F_{2})_{(n,1)} \rangle_{(n,1)} }_{n}^{\tilde{}} 
	-  
	\expec{(H_{1}'')_{n}  
	\expec{Z''_{(n,1)} 
	\langle (F_{2})_{(n,2)} \rangle_{(n,2)} 
	}_{(n,1)} }_{n}^{\tilde{}} , 
\enalign 
and we continue expanding the term containing $F_1$, to infinite order. 
The expansion to infinite order will be justified by the diagrammatic
estimates of the next section.
This leads to an identity that can be abbreviated as
\eq
	\expec{(H_{1})_{n}}_{n}^{\tilde{}} 
	= \hat{u}^{(n)}_h(0) + \hat{v}^{(n)}_h(0) \chi_h
	+ \hat{e}^{(n)}_h(0),
\en
where the first and second terms respectively comprise
the terms with innermost expectation involving $Z'$ and $Z''$, and
the last term comprises those involving $F_2$.  Explicitly, 
\eqalign
\lbeq{vhat.def}
	\hat{v}_{h}^{(n)}(0) 
	& = 
	\sum_{\ell =0}^{\infty} (-1)^{\ell} \hat{v}_{h}^{(n,\ell)}(0),
	\quad \quad
	\hat{v}_{h}^{(n,\ell)}(0) = 
	\tilde{\Ebold}_{n} (H_{1}'')_{n} \, 
	{\Ebold}_{(n,1)} Z_{(n,1)}'' \, 
	\cdots 
	{\Ebold}_{(n,\ell)} Z_{(n,\ell)}'' \, 
	\\
\lbeq{ehatdef}
	\hat{e}_{h}^{(n)}(0) 
	& = 
	\sum_{\ell =1}^{\infty} (-1)^{\ell} \hat{e}_{h}^{(n,\ell)}(0),
	\quad \quad
	\hat{e}_{h}^{(n,\ell)}(0) =
	\tilde{\Ebold}_{n} (H_{1}'')_{n} \, 
	{\Ebold}_{(n,1)} Z_{(n,1)}'' \, 
	\cdots 
	{\Ebold}_{(n,\ell-1)} Z_{(n,\ell-1)}'' \, 
	{\Ebold}_{(n,\ell)} (F_{2})_{(n,\ell)} ,	
\enalign	
and $\hat{u}_h^{(n)}(0)$ is defined as in \refeq{vhat.def} with 
$Z_{(n,\ell)}'$ replacing $Z_{(n,\ell)}''$, and $H_1'$ replacing $H_1''$
for $\ell =0$.  In the $\ell =0$ term, only the leftmost expectation occurs.

Defining
\eq
\lbeq{VNnelldef}
	\hat{\Vcal}_{h}^{(N,n,\ell)}(0)
	= 
	{\Ebold}_{0} I[E_{0}''] \, 
	{\Ebold}_{1} Y_{1}'' \, 
	\cdots 
	{\Ebold}_{n-1} Y_{n-1}'' \, 
	\tilde{\Ebold}_{n+1} I[(F_{0})_{n+1} ] \, 
	\tilde{\Ebold}_{n}  \hat{v}^{(n,\ell)}_h(0) \, 
	{\Ebold}_{n+2}  Y_{n+2}'' \, 
	\cdots 
	{\Ebold}_{N}  Y_{N}'' ,
\en 
this gives \refeq{PhiNdif} with
\eq
\lbeq{Vcaldef}
	\hat{\Vcal}_{h}^{(N)}(0)
	= \sum_{n=0}^N \sum_{\ell =0}^\infty 
	(-1)^{N} \hat{\Vcal}_{h}^{(N,n,\ell)}(0) 
\en 
and analogous expressions for $\hat{\Ucal}^{(N)}_h(0)$ 
and $\hat{\Ecal}^{(N)}_h(0)$.  The right side of \refeq{VNnelldef}
requires special interpretation for the terms $n=0,N$.

%%%%%%%%%%%%%%%%%%%%%%%%%%%%%%%%%%%%%%%%%%%%%%%%%%%%%%%%%%%%%%%%%%%%%%%%%%%%%%%
%%%%%%%%%%%%%%%%%%%%%%%%%%%%%%%%%%%%%%%%%%%%%%%%%%%%%%%%%%%%%%%%%%%%%%%%%%%%%%%
\subsection{Proof of Proposition~\protect\ref{prop-Phi}}
\label{sub-Phi.diff.bounds}

Proposition~\protect\ref{prop-Phi} follows from \refeq{PhiNdif} and 
the following lemma.  Section~\ref{sub-Phi.diff.bounds} is devoted
to proving the lemma.  
The constant $K_1$ of Proposition~\protect\ref{prop-Phi} is given by
$K_1 = \hat{\Vcal}_0(0)$, where $\hat{\Vcal}_0(0)$ appears in the lemma.

\begin{lemma}
\label{lem-Psih}
The series $\hat{\Ucal}_h(0) = \sum_{N=0}^\infty \hat{\Ucal}_h^{(N)}(0)$,
$\hat{\Vcal}_h(0) = \sum_{N=0}^\infty \hat{\Vcal}_h^{(N)}(0)$, and
$\hat{\Ecal}_h(0) = \sum_{N=0}^\infty \hat{\Ecal}_h^{(N)}(0)$ converge 
absolutely.  
Moreover, $\hat{\Ucal}_h(0) = O(1)$, $\hat{\Ecal}_h(0) = \oh \chi_h$,
and
\eq
	\hat{\Vcal}_h(0) = \hat{\Vcal}_0(0) + \oh , 
	\qquad 
	\hat{\Vcal}_0(0) = 1 + O(\oneOd). 
	\lbeq{Vcalh-Vcal0}
\en
\end{lemma}   

\Proof  The analysis of $\hat{\Ucal}_h(0)$ is almost
identical to that of $\hat{\Vcal}_h(0)$, so we discuss only 
$\hat{\Vcal}_h(0)$ and $\hat{\Ecal}_h(0)$.
The proof consists of obtaining suitable diagrammatic
estimates on $\hat{\Vcal}_{h}^{(N,n,\ell)}(0)$ and
$\hat{\Ecal}_{h}^{(N,n,\ell)}(0)$ (see \refeq{VNnelldef}).  
Roughly speaking,
the bounds will involve horizontal ``ladder'' diagrams like those encountered
in the bounds on $\hat{\Phi}_h(k)$, with an additional vertical ladder
resulting from the nested expectation of $\hat{v}_{h}^{(n,\ell)}(0)$ or 
$\hat{e}_{h}^{(n,\ell)}(0)$.  We will estimate these diagrams with the
help of the power
counting methodology of Appendix~\ref{app}.  

We first consider the case $\ell =0$, and then move on to $\ell \geq 1$.
Finally, we will prove \refeq{Vcalh-Vcal0}.

\medskip \noindent
{\bf The case $\ell = 0$} 

\smallskip
\noindent
There is no contribution to $\hat{\Ecal}_{h}^{(N)}(0)$ arising from
$\ell =0$, so we are concerned here with $\hat{\Vcal}_{h}^{(N)}(0)$.
Our goal is a diagrammatic estimate for
\eq
	\hat{\Vcal}_{h}^{(N,n,0)}(0)
	= 
	{\Ebold}_{0} I[E_{0}''] \, 
	{\Ebold}_{1} Y_{1}'' \, 
	\cdots 
	\tilde{\Ebold}_{n-1} Y_{n-1}' \, 
	\tilde{\Ebold}_{n+1} I[(F_{0})_{n+1} ] \, 
	\tilde{\Ebold}_{n}  (H_{1}'')_{n} \, 
	{\Ebold}_{n+2}  Y_{n+2}'' \, 
	\cdots 
	{\Ebold}_{N}  Y_{N}'' .
\en 
(We omit any discussion of the special cases $n=0,N$, which can
be handled similarly.)
This estimate will involve a modification of the diagrams used to
estimate $\hat{\Phi}_h(0)$, obtained by ``growing'' a vertical
diagram from the horizontal diagrams encountered in bounding
$\hat{\Phi}_h(0)$.

We first note that, by \refeq{F1F0Hcut} and \refeq{H1def}--\refeq{H1ppdef}, 
the intersection $(F_{0})_{n+1} \cap (H_{1}'')_{n}$ is a subset of
\eq
	\event{a_0 \conn v_{n-1} } \cap 
	\event{a_0 \dbc \{v_{n-1}, u_{n}\} \cup \Bcal_{n+1} } \cap 
	F_{1}'(v_{n-1}, u_{n}; C_{n-1})_{n,b} \cap
	F_{1}'(v_{n}, u_{n+1}; C_{n})_{n+1,b} ,
\en
where the subscripts $b$ denote bond events with site conditions
relaxed.  Thus the bond connections that gave rise to the diagrammatic
estimates on $\hat{\Phi}_h(k)$ in Section~\ref{sub-bound.basic}, due to
$(F_1')_{n,b}$ and $(F_1')_{n+1,b}$, remain present after differentiation.
The event 
\eq
\lbeq{a-cond.Phi}
	\event{a_0 \conn v_{n-1} } \cap 
	\event{a_0 \dbc \{v_{n-1}, u_{n}\} \cup \Bcal_{n+1} }
\en 
provides additional
connections that can be bounded using the BK inequality. 
In fact, \refeq{a-cond.Phi} implies the existence of either (i)
two disjoint level-$n$ paths $a_0 \conn \{v_{n-1}, u_{n}\}$,
or (ii) disjoint level-$n$  paths $a_0 \conn \{v_{n-1}, u_{n}\}$ and 
$a_0 \conn \Bcal_{n+1}$.  We now consider the diagrammatic implications
of these two cases.

\begin{figure}
\begin{center}
\includegraphics[scale = 0.3]{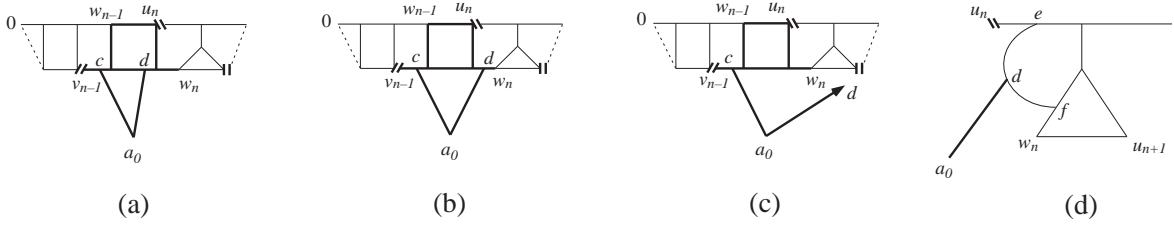} 
\end{center}
\caption{(a,b) Examples of diagrams arising from  
case (i). (c,d)  An example of the diagrammatic bound of case (ii).
Thick lines represent level-$n$ connections and thin lines represent
levels-$(n \pm 1)$.}
\label{fig-Phi0-adding}
\end{figure}

In case (i), the two connections implied by
$a_0 \dbc \{v_{n-1}, u_{n}\}$ are added
to the connections due to $(F_1')_{n,b}$ that arise in the diagrams
bounding this part of $\hat{\Phi}_h(k)$.
This can be bounded, using the BK inequality, by adding two lines
to the lines appearing already in the diagrammatic estimate for
this part of $\hat{\Phi}_h(k)$ (the ``old'' lines), 
with the lines going from $a$ to two new vertices, say $c$ and $d$.
Thus the resulting new diagrams are obtained by performing 
construction~1 followed by construction~2 
of Section~\ref{sub-ind.1} to the old diagrams.  Examples of this construction
are given in Figure~\ref{fig-Phi0-adding}~(a--b).  
 
%%%%%FIGFIGFIGFIGFIGFIGFIGFIGFIGFIGFIGFIGFIGFIGFIGFIGFIGFIGFIGFIG
%%%%%FIGFIGFIGFIGFIGFIGFIGFIGFIGFIGFIGFIGFIGFIGFIGFIGFIGFIGFIGFIG
%%%%%FIGFIGFIGFIGFIGFIGFIGFIGFIGFIGFIGFIGFIGFIGFIGFIGFIGFIGFIGFIG
%%%%%FIGFIGFIGFIGFIGFIGFIGFIGFIGFIGFIGFIGFIGFIGFIGFIGFIGFIGFIGFIG

In case (ii), we add disjoint connections $a_0 \conn \{v_{n-1}, u_{n}\}$ and 
$a_0 \conn \Bcal_{n+1}$.
We need only consider the case where the second of these paths 
is disjoint from the paths extracted from the event $(F_1')_n$,
because otherwise the situation reduces
to case (i). 
For the path resulting from $a_0 \conn \{v_{n-1}, u_{n}\}$,
we add a new vertex $c$ on an 
existing level-$n$ line of the $\Phi$ diagram, 
and then connect this vertex $c$ and $a_0$ with a new line.
This takes care of the level-$n$ connection, and the situation
is depicted in Figure~\ref{fig-Phi0-adding}~(c).  
For the second path, we must have
a path from $a_0$ to some point $d$ in $\Bcal_{n+1}$.   
We now have to ask how this $d$ is connected, via level-$(n+1)$
connections, to the rest of a level-$(n+1)$ diagram of $\Phi_h(k)$.  
For this purpose, we recall that $\Bcal_{n+1}$  
is the set of all points which are on a 
path from $v_{n}$ to $u_{n+1}$.  As an upper bound, we just require
$d \dbc \{v_n, u_{n+1}\}$, which brings us back to the situation of
case (i).  Thus there are sites $e,f$ on the level-$(n+1)$ lines, with
lines from $d$ to each of these sites.
This is depicted in Figure~\ref{fig-Phi0-adding}~(d). 
The net result is an application of construction~1 (at $e$)
followed by two applications of construction~2.

In summary, the resulting diagrams can be obtained by applying
construction~1, followed by one or two applications of 
construction~2, to the diagrams used previously
to bound $\hat{\Phi}_h(k)$, with the construction applied either
at level-$n$ or on levels-$n$ and $n+1$.
Now we bound these diagrams, in several steps.  

\begin{figure}
\begin{center}
\includegraphics[scale = 0.5]{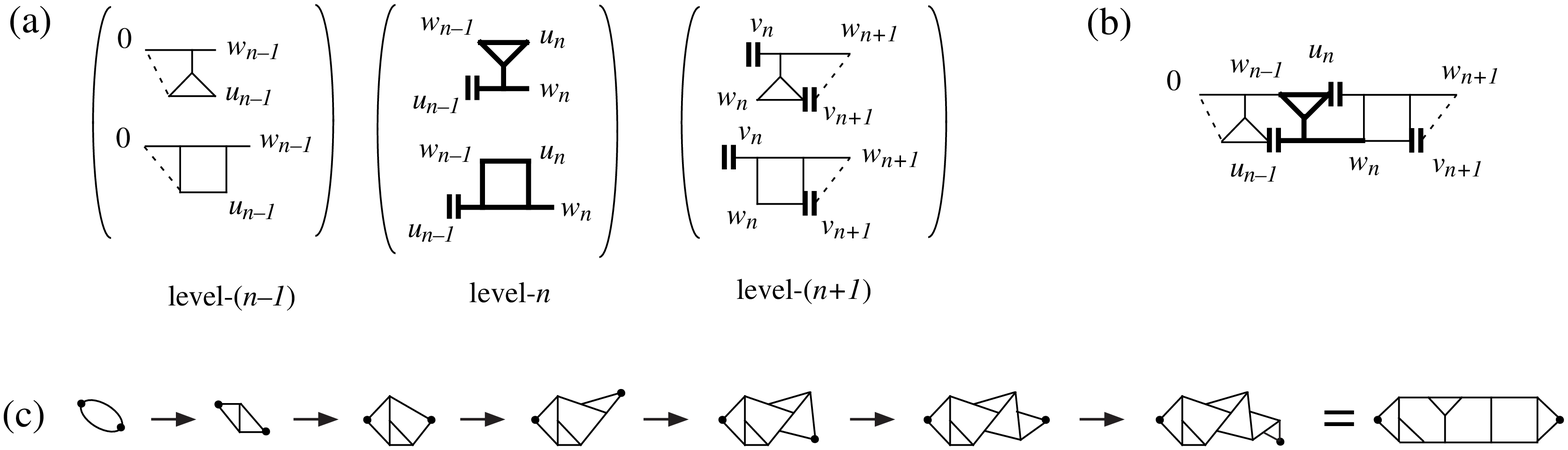} 
\end{center}
\caption{(a) The combinations producing the central diagram remaining 
after decomposition.
(b) An example of a combination in (a).
(c) Construction of the diagram in (b) via application of construction~2 
to the bubble.}
\label{fig-peel-before}
\end{figure}

We begin by decomposing the diagrams
as in Section~\ref{sub-bound.basic}, both from the side of level-$0$
and level-$N$.  These estimates produce triangles, with corresponding
factors of $\oneOd$.  The decomposition from level-$0$
stops at level-$(n-2)$, and that from level-$N$ stops at level-$(n+2)$,
with levels $n-1$, $n$ and $n+1$ remaining to be handled.   
A diagram corresponding to these levels will be open at its two 
ends, with a supremum over the displacement corresponding to the
opening.  This can be bounded above by the diagram obtained
by closing the two ends and by closing the small openings corresponding
to pivotal bonds.  The possible results, before application of 
constructions~1 and 2, are depicted in Figure~\ref{fig-peel-before}~(a), 
with the dashed lines representing the lines which are ``moved'' to close
the diagram as an upper bound.  There are eight possible combinations in all,
with an example depicted in Figure~\ref{fig-peel-before}~(b).

For the nearest-neighbour model in sufficiently high dimensions, 
we may employ squares, pentagons, etc.\ in our estimates, and it
is not difficult to see that the diagrams obtained after applying
constructions~1 and 2 are all $O(1)$.  We therefore restrict attention now
to the spread-out model.
It can be checked that each of the eight diagrams can be obtained
by applying construction~2 to the bubble, as depicted in 
Figure~\ref{fig-peel-before}~(c).  
By Lemma~\ref{lem-Ginduction}, for $d>6$,
each of these diagrams therefore has infrared degree 
at least that of the bubble diagram, which has $\underline{\deg}_0 = d-4$.
Thus, by \refeq{cor-Gind},  the diagrams applied by a subsequent application
of constructions~1 and 2 will have $\underline{\deg}_0 \geq d-6 > 0$.
By Theorem~\ref{thm-R1}, these diagrams are therefore convergent and $O(1)$.

In conclusion, we obtain the bound
\eq
	|\hat{\Vcal}_{h}^{(N,n,0)}(0)| \leq \min \{O(1),O(\oneOd^{N-3})\} .
\en
  
%%%%%FIGFIGFIGFIGFIGFIGFIGFIGFIGFIGFIGFIGFIGFIGFIGFIGFIGFIGFIGFIG
%%%%%FIGFIGFIGFIGFIGFIGFIGFIGFIGFIGFIGFIGFIGFIGFIGFIGFIGFIGFIGFIG
%%%%%FIGFIGFIGFIGFIGFIGFIGFIGFIGFIGFIGFIGFIGFIGFIGFIGFIGFIGFIGFIG
%%%%%FIGFIGFIGFIGFIGFIGFIGFIGFIGFIGFIGFIGFIGFIGFIGFIGFIGFIGFIGFIG

%%%%%%%%%%%%%%%%%%%%%%%%%%%%%%%%%%%%%%%%%%%%%%%%%%%%%%%%%%%%%%%%%%%%%%%%%%%%%%%
\bigskip \noindent {\bf The case $\ell \geq 1$}

\smallskip \noindent
We now consider the case $\ell \geq 1$, and obtain the bounds 
\eq
\lbeq{VEcallbds}
	|\hat{\Vcal}_{h}^{(N,n,\ell)}(0)| 
	\leq \min \{ O(1),O(\oneOd^{N+\ell-4}) \},
	\quad \quad
	|\hat{\Ecal}_{h}^{(N,n,\ell)}(0)| 
	\leq \min \{O(1), O(\oneOd^{N+\ell-4}) \} o_h(1) \chi_h.
\en
This is sufficient to prove Lemma~\ref{lem-Psih}, apart from 
\refeq{Vcalh-Vcal0} which we will prove later.
The quantities on the left side of \refeq{VEcallbds} 
will be bounded in terms of diagrams
like those encountered above for $\ell =0$, but with further growth
in the ``vertical'' direction.  For $\Vcal$, this vertical growth arises from
additional expectations of $F_1''$, 
whereas for $\Ecal$, the expectation at level-$(n,\ell)$
is $F_2''$.  These modifications to the $\ell =0$ diagrams do not depend
on levels $0$ to $n-2$ or on levels $n+2$ to $N$, and the
expectations corresponding to these levels 
can be bounded by triangles as before to give rise to
a factor $\oneOd^{N-3}$ multiplied by a diagram with two ends closed
as in the example shown in Figure~\ref{fig-peel-before}~(b).  
Our task now is to
understand the structure of this remaining diagram, with its vertical
growth, and to bound it
appropriately.  We begin by considering the case $\ell = 1$ of
\refeq{VEcallbds}.

\begin{figure}
\begin{center}
\includegraphics[scale = 0.5]{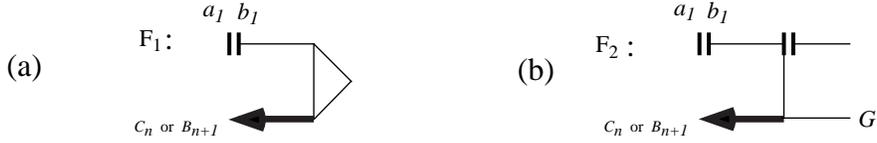} 
\end{center}
\caption{(a) The connections arising from level-$(n,1)$ in
$\hat{\Vcal}_h^{(N,n,1)}(0)$.
(b) The connections arising from level-$(n,1)$ in
$\hat{\Ecal}_h^{(N,n,1)}(0)$.}
\label{fig-n1}
\end{figure}

First, we consider $\hat{\Vcal}_{h}^{(N,n,1)}(0)$.  In view of 
\refeq{H1F12}--\refeq{secexp-Phi.Hnexp}, this requires estimation of
\eq 
\lbeq{H1F1pp}
	\bigl \langle I[ H_{1}''(a_0,b_0)_{n}] 
	\langle I[F_{1}''(b_0, y; \tilde{C}_n^{\{a_0,b_0\}}
	(\{v_{n-1}, u_{n} \} \cup \Bcal_{n+1})) 
	\rangle_{(n,1)} 
	\bigr \rangle_{n}^{\tilde{}}.
\en
In a similar fashion to the case $\ell =0$ already treated, 
the $H_1''$ leads to application of constructions~1 and 2 applied to the 
$\hat{\Phi}_h(k)$-diagram reduced as in Figure~\ref{fig-peel-before}~(b).  
This construction
gives an infrared degree $\underline{\deg}_0 \geq d-6 >0$, 
as before.
However, there are
now additional new connections arising from $F_1'$ in the level-$(n,1)$
expectation.  These connections are as depicted in Figure~\ref{fig-n1}~(a).
They correspond to construction~3, which does not lower the infrared
degree.  It is not difficult to see that \refeq{H1F1pp} is $O(1)$ 
for the nearest-neighbour model in sufficiently high dimensions.  For the
spread-out model, it is also $O(1)$, by power counting.  This
gives to an overall
bound of order $\min \{ 1, \oneOd^{N-3} \}$ for $\hat{\Vcal}_h^{(N,n,1)}(0)$.

%%%%%FIGFIGFIGFIGFIGFIGFIGFIGFIGFIGFIGFIGFIGFIGFIGFIGFIGFIGFIGFIG
%%%%%FIGFIGFIGFIGFIGFIGFIGFIGFIGFIGFIGFIGFIGFIGFIGFIGFIGFIGFIGFIG
%%%%%FIGFIGFIGFIGFIGFIGFIGFIGFIGFIGFIGFIGFIGFIGFIGFIGFIGFIGFIGFIG
%%%%%FIGFIGFIGFIGFIGFIGFIGFIGFIGFIGFIGFIGFIGFIGFIGFIGFIGFIGFIGFIG

Consider now $\hat{\Ecal}_h^{(N,n,1)}(0)$, for which
$F_{2}$ occurs on level-$(n,1)$.  
We first remove triangles as above, obtaining a factor of order
$\min \{ 1, \oneOd^{N-3} \}$.  We then 
apply the cut-the-tail Lemma~\ref{lem-tail.tau1}  
in the usual way, extracting
a factor $\chi_h$.  The tail and remaining connections due to $F_2$ are depicted
in Figure~\ref{fig-n1}~(b).  A factor $M_h$ arises from the connection
to $G$, and the three remaining lines due to $F_2$ can be estimated
by a factor of the triangle diagram.  The remaining diagram is
obtained from a diagram from the $\ell =0$ case by addition of
a vertex on one of the lines corresponding to
level-$n$ or level-$(n+1)$.  For the nearest-neighbour model in
sufficiently high dimensions, we may employ the square and larger diagrams
and conclude an overall bound here of $O(1)M_h\chi_h = O(1)$.
However, as we now explain, the spread-out model in dimensions $d>6$
requires more care.

Arguing as in Example~\ref{ex-R1Sh}, using Lemma~\ref{lem-onehline}
we may choose any one of the diagrammatic lines arising in any
expectation other than level-$(n,1)$ to be $G$-free.  We may
therefore regard the additional vertex mentioned in the previous
paragraph as residing on a massive line, with $\mu^2 = \chi_h^{-1}$.
Thus this extra vertex at worst reduces $\underline{\deg}_0$ by 2 
to $d-8$ (according
to Lemma~\ref{lem-Ginduction}),
but does not change $\underline{\deg}_\mu$.
Hence the diagram is convergent for $h>0$, and by Theorem~\ref{thm-R2},
its rate of divergence as $h \to 0$ is bounded above by
$O(\mu^{d-8} | \log \mu | ^L ) = O( \chi_h^{(6-d)/2} |\log \chi_h|^L) \chi_h
= o_h(1) \chi_h$, consistent with \refeq{VEcallbds}.

Now we turn to the case $\ell \geq 2$, beginning with 
$\hat{\Vcal}_h^{(N,n,\ell)}(0)$.  Again we bound the expectations corresponding
to levels-0 to $n-2$ and $n+2$ to $N$ by triangles, and close up the
ends of the resulting diagram.  Each expectation from levels-$(n,1)$
to $(n,\ell)$ corresponds diagrammatically to construction~2 applied
to the diagrams encountered for the case $\ell =0$, and does not
decrease the infrared degree.  We may estimate each of these expectations
with a triangle (each providing a factor $\oneOd$), leaving a bounded
diagram.  This gives the desired bound for $\hat{\Vcal}_h^{(N,n,\ell)}(0)$.
For $\hat{\Ecal}_h^{(N,n,\ell)}(0)$, we combine the method used for
$\hat{\Vcal}_h^{(N,n,\ell)}(0)$ with that employed for
$\hat{\Ecal}_h^{(N,n,1)}(0)$.

%%%%%%%%%%%%%%%%%%%%%%%%%%%%%%%%%%%%%%%%%%%%%%%%%%%%%%%%%%%%%%%%%%%%%%%%%%%%%%%
\bigskip \noindent {\bf Proof of \protect\refeq{Vcalh-Vcal0}}

\smallskip \noindent
First we consider $\hat{\Vcal}_{h} - \hat{\Vcal}_{0}$.  
As we have seen, diagrams contributing to $\hat{\Vcal}_h(0)$ are finite.
As in \refeq{Phiz(2)}, the difference gives rise to a connection to $G$.
This is bounded pointwise by $M_h$, and hence the dominated convergence
theorem can be applied to conclude that 
$\hat{\Vcal}_{h}(0) - \hat{\Vcal}_{0}(0) = o_h(1)$.
    
Finally, we argue that $\hat{\Vcal}_{0}(0) = 1 + O(\oneOd)$. 
Consider first the nearest-neighbour model.  Using the square and
higher diagrams if necessary, we can bound 
$\sum_{N,n,\ell}\hat{\Vcal}_{h}^{(N,n,\ell)}(0)$
by $O(\oneOd)$ for all terms except $N=n=\ell=0$.  This can be seen
from the fact that these terms
all include at least one expectation having a pivotal bond, and the occurrence
of a pivotal bond implies a bound $O(\oneOd)$.  The remaining term is
the first term of \refeq{Vcal0def}.  In that term, the contribution
due to $u_0=0$ is readily seen to be $1+O(\oneOd)$, while the contribution
due to $u_0 \neq 0$ is $O(\oneOd)$.  This gives the desired result for
the nearest-neighbour model.

For the spread-out model, we argue similarly.  However, in this case
there are contributions from diagrams that we are unable to bound
using a factor of the triangle that is clearly $O(\oneOd)$.  We have
used power counting, previously, to bound these contributions by
$O(1)$.  To improve these bounds to
$O(\oneOd)$, we appeal to dominated convergence via the
following argument (similar to \cite[Lemma~5.9]{HS90a}).  First we observe
that by \cite[(5.36)]{HS90a}, if $k \neq 0$ then 
$\lim_{L \to \infty} 1-\hat{D}(k) = 1$.  
It follows that the
limit of any convergent diagram containing summation over a pivotal bond
$\{u,v\}$ 
is the corresponding integral of $e^{ik\cdot (v-u)}$, integrated over 
$[-\pi,\pi]^d$.  This integral is zero.
Since all the diagrams that were bounded using power
counting do contain such a pivotal bond, we obtain the desired bound
$O(\oneOd)$.

\qed

%%%%%%%%%%%%%%%%%%%%%%%%%%%%%%%%%%%%%%%%%%%%%%%%%%%%%%%%%%%%%%%%%%%%%%%%%%%%%%%
%%%%%%%%%%%%%%%%%%%%%%%%%%%%%%%%%%%%%%%%%%%%%%%%%%%%%%%%%%%%%%%%%%%%%%%%%%%%%%%
\subsection{Differentiation of $\hat{\Xi}_h(0)$}
\label{sub-Xi.main}

In this section, we discuss the second expansion used in the proof of 
Proposition~\ref{prop-Xi}.  Our analysis of $\hat{\Xi}_h(0)$
has much in common with the above analysis of $\hat{\Phi}_h(0)$,
but there are also important differences.  For $\hat{\Phi}_h(0)$,
we used the fact that all connections involved in the
definition of $\hat{\Phi}_h(0)$ were $G$-free to show 
that its derivative gave rise to a ``tail''
corresponding, after a second expansion to cut off the tail,
to a factor of $\chi_h$.  Integration of this factor of $\chi_h$
then gave rise to the $M_h$ appearing in \refeq{denom.M}.
For $\hat{\Xi}_h(0)$, on the other hand, there is a connection to $G$
explicitly demanded in the expectation containing $W'$.
Although this easily gives rise to a bound involving $M_h$,
we need to extract a factor of the magnetization in an 
asymptotic relation.  We do not have an expansion that can
be used to ``cut off'' a factor of $M_h$, so we will differentiate
in this expectation to convert this connection to $G$ to a $G$-free tail.
This tail can then be cut off, as a factor of the susceptibility,
by means of a second expansion.

By the fundamental theorem of calculus, $\hat{\Xi}_h^{(n,m)}(0)$ 
can be written as
\eqarray
\lbeq{Xi.diff.0}
	\hat{\Xi}_h^{(n,m)}(0)   & =  &(-1)^{m+n-1}
	\int_0^h du {\Ebold}_{0,h} I[E_{0}''] {\Ebold}_{1,h} Y_{1}''  
	\cdots \tilde{\Ebold}_{n-1,h} Y_{n-1}'
	\frac{d}{du} 
	\tilde{\Ebold}_{n,u}
	W_{n}' 
	\nonumber \\ && \hspace{28mm}
	\times \; \tilde{\Ebold}_{n+1,h} Y_{n+1}'  
	{\Ebold}_{n+2,h} Y_{n+2}''  \cdots {\Ebold}_{n+m,h} Y_{n+m}'',
\enarray
where subscripts indicate the value of the magnetic field for each
expectation. 
The factor $W_n'$ involves the events $F_3'$ and $F_5'$, which
require a connection to $G$, while the factors $Y_j$ involve
only $G$-free connections.

%%%%%FIGFIGFIGFIGFIGFIGFIGFIGFIGFIGFIGFIGFIGFIGFIGFIGFIGFIGFIGFIG
%%%%%FIGFIGFIGFIGFIGFIGFIGFIGFIGFIGFIGFIGFIGFIGFIGFIGFIGFIGFIGFIG
\begin{figure}
\begin{center}
\includegraphics[scale = 0.4]{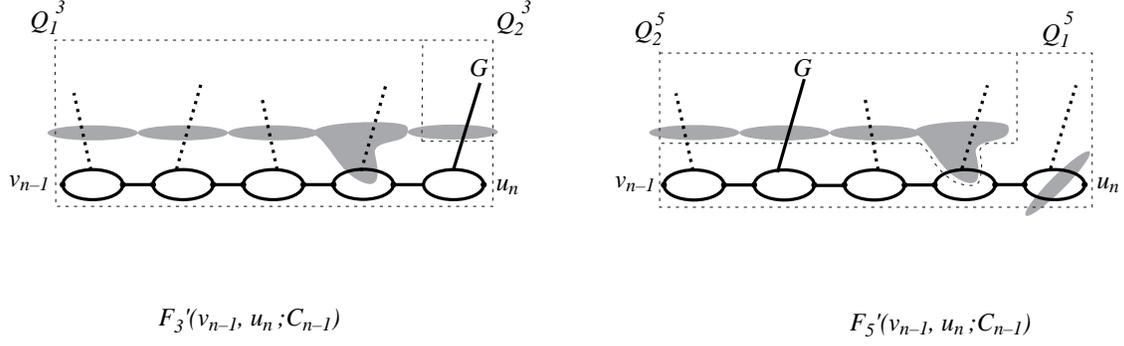} 
\end{center}
\caption{Schematic depiction of $Q^3_1, Q^3_2,Q^5_1,Q^5_2$.
Crosshatched regions represent $C_{n-1}$ and dotted lines represent
possible but not mandatory connections in $C(v_{n-1})$.}
\label{fig-Q35.def}
\end{figure}
%%%%%FIGFIGFIGFIGFIGFIGFIGFIGFIGFIGFIGFIGFIGFIGFIGFIGFIGFIGFIGFIG
%%%%%FIGFIGFIGFIGFIGFIGFIGFIGFIGFIGFIGFIGFIGFIGFIGFIGFIGFIGFIGFIG

To understand the derivative here, we introduce the clusters 
$Q^j_1, Q^j_2$ ($j=3,5$) depicted in 
Figure~\ref{fig-Q35.def}.  Explicitly, these are defined 
in conjunction with the occurrence of the event
$F_j'(v_{n-1},u_n;\tilde{C}_{n-1})$, as follows:
\eqarrstar
	Q^3_2 & = & \{ y \in \mbox{ last sausage of } {C}(v_{n-1})
	: u_n \ct{{C}_{n-1}} y \},
	\\
	Q^3_1 & = & C(v_{n-1}) \backslash Q^3_2 ,
	\\
	Q^5_2 & = & \{ y \in {C}(v_{n-1}) \backslash
	[\mbox{last sausage of } {C}(v_{n-1})]
	: v_{n-1} \ct{{C}_{n-1}} y \} ,
	\\
	Q^5_1 & = & C(v_{n-1}) \backslash Q^5_2	.
\enarrstar
The $u$-dependence of $F_3'$ and $F_5'$ is then given by 
\eq
	e^{-u |Q^j_1|} \, (1 - e^{-u |Q^j_2|}) 
	\quad (j=3,5). 
\en
Its derivative is
\eq
\lbeq{Qder}
	- |Q^{j}_1| e^{-u |Q^j_{1}|} \, (1 - e^{-u |Q^j_2|}) 
	+  |Q^j_2| e^{-u (|Q^j_{1}| + |Q^j_2|)}.
\en 
The second term will turn out to be the main term, for both $j=3,5$.
The first term
will give rise to error terms that can be handled more easily,
using bounds which we defer to Section~\ref{sub-Xi.diff.bounds}.
We divide these contributions as
\eq
\lbeq{XiMU}
	\hat{\Xi}_h^{(n,m)}(0) = \int_0^h du 
	\left( \hat{\Mcal}_{h,u}^{(n,m)}(0) 
	+ \hat{\Ncal}_{h,u}^{(n,m)}(0) \right) ,
\en
where $\hat{\Mcal}_{h,u}^{(n,m)}(0)$ comprises the main terms due
to the second term of \refeq{Qder},
and $\hat{\Ncal}_{h,u}^{(n,m)}(0)$ contains the terms corresponding
to the first term of \refeq{Qder}.  In the remainder of this section,
we consider only  $\hat{\Mcal}_{h,u}^{(n,m)}(0)$, because this is the 
term for which we apply a second expansion.  

We first discuss the contribution to $\hat{\Mcal}_{h,u}^{(n,m)}(0)$ 
arising from
the $j=3$ case of \refeq{Qder}.  The result of
the differentiation is $u$-dependence of the form 
which corresponds to the cluster $C_{n}$ being entirely
$G$-free.  We write the prefactor $|Q^3_2|$ as $\sum_{y}I[y \in Q^3_2]$.
Our goal is to use a second expansion to cut off the connection
to $y$, as in Section~\ref{sub-Phidiff}.	 

\begin{figure}
\begin{center}
\includegraphics[scale = 0.5]{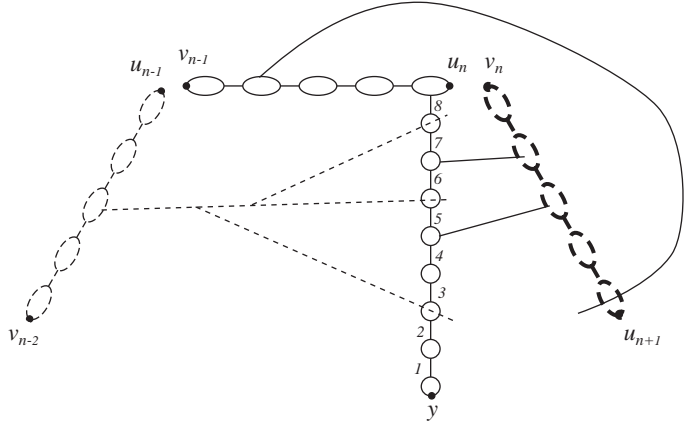} 
\end{center}
\caption{Schematic depiction of the choice of
cutting bond for $F_{3}$.  Solid lines represent 
$C_{n}$, dashed lines $C_{n-1}$, and bold dashed lines represent 
$\Bcal_{n+1}$.  Pivotal bonds in the last sausage are numbered  1 to 
8.  In this example, the cutting bond is bond~4.}
\label{fig-XiF3-cut}
\end{figure}

As in Section~\ref{sub-Phidiff}, the first step is the identification
of a suitable pivotal bond at which to sever the connection to $y$.
Again we apply Fubini's theorem to interchange the $n^{{\rm th}}$ and 
$(n+1)^{{\rm st}}$ (bond/site) expectations, 
and regard the clusters $C_{n \pm 1}$ as being fixed.  
The occurrence of the events $(F_1')_l$, $l=n\pm 1$ 
and the event $(F_{3}')_{n}$ enforces
a compatibility between $C_n$ and $C_{n \pm 1}$, in the sense that 
certain connections are required to occur for $C_n$.  
These connections
are depicted schematically in Figure~\ref{fig-XiF3-cut}.
In particular, the site $y$ is in the \emph{last} sausage for
$v_{n-1} \longrightarrow u_n$, and is 
connected to $v_{n-1}$ and $u_{n}$ 
\emph{through} $\tilde{C}_{n-1}$, since $y \in Q^{3}_2$.  

%%%%%FIGFIGFIGFIGFIGFIGFIGFIGFIGFIGFIGFIGFIGFIGFIGFIGFIGFIGFIGFIG
%%%%%FIGFIGFIGFIGFIGFIGFIGFIGFIGFIGFIGFIGFIGFIGFIGFIGFIGFIGFIGFIG
%%%%%FIGFIGFIGFIGFIGFIGFIGFIGFIGFIGFIGFIGFIGFIGFIGFIGFIGFIGFIGFIG
%%%%%FIGFIGFIGFIGFIGFIGFIGFIGFIGFIGFIGFIGFIGFIGFIGFIGFIGFIGFIGFIG

To define the cutting bond, we let $\Pcal(y)$ be the set
of occupied pivotal bonds for $y \to u_{n}$, and given $b \in \Pcal(y)$,
we let $b_+$ be the endpoint of $b$ such that $u_n \in \tilde{C}_n^b(b_+)$.
We define the following two subsets of $\Pcal(y)$: 
\eqarrstar 
    \Pcal_{n+1}(y) & = & \{ b \in \Pcal(y) : b 
    \textnormal{ is an occupied pivotal for } y \to \Bcal_{n+1} \},
    \\ 
    \Pcal_{n-1}(y) 
    & = & \{ b \in \Pcal(y) : b_{+} \ct{{C}_{n-1}} u_{n} \}.  
\enarrstar
In  Figure~\ref{fig-XiF3-cut}, 
$\Pcal_{n+1}(y) = \{1, 2, 3, 4\}$, while 
$\Pcal_{n-1}(y) = \{1, 2, 3, 4, 5, 6, 7\}$. 
The cutting bond is then defined to be the last 
element of $\Pcal_{n-1}(y) \cap \Pcal_{n+1}(y)$, in the direction $y \to u_n$.  
In the example of Figure~\ref{fig-XiF3-cut}, 
the cutting bond is bond~4. 
It is possible that no such pivotal bond exists, and in 
that case, no expansion will be required.

The reason for the above choice for the cutting bond is
as in Section~\ref{sub-Phidiff}.  We require the cutting bond
to be pivotal for
$\Bcal_{n+1}$ to ensure that we do not cut off as a tail something
which may be needed to insure
the intersection/avoidance properties 
between $C_{n}$ and $\Bcal_{n+1}$ imposed by $(F_{3}')_{n}$ and 
$(F_{1}')_{n+1}$.  
We also want to maintain the connections through $C_{n-1}$ on the last 
sausage of $C_{n}$, and our choice does preserve these
connections.  This is analogous to the first expansion, where we 
cut at the first pivotal bond after the condition $v \ct{A} x$ has
been satisfied. 

Having chosen the cutting bond, we next examine the
overall conditions present in the level-$n$ expectation.
In addition to $F_{3}'(v_{n-1}, u_{n}; C_{n-1})_{n}$ itself,
there are conditions arising from the event
$F_{1}'(v_{n}, u_{n+1}; C_{n})_{n+1}$.  Recalling the definitions in 
Section~\ref{sub-Phidiff}, the latter event can be decomposed as
\eq
	F_{1}'(v_{n}, u_{n+1}; C_{n})_{n+1} 
	= F_{0}(v_{n}, u_{n+1})_{n+1} \cap H_{\rm cut}(\Bcal_{n+1})_{n}. 
\en 
Our task now is to rewrite the overall level-$n$ condition
$F_{3,b}'(v_{n-1}, u_{n}; C_{n-1})_{n} \cap \{v_{n-1} \nc G\}_n \cap
H_{\rm cut}(\Bcal_{n+1})_{n} \cap \event{y \in Q^{3}_2}_{n}$
into a form suitable for generating the expansion. 
Here, $F_{3,b}'(v_{n-1}, u_{n}; C_{n-1})_{n}$ denotes the
intersection of $\{ v_{n-1} \conn u_n \}$ with the event that the last
sausage of $v_{n-1} \to u_n$ is connected to $C_{n-1}$.

As in Section~\ref{sub-Phidiff}, we define the events:  
\begin{align}
\lbeq{H3def}
	H_{3}(y)_{n}
	= \,  & \event{y \in Q^{3}_2} \cap 
	F_{3,b}'(v_{n-1}, u_{n}; C_{n-1})_{n} \cap 
	\{v_{n-1} \nc G\}_n  \cap 
	H_{\rm cut}(\Bcal_{n+1})_{n} ,
	\\
	H_{3}'(y)_{n}
	= \,  & H_{3}(y)_{n} \cap 
	\event{\Pcal_{n-1}(y) \cap \Pcal_{n+1}(y) = \emptyset} ,
%	\event{y \dbc C_{n-1} \cup \Bcal_{n+1} } ,
	\\ 
\lbeq{H3ppdef}
	H_{3}''(a, a')_{n} 
	= \, & 
	\{ H_{3}'(a)_n \ON  
	\tilde{C}_{n}^{\{a,a'\}}(\{v_{n-1}, u_{n}\} \cup \Bcal_{n+1}) 
	\} ,
	\\
	H_{3}(a, a', y)_{n}
	= \, & 
	H_{3}(y)_{n} \cap 
	\event {(a', a) \textrm{ is the last occupied pivotal bond in }
	\Pcal_{n-1}(y) \cap \Pcal_{n+1}(y) } .
\end{align}
Classifying configurations in $H_3(y)_n$ 
according to the last pivotal bond $(a, a')$ (if there is one) yields
the disjoint union
\eq
\lbeq{H3decomp}
	H_{3}(y)_{n} = 
%	F_{1}'(v_{n-1}, u_{n}; C_{n-1})_{n} \cap 
%	H_{\rm cut}(\Bcal_{n+1})_{n} \cap \event{y \in C_{n}} 
%	= 
	H_{3}'(y)_{n} \bigcup^{\cdot}
	\biggl ( \bigcup_{(a,a')}^{\cdot} 
	H_{3}(a, a', y)_n \biggr ) . 
\en
For $H_{3}'(y)_{n}$, no second expansion is required.
For the configurations in which there is a pivotal bond, we will use
the following lemma. 

\begin{lemma}
\label{lem-F3cut}
The events $H_{3}(a, a', y)_n$ and $H_{3}''(a, a')_{n}$ obey 
\begin{align}
	H_{3}(a, a', y)_n
	= \, & H_{3}''(a, a')_{n}
	\cap 
	\event{ ( y \conn a' \AND y \nc G) \INSIDE \Zd \backslash 
         \tilde{C}_n^{\{a,a'\}}(\{v_{n-1}, u_{n}\} \cup \Bcal_{n+1}) } 
	 \nnb 
	 & 
	 \cap \event{\{a, a'\} \textnormal{ is occupied}}.
\lbeq{F3cuteq}
\end{align}
\end{lemma}

We omit the proof of Lemma~\ref{lem-F3cut}, since
it proceeds in the same way as the proof
of Lemma~\ref{lem-F1cut}.  However, there is one
respect which is somewhat different.  Unlike the analysis for
$H_1(y)$, for $H_3'$ it is not possible to write the
``no pivotal'' condition as $\{y \dbc C_{n-1} \cup \Bcal_{n+1}\}$.  
It is true that every configuration in $H_3'$ obeys
$\{y \dbc C_{n-1} \cup \Bcal_{n+1}\}$, but the converse is not true.  
For example, in 
the configuration of Figure~\ref{fig-XiF3-cut}, the true 
cutting bond is bond~4, even though 
$\{y \dbc C_{n-1} \cup \Bcal_{n+1}\}$ occurs already after bond~2.
But this difference from the situation in Lemma~\ref{lem-F1cut} is
a minor one, and because our choice of the 
cutting bond imposes no $C_{n-1}$-related conditions on the $a'$-side of 
the connection $y \conn a'$, the analysis can proceed as before. 

Now we note that together with \refeq{H3decomp} and
Lemma~\ref{lem-cond.0}, Lemma~\ref{lem-F3cut} implies the identity
\eq
\lbeq{H3-cut.2}
	\expec{I[H_{3}(y)_{n}]}_{n}^{\tilde{}} = 
	\expec{I[H_{3}'(y)_{n}]}_{n}^{\tilde{}} +
	p_c \sum_{(a,a')}  
	\langle I[ H_{3}''(a, a')_{n}] \, 
	\tau_{u}^{\tilde{C}_n^{\{a,a'\}}
	(\{v_{n-1}, u_{n}\} \cup \Bcal_{n+1})}(a', y) \rangle_{n}^{\tilde{}} . 
\en 
As in \refeq{H1-cut.2}, the restricted
two-point function in \refeq{H3-cut.2} is with respect to 
the ordinary unconditional expectation.
The identity \refeq{H3-cut.2}
is exactly analogous to \refeq{H1-cut.2}, and the second expansion
can be derived for \refeq{Xi.diff.0} using the one-$M$ scheme,
exactly as was done in Section~\ref{sub-Phidiff}.
A minor difference here is that the
level-$n$ expectation has magnetic field $u$.
As a result, the expansion for the level-$n$ expectation is 
\refeq{secexp-Phi.Hnexp} with all the $H_{1}$ replaced by $H_{3}$, and 
with all the expectations of levels-$(n,m)$ having magnetic 
field $u$.  Then we substitute this result back into \refeq{Xi.diff.0}. 
The result of the expansion will be given below, after we consider 
the case of $F_5$.

We now consider the main contribution to the $F_{5}$ case, which
arises from the $j=5$ case of the seond term of \refeq{Qder}.
The choice of cutting bond is defined in exactly the same
way as it was for the case of $F_3$, and the second expansion
proceeds in the same way.  We define $H_5$ events as in 
\refeq{H3def}--\refeq{H3ppdef}, with $F_3'$ and $\{y \in Q^3_2\}$
in \refeq{H3def} respectively replaced by $F_5'$ and $\{y \in Q^5_2\}$.
This leads, as above, to the identity
\eq
\lbeq{H5-cut.2}
	\expec{I[H_{5}(y)_{n}]}_{n}^{\tilde{}} = 
	\expec{I[H_{5}'(y)_{n}]}_{n}^{\tilde{}} +
	p_c \sum_{(a,a')}  
	\langle I[ H_{5}''(a, a')_{n}] \, 
	\tau_{u}^{\tilde{C}_n^{\{a,a'\}}
	(\{v_{n-1}, u_{n}\} \cup \Bcal_{n+1})}(a', y) \rangle_{n}^{\tilde{}} . 
\en
The second expansion then proceeds as usual, via the one-$M$ scheme.
We perform the second expansion to infinite order, as will be justified
by the bounds of the next section.

To summarize, the second expansion yields a result of the form
\eq
\lbeq{MGHR}
	\hat{\Mcal}_{h,u}^{(n,m)}(0) = \hat{\Gcal}_{h,u}^{(n,m)}(0)
	+ \hat{\Hcal}_{h,u}^{(n,m)}(0) \chi_u + \hat{\Rcal}_{h,u}^{(n,m)}(0) .
\en
The terms on the right side are sums of doubly nested expectations,
with the second nesting occurring at level $n$ of the original
nested expectation defining $\hat{\Xi}^{(n,m)}_h(0)$.
The term $\hat{\Gcal}_{h,u}^{(n,m)}(0)$ contains the terms in which the
innermost expectation in the second nesting carries a single prime, and
is analogous to $\Ucal$ of \refeq{PhiNdif}.  
The term $\hat{\Hcal}_{h,u}^{(n,m)}(0)$ contains the terms 
in which the
innermost expectation in the second nesting carries a double prime, and 
is analogous to $\Vcal$ of \refeq{PhiNdif}. 
The term $\hat{\Rcal}_{h,u}^{(n,m)}(0)$ contains the terms 
in which the
innermost expectation in the second nesting involves $F_2$, and 
is analogous to $\Ecal$ of \refeq{PhiNdif}. 
Each of these three quantities can be written as a sum over $\ell$,
where $\ell$ is the number of expectations nested within the $n^{\rm th}$
expectation.

%%%%%%%%%%%%%%%%%%%%%%%%%%%%%%%%%%%%%%%%%%%%%%%%%%%%%%%%%%%%%%%%%%%%%%%%%%%%%%%
%%%%%%%%%%%%%%%%%%%%%%%%%%%%%%%%%%%%%%%%%%%%%%%%%%%%%%%%%%%%%%%%%%%%%%%%%%%%%%%
\subsection{Proof of Proposition~\protect\ref{prop-Xi}}
\label{sub-Xi.diff.bounds}

%Proposition~\protect\ref{prop-Xi} is a consequence of the following lemma.

\begin{lemma}
\label{lem-Xih}
Let $u \in [0,h]$.
The series $\hat{\Gcal}_{h,u}(0) = 
\sum_{n,m=1}^\infty \hat{\Gcal}_{h,u}^{(n,m)}(0)$,
$\hat{\Hcal}_{h,u}(0) = \sum_{n,m=1}^\infty \hat{\Hcal}_{h,u}^{(n,m)}(0)$, 
$\hat{\Rcal}_{h,u}(0) = \sum_{n,m=1}^\infty \hat{\Rcal}_{h,u}^{(n,m)}(0)$,
and $\hat{\Ncal}_{h,u}(0) = \sum_{n,m=1}^\infty \hat{\Ncal}_{h,u}^{(n,m)}(0)$
converge absolutely.
Moreover, $\hat{\Gcal}_{h,u}(0) = O(1)$, $\hat{\Rcal}_{h,u}(0) = \oh \chi_u$,
$\hat{\Ncal}_{h,u}(0) = \oh \chi_u$, and 
\eq
\lbeq{Hcal.value}
	\hat{\Hcal}_{h,u}(0) = \hat{\Hcal}_{0,0}(0) + \oh ,
	\quad	\quad    \hat{\Hcal}_{0,0}(0) = O(\oneOd).
\en
\end{lemma}   

Combining \refeq{XiMU},
\refeq{MGHR}, and Lemma~\ref{lem-Xih}, we have
\eq
\lbeq{Xi.ex-by-Hcal}
	\hat{\Xi}_{h}(0) = \int_0^h du \left( 
	\hat{\Hcal}_{0,0}(0) \chi_u +\oh \chi_u
	\right) = \left( 
	\hat{\Hcal}_{0,0}(0)   + \oh \right) M_h .
\en
This gives Proposition~\ref{prop-Xi} with
with $K_2 = - \hat{\Hcal}_{0,0}(0)$.

\medskip \noindent
{\bf Proof of Lemma~\ref{lem-Xih}}.  
The bounds on $\hat{\Gcal}_{h,u}^{(n,m)}(0)$ are similar to those
on $\hat{\Hcal}_{h,u}^{(n,m)}(0)$, and will not be discussed further.
Also, the arguments required for the nearest-neighbour and spread-out
models are slightly different, as in the proof of Lemma~\ref{lem-Psih},
and for simplicity
we restrict attention in what follows to the spread-out model.

\begin{figure}
\begin{center}
\includegraphics[scale = 0.4]{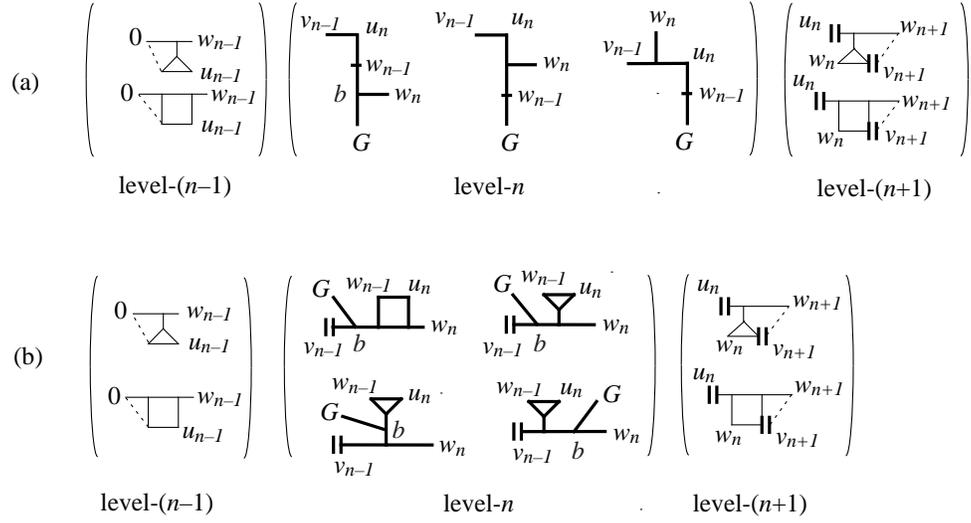} 
\end{center}
\caption{The truncated diagrams contributing to $\Xi$
due to (a) $F_3$ and (b) $F_5$.}
\label{fig-peel-before-Xi}
\end{figure}

Before beginning the proof in earnest, we examine the diagrams
used to bound $\hat{\Xi}_h^{(n,m)}(0)$ in more detail.
These differ from the diagrams of $\hat{\Phi}_h^{(n+m)}(0)$
only at levels $n-1, n, n+1$.  When performing diagrammatic estimates,
the other expectations can be bounded using triangles.  These
triangles give rise
to a factor $\min\{1,\oneOd^{n+m-3}\}$.
Recalling \refeq{Wppw-event.2}--\refeq{Wppbd.2} and 
Figure~\ref{fig-peel-before}~(a), we are left with the 
truncated diagrams depicted schematically
in Figure~\ref{fig-peel-before-Xi}.  The factor $\oneOd^{n+m-3}$ controls the
sums over $n,m$, and it suffices to obtain an appropriate bound
on the truncated diagram, modified to take into account the
diagrammatic changes arising in $\Ncal, \Hcal, \Rcal$.
 
%%%%%FIGFIGFIGFIGFIGFIGFIGFIGFIGFIGFIGFIGFIGFIGFIGFIGFIGFIGFIGFIG
%%%%%FIGFIGFIGFIGFIGFIGFIGFIGFIGFIGFIGFIGFIGFIGFIGFIGFIGFIGFIGFIG
%%%%%FIGFIGFIGFIGFIGFIGFIGFIGFIGFIGFIGFIGFIGFIGFIGFIGFIGFIGFIGFIG
%%%%%FIGFIGFIGFIGFIGFIGFIGFIGFIGFIGFIGFIGFIGFIGFIGFIGFIGFIGFIGFIG

It is helpful to examine the infrared degree of divergence of the
diagrams of Figure~\ref{fig-peel-before-Xi}.
Consider first the diagrams of Figure~\ref{fig-peel-before-Xi}~(a)
with the line
terminating at $G$ and the vertex at $b$ (for the first diagram)
or $w_{n-1}$ (for the second and third diagrams) omitted, and those of
Figure~\ref{fig-peel-before-Xi}~(b)
with the line
terminating at $G$ and the vertex at $b$ omitted.  We call these
the amputated truncated diagrams.
As in Figure~\ref{fig-peel-before}~(c), it can be seen that the
infrared degree of divergence of the amputated
truncated diagrams is at least $d-4$.
If we then restore the vertex at $w_{n-1}$ or $b$ to an amputated 
truncated diagram,
this is construction~1 and hence the resulting diagram has infrared
degree at least $d-6$, by Lemma~\ref{lem-Ginduction}.

We divide the proof of Lemma~\ref{lem-Xih} into several parts.  
First, we consider the error term $\Ncal$.  We then consider the $\ell =0$ 
contribution to $\Hcal$.  Then we move on to the contributions of 
$\ell \geq 1$ to $\Hcal$ and $\Rcal$.  Finally, we prove \refeq{Hcal.value}. 
Our discussion will be brief at points where it does not differ
substantially from the proof of Lemma~\ref{lem-Psih}.

%%%%%%%%%%%%%%%%%%%%%%%%%%%%%%%%%%%%%%%%%%%%%%%%%%%%%%%%%%%%%%%%%%%%%%%%%%%%%%%
\smallskip \noindent
{\bf Bounds on $\Ncal$}

\noindent
The error term $\Ncal$ 
is generated by the first term of \refeq{Qder}, which 
involves adding a connection to $y \in Q_{1}^{3}$ or $y \in Q_{1}^{5}$,
and then summing over $y$.  

Consider first the case of an added connection to $y \in Q_{1}^{5}$.
If we ignore the connection to $G$ in
an upper bound, this corresponds to adding a vertex at level-$n$ to
a truncated amputated diagram, with
a line emanating to $y$.  We can use the cut-the-tail Lemma~\ref{lem-tail.tau1}
to extract a factor of $\chi_u$, multiplied by a diagram with infrared
degree $d-6$.  (This also involves an application of construction~2, which
does not decrease the infrared degree.)
However, if we recall that this diagram actually has
a connection to $G$, we can use the dominated convergence theorem to
obtain an overall bound $o_h(1) \chi_u$.  The sum over $m,n$ can then
be performed, thanks to the factor $\oneOd^{n+m-3}$ mentioned previously.

Now consider the case of an added connection to $y \in Q_{1}^{3}$.
In this case, the connection to $G$ plays an essential role in maintaining
diagram connectivity, and it cannot be ignored.  However, we can
use Lemma~\ref{lem-tail.tau1} to produce a factor $\chi_u$, we can extract
a factor $M_u$ from the connection to $G$, and by employing a $G$-free
line from the expectation at level-$(n-1)$ or level-$(n+1)$ we can bound
the remaining diagram by an analysis similar to that used to bound
$\hat{\Ecal}_h^{(N,n,1)}(0)$ in Section~\ref{sub-Phi.diff.bounds}.
The overall result is a bound $\chi_u M_u h^{(d-8)/4} \leq \oh \chi_u$.

%%%%%%%%%%%%%%%%%%%%%%%%%%%%%%%%%%%%%%%%%%%%%%%%%%%%%%%%%%%%%%%%%%%%%%%%%%%%%%%
\bigskip \noindent
{\bf The case $\ell =0$}

\smallskip \noindent
We now bound the diagrams contributing to the
lowest order ($\ell =0$) contributions to $\Hcal$, i.e.,
\eq
\lbeq{Hjpp}
	{\Ebold}_{0} I[E_{0}''] \, 
	{\Ebold}_{1} Y_{1}'' \, 
	\cdots 
	\tilde{\Ebold}_{n-1} Y_{n-1}' \, 
	\tilde{\Ebold}_{n+1} I[(F_{0})_{n+1} ] \, 
	\tilde{\Ebold}_{n}  (H_{j}'')_{n} \, 
	{\Ebold}_{n+2}  Y_{n+2}'' \, 
	\cdots 
	{\Ebold}_{n+m}  Y_{n+m}'' ,
\en 
where $j=3$ or $j=5$.  The discussion parallels the corresponding
part of the proof of Lemma~\ref{lem-Psih}.  The diagram is truncated
as described above, and we bound the modification of the amputated truncated
diagram which takes into account the additional connections implied
by $H_j''$.  We will argue that these connections arise from
application of construction~1 
followed by constructions~2 or 3.  Thus the infrared degree of divergence
is reduced from at least $d-4$ to at worst $d-6$, by 
\refeq{Gind}--\refeq{cor-Gind}, and hence the diagrams are $O(1)$.

Consider first the case $j=3$.
In addition to the bond connections required by $F_3'(v_{n-1},u_n;C_{n-1})$,
we add connections due to
$a_0 \in Q^3_2$ and $\Pcal_{n-1}(a_0) \cap \Pcal_{n+1}(a_0) = \emptyset$.
The site $a_0$ is located in the place of $G$ of 
Figure~\ref{fig-peel-before-Xi}~(a).  
By definition, it is either the case that 
$\Pcal_{n-1}(a_0) \subset \Pcal_{n+1}(a_0)$ or
$\Pcal_{n+1}(a_0) \subset \Pcal_{n-1}(a_0)$.
Thus we need only consider the two cases 
(i) $\Pcal_{n-1}(a_0) = \emptyset$ and (ii) $\Pcal_{n+1}(a_0) = \emptyset$.

\begin{figure}
\begin{center}
\includegraphics[scale = 0.2]{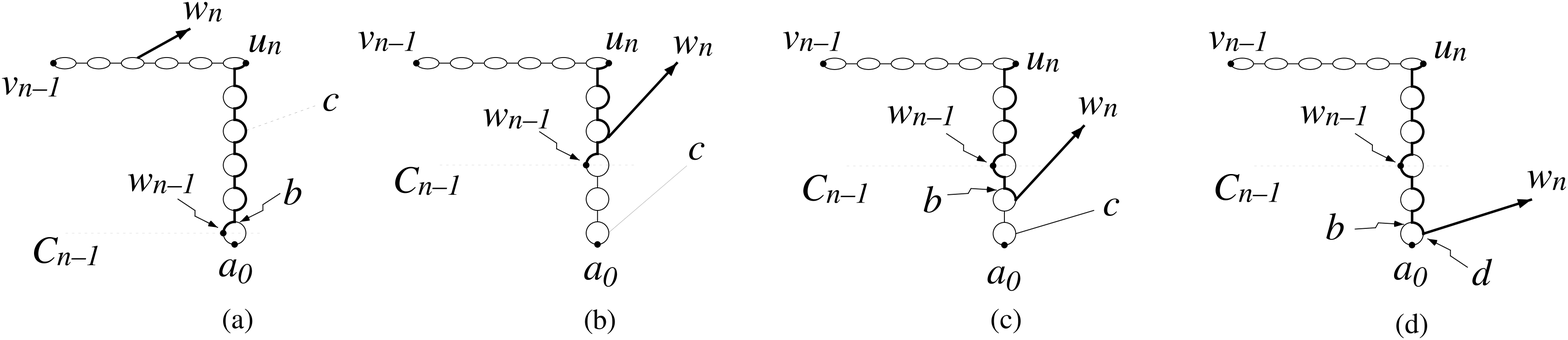} 
\end{center}
\caption{Original configurations for $\Xi$ ($F_{3}$ case), and new 
connections required for $a_0$.}
\label{fig-Xi0Q3}
\end{figure}

We first consider case (i).
This is depicted in Figure~\ref{fig-Xi0Q3}~(a).  The extra required
(disjoint) connections are $a_0 \conn w_{n-1}$ and $a_0 \conn b$.
Here $b$ is a new vertex, while $w_{n-1}$ was existing.  Thus this
is an application of construction~2, which does not reduce the infrared 
degree.  (A similar construction can be applied in the case where
the connection to $w_n$ emerges from the sausage containing $w_{n-1}$.)
 
%%%%%FIGFIGFIGFIGFIGFIGFIGFIGFIGFIGFIGFIGFIGFIGFIGFIGFIGFIGFIGFIG
%%%%%FIGFIGFIGFIGFIGFIGFIGFIGFIGFIGFIGFIGFIGFIGFIGFIGFIGFIGFIGFIG
%%%%%FIGFIGFIGFIGFIGFIGFIGFIGFIGFIGFIGFIGFIGFIGFIGFIGFIGFIGFIGFIG
%%%%%FIGFIGFIGFIGFIGFIGFIGFIGFIGFIGFIGFIGFIGFIGFIGFIGFIGFIGFIGFIG

We next consider case (ii).
There are several geometries to consider, three of which are
depicted in Figure~\ref{fig-Xi0Q3}~(b-d).  In (b) and (c), there
are additional disjoint paths from an existing vertex $b$ in
the $\Xi$-diagram to $a_0$ (in (b), we take 
$b = w_{n-1}$) and from $a_0$ to a new vertex $c \in \Bcal_{n+1}$.
We may then argue as we did in the proof of Lemma~\ref{lem-Psih} that
$c$ is then connected to existing diagram lines from level-$(n+1)$
in such a way that the overall additional lines are of the form of
construction~3.  The geometry of Figure~\ref{fig-Xi0Q3}~(d), in which
the connection to $w_n$ emerges from the sausage of $a_0$, deserves
special comment.  In this case, we may neglect any additional connection
to $\Bcal_{n+1}$ beyond that already present due to the connection
from $b$ to $w_n$.  The new lines to be added are those from $b$ to $a_0$
and from $a_0$ to $d$ (new),
which is an application of construction~2.

Next, we move on to the case $j=5$ in \refeq{Hjpp}, which involves
$Q_{2}^{5}$ shown schematically in Figure~\ref{fig-Q35.def}. 
%%%%%FIGFIGFIGFIGFIGFIGFIGFIGFIGFIGFIGFIGFIGFIGFIGFIGFIGFIGFIGFIG
%%%%%FIGFIGFIGFIGFIGFIGFIGFIGFIGFIGFIGFIGFIGFIGFIGFIGFIGFIGFIGFIG
\begin{figure}
\begin{center}
\includegraphics[scale = 0.2]{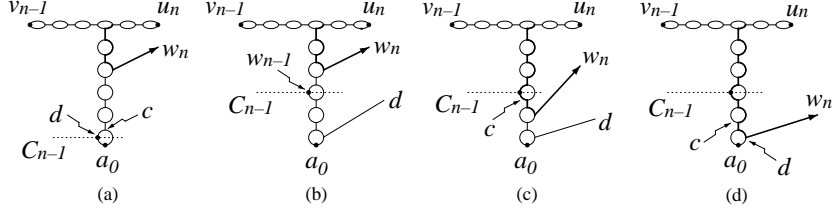} 
\end{center}
\caption{Original configurations for $\Xi$ ($F_{5}$ case), and new 
connections required for $a_0$. }
\label{fig-Xi0Q5}
\end{figure}
%%%%%FIGFIGFIGFIGFIGFIGFIGFIGFIGFIGFIGFIGFIGFIGFIGFIGFIGFIGFIGFIG
%%%%%FIGFIGFIGFIGFIGFIGFIGFIGFIGFIGFIGFIGFIGFIGFIGFIGFIGFIGFIGFIG
Four of the relevant geometries are depicted in Figure~\ref{fig-Xi0Q5}.
In (a), the new connections link $a_0$ to $d$ (existing) and $a_0$ to
$c$ (new), which is construction~2.  In (b), the new connections link 
$a_0$ to $w_{n-1}$ and $a_0$ to $d$ (new), which is construction~3.
In (c), the new connections link $a_0$ to $c$ (existing) and $a_0$ to $d$ (new),
which is construction~3.  Finally, in (d), the new connections link
$a_0$ to $c$ (existing) and $a_0$ to $d$ (new), which is construction~2.
None of these constructions decrease the infrared degree from its
original value of $d-6$, so the diagrams are $O(1)$.

%%%%%%%%%%%%%%%%%%%%%%%%%%%%%%%%%%%%%%%%%%%%%%%%%%%%%%%%%%%%%%%%%%%%%%%%%%%%%%%
\bigskip \noindent
{\bf The case $\ell \geq 1$}

\smallskip \noindent
This case is bounded exactly as was done in the proof of Lemma~\ref{lem-Psih}.
Note that we require massive lines to estimate $\Rcal$, but after
differentiation the level-$n$ expectation becomes $G$-free and
the massive lines can be obtained as before.
One new ingredient here is that we have two magnetic fields $h$ and $u$,
but the simple inequalities $u \leq h$, $M_u \leq M_h$, and 
$\chi_u \geq \chi_h$ can be applied.

%%%%%%%%%%%%%%%%%%%%%%%%%%%%%%%%%%%%%%%%%%%%%%%%%%%%%%%%%%%%%%%%%%%%%%%%%%%%%%%
\bigskip \noindent
{\bf Proof of \refeq{Hcal.value}}

\smallskip \noindent
The proof that $\hat{\Hcal}_{h,u}(0) = \hat{\Hcal}_{0,0}(0) + \oh$
proceeds as in the portion of the proof of Lemma~\ref{lem-Psih}
showing that $\hat{\Vcal}_h(0) - \hat{\Vcal}_0(0) = \oh$, 
apart from the fact that now we have two magnetic fields $h$ and $u$. 
We again write the difference as a telescopic sum 
of differences at each level, and each difference implies a connection to 
$G$.  We further decompose this sum of nested expectations as plus or
minus sums 
of positive expectations. The difference can then be bounded above  
by setting $u=h$, since increasing the magnetic field increases the 
difference.   
Then our previous discussion applies.  

To prove $\hat{\Hcal}_{0,0}(0) = \oneOd$, we use the dominated convergence
theorem as in the proof that $\hat{\Vcal}_0(0) = 1+O(\oneOd)$. 
However, each term in $\hat{\Hcal}_{0,0}(0)$ contains at least one
loop with a pivotal bond, and the $O(1)$ contribution does not occur.
\qed

%%%%%%%%%%%%%%%%%%%%%%%%%%%%%%%%%%%%%%%%%%%%%%%%%%%%%%%%%%%%%%%%%%%%%%%%%%%%%%%
%%%%%%%%%%%%%%%%%%%%%%%%%%%%%%%%%%%%%%%%%%%%%%%%%%%%%%%%%%%%%%%%%%%%%%%%%%%%%%%
%%%%%%%%%%%%%%%%%%%%%%%%%%%%%%%%%%%%%%%%%%%%%%%%%%%%%%%%%%%%%%%%%%%%%%%%%%%%%%%
%%%%%%%%%%%%%%%%%%%%%%%%%%%%%%%%%%%%%%%%%%%%%%%%%%%%%%%%%%%%%%%%%%%%%%%%%%%%%%%
\renewcommand{\thesection}{\Alph{section}}
\setcounter{section}{0}
\section{Power counting for Feynman diagrams}
\label{app}
\setcounter{equation}{0} 

In Section~\ref{sub-pc}, we summarize results of \cite{Reis88b,Reis88a} 
concerning the estimation of Feynman diagrams using the quantum
field theoretic technique of power counting.  
Then in Section~\ref{sub-ind.1}, we
provide a lemma which allows for an efficient application
of these results for the Feynman diagrams arising in this paper.

%%%%%%%%%%%%%%%%%%%%%%%%%%%%%%%%%%%%%%%%%%%%%%%%%%%%%%%%%%%%%%%%%%%%%%%%%%%%%%%
%%%%%%%%%%%%%%%%%%%%%%%%%%%%%%%%%%%%%%%%%%%%%%%%%%%%%%%%%%%%%%%%%%%%%%%%%%%%%%%
\subsection{Power counting}
\label{sub-pc}

Consider a Feynman diagram $G$ consisting of  
$N$ internal lines, no external lines, and $V$ vertices.  
Each (internal) line carries a $d$-dimensional 
momentum $p_{i}$ ($i = 1, 2, \ldots, N$), and 
represents a propagator  
\eq
\lbeq{Fdprop}
	\frac{1}{p_i^2 + \mu_i^2} ,
\en 
where the mass $\mu_i$ of the $i^{\rm th}$ line can be either
$0$ or $\mu >0$, with $\mu$ not depending on $i$.  
The massless ($\mu_i=0$) and massive $(\mu_i = \mu$) lines are fixed in $G$.

The Feynman diagrams encountered in this paper have propagator
$\hat{\tau}_{h,p_c}(k)$.  By Proposition~\ref{prop-taubd}, this propagator
is bounded above by a constant multiple of $([1-\hat{D}(k)] + h^{1/2})^{-1}$,
with the constant independent of $\Omega$.  For both the nearest-neighbour
and spread-out models, $1-\hat{D}(k)$ is bounded below by a universal
constant multiple of $k^2/d$ (see \cite[Appendix~A]{MS93}).  
Therefore $\hat{\tau}_{h,p_c}(k)$ is bounded
above by a propagator of the form \refeq{Fdprop} times a factor $d$ which
should be taken into account in bounding diagrams for the nearest-neighbour
model, but which is unimportant for the spread-out model.  Diagram lines
with $h > 0$ have mass $\mu = h^{1/4}$.

Each of the $V$ vertices of $G$ imposes a 
momentum conservation condition, according to Kirchoff's law.  
Of these, $V-1$ are independent, with
the momentum conservation at the other vertex then guaranteed by overall
momentum conservation.  As a result, we 
have $V-1$ independent momentum constraints.
This leaves $L = N-V+1$ independent momenta $k_j$ ($j=1, 2, \ldots , L$), 
called loop momenta, which can be chosen from $\{p_i\}_{i=1}^{N}$. 
The choice of loop momenta is not uniquely determined by $G$.
Given a choice of loop momenta $\{k_j\}_{j=1}^{L}$, 
each $p_i$ can be written as a linear combination 
of the $k_j$.  
The Feynman integral $I_G$ giving the value of the Feynman diagram $G$
is then 
\eq
	I_G =  \int_{[-\pi, \pi]^d} \frac{d^dk_1}{(2\pi)^d} \cdots 
	\int_{[-\pi, \pi]^d} \frac{d^dk_L}{(2\pi)^d} \prod_{i=1}^{N} 
	\frac{1}{p_i^2 + \mu_i^2} .
\en 
The value of $I_G$ is independent of the choice of loop momenta.

Our goals are (i) to provide a sufficient condition for convergence 
of $I_{G}$ when $\mu \geq 0$, 
and (ii) to determine the rate of divergence of $I_{G}$, as $\mu \to 0$, 
in the case where $I_{G}$ is convergent for $\mu >0$ but not for $\mu =0$.
For this, we will use the infrared degree of divergence 
$\underline{\deg}_\mu(G)$,
which is defined as follows.
First, given a set $\Gcal$ of $L$ loop momenta, 
and a subset $\Hcal \subset \Gcal$ of cardinality $\ell$, 
we define the infrared degree of divergence of $\Hcal$ by 
\eq
\lbeq{degIRHdef}
		\underline{\deg}_\mu(\Hcal) = d \ell 
		- 2 \# \{
		\textrm{massless line momenta determined by $\Hcal$}
		\}.
\en
Note that \refeq{degIRHdef} makes sense for $d \in {\mathbb R}$.
For $\underline{\deg}_0$, all lines are regarded as massless, whereas for 
$\underline{\deg}_\mu$,
only the lines for which $\mu_i =0$ are massless.
In \refeq{degIRHdef}, 
a line momentum $p_i$ is said to be determined by $\Hcal$ if
$p_i$ is in the linear span of $\Hcal$.
Then we define the infrared degree of divergence of the full graph $G$ by
\eq
	\lbeq{degIR.def1}
		\underline{\deg}_\mu(G) = 
		\min_{\Gcal} \, 
		\min_{\Hcal:  \Hcal \subset \Gcal, \Hcal \neq \emptyset} \, 
		\underline{\deg}_\mu(\Hcal),
\en 
where the minimum is taken over all choices 
$\Gcal$ of loop momenta for $G$ and over all 
nonempty subsets $\Hcal \subset \Gcal$. 
The following theorem, which is \cite[Theorem~1]{Reis88b},
asserts that a Feynman integral is finite if
its infrared degree of divergence is positive.

\begin{theorem}
\label{thm-R1}
The Feynman integral $I_{G}$ converges if $\underline{\deg}_\mu(G) > 0$.  
\end{theorem}

Theorem~\ref{thm-R1} implies that a diagram is more likely to be 
convergent in high dimensions.  We define $d_c(G)$, the \emph{critical 
dimension} of a diagram $G$, by 
\eq
\lbeq{dcdef}
	d_c(G) = \inf \{  d \in {\mathbb R} : \underline{\deg}_0(G) > 0 \} . 
\en 
By definition, $I_G$ converges if $d > d_c(G)$. 

We now consider the situation where $I_{G}$ is convergent for
$\mu >0$ but not for $\mu =0$.  In this case, the following theorem, which
we will show to be 
a consequence of \cite[Theorem~2]{Reis88b}, indicates that the infrared
degree of divergence $\underline{\deg}_0(G)$ bounds the rate of divergence 
of $I_{G}$
in the limit $\mu \to 0$.

\begin{theorem}
\label{thm-R2}
Suppose $\underline{\deg}_\mu(G)>0$ but $\underline{\deg}_0(G) \leq 0$.  
Then the Feynman integral $I_{G}$ is finite for $\mu >0$, and, 
as $\mu \to 0$, obeys the bound
\eq
	I_{G} \leq \mbox{const.} \mu^{\underline{\deg}_0(G)} |\log \mu|^L .
\en 
\end{theorem}

\Proof
Making the change of variables
$\tilde{k}_i = \mu^{-1} k_i , \; \tilde{p}_j = \mu^{-1} p_j $ 
gives
\eq
	I_G  
	= \mu^{d L - 2 N} \int_{[-\pi/\mu, \pi/\mu]^d} 
	\frac{d^d \tilde{k}_1}{(2\pi)^d} \cdots 
	\int_{[-\pi/\mu, \pi/\mu]^d} 
	\frac{d^d \tilde{k}_L}{(2\pi)^d} \prod_{i=1}^{N} 
	\frac{1}{\tilde{p}_i^2 + \tilde{\mu}_i^2}  
	\equiv \mu^{d L - 2 N} J_G, 
	\lbeq{intUV-rewrite}
\en 
where $\tilde{\mu}_i$ is zero or one, depending on whether $\mu_i$ is 
zero or $\mu$.  
The rate of divergence of $J_G$ is given in  
\cite{Reis88b} in terms of the 
\emph{ultraviolet}\/ degree of divergence.  This is defined as follows. 
Given a set of loop momenta $\Gcal$ for $G$, 
and a subset $\Hcal \subset \Gcal$ of cardinality $\ell$, 
we define  
\eq
\lbeq{deguvdef}
	\overline{\deg}(G) 
	= \max_\Gcal 
	\max_{\Hcal : \Hcal \subset \Gcal, \Hcal \neq \emptyset} \, 
	\overline{\deg}(\Hcal), 
	\; \; \mbox{ where} \; \;
	\overline{\deg}(\Hcal) = d \ell - 2 \# \{
	\textrm{line momenta depending on $\Hcal$}
	\}	.
\en 
Here, a line momentum $p_i$ is said to be 
depending on $\Hcal$ if it is not determined by $\Gcal \backslash \Hcal$.  
%It is known that $\overline{\deg}(G)$ is independent of the choice of 
%$\Gcal$.  
Also, since
there is no mention of massless lines in the definition of the ultraviolet
degree of divergence, there is no need to distinguish $\overline{\deg}_\mu$ and
$\overline{\deg}_0$.
It then follows from \cite[Corollary to Theorem~2]{Reis88b} that
\eq
	J_G \leq \begin{cases} 
		\textrm{const}. 	& (\overline{\deg}(G) < 0) \\
		\textrm{const}. \, \mu^{- \overline{\deg}(G)} 
		| \log \mu |^L \quad 	& (\overline{\deg}(G) \geq 0) .
	\end{cases} 
	\lbeq{int-div-UV}
\en 
Therefore
\eq
\lbeq{IG1}
	I_G \leq \mbox{const.}
	\mu^{dL-2N - \max \{ 0, \overline{\deg}(G) \}} | \log \mu |^L.
\en

It remains to relate the exponent of $\mu$ on the right side to 
$\underline{\deg}_0(G)$.
First, we claim that for any subset $\Hcal \subset \Gcal$,
\eq
	\underline{\deg}_0(\Hcal) 
	= d L - 2N  - \overline{\deg}(\Gcal\backslash \Hcal) .
	\lbeq{rel-IRUV2}
\en 
Here, we employ the convention 
$\underline{\deg}_0(\emptyset) = \overline{\deg}(\emptyset) 
= 0$.  The claim follows from the fact that the set $\Gcal$ of all line momenta
is the disjoint union of the set of line momenta determined by $\Hcal$
and the set of line momenta depending on $\Gcal \backslash \Hcal$. 
Now, we take the minimum of \refeq{rel-IRUV2} 
over all $\Gcal$ and nonempty $\Hcal \subset \Gcal$.  This leads to 
\eqalign
\lbeq{rel-IRUV3}
	\underline{\deg}_0(G) 
	& = d L - 2 N - \max_{\Gcal} 
	\max_{\Hcal : \Hcal \subset \Gcal, \Hcal \neq \emptyset} 
	\overline{\deg}(\Gcal \backslash \Hcal) 
	= d L - 2 N - \max_{\Gcal} 
	\max_{\Hcal :  \Hcal \subset \Gcal, \Hcal \neq \Gcal}
	\overline{\deg}(\Hcal) 
	\nnb
	& = d L - 2 N - \max \biggl \{ 0, 
	\max_{\Gcal} 
	\max_{\Hcal :  \Hcal \subset \Gcal; \Hcal \neq \emptyset, \Gcal}
	\overline{\deg}(\Hcal)
	\biggr \} . 
\enalign

We first consider the case where the maximum in the definition 
\refeq{deguvdef} of
$\overline{\deg}(G)$ is attained at some $\Hcal \neq \Gcal$.  Then
\refeq{rel-IRUV3} implies that 
$dL-2N - \max \{ 0, \overline{\deg}(G) \} = \underline{\deg}_0(G)$,
and the bound of the theorem follows from \refeq{IG1}.

This leaves the case where the maximum is attained at $\Hcal = \Gcal$,
which implies $\overline{\deg}(G) = dL-2N$.  Suppose first that
$dL-2N <0$.  Then we may include the case $\Hcal = \Gcal$ in the right
side of \refeq{rel-IRUV3} without changing its value, so again we have
$dL-2N -\max\{0,\overline{\deg}(G)\}= \underline{\deg}_0(G)$ 
and the desired result follows
from \refeq{IG1}.
Finally, suppose that $\overline{\deg}(G) = dL-2N \geq 0$.  
Relaxing the condition $\Hcal \neq \Gcal$ on the right side of
\refeq{rel-IRUV3} gives
$\underline{\deg}_0(G) \geq d L - 2 N - \max  \{ 0, 
\overline{\deg}(G)\}  = d L - 2 N  - (d L - 2 N ) = 0$.  Since 
$\underline{\deg}_0(G) \leq 0$ by hypothesis, we may assume that
$\underline{\deg}_0(G) =  0$.  The desired result then follows
from \refeq{IG1}, because the exponent of $\mu$ in \refeq{IG1} is zero 
(as we have just shown), and thus 
$I_{G} \leq \textrm{const}. \mu^{0} | \log \mu |^{L}
= \textrm{const}. \mu^{\underline{\deg}_0(G)} | \log \mu |^{L}$. 
\qed

\medskip

Theorems~\ref{thm-R1} and \ref{thm-R2} reduce the 
analysis of Feynman integrals to the evaluation of the
infrared degree of divergence. 
In the next section, we give a practical method for calculating the
infrared degree of divergence of some of the Feynman diagrams arising in this
paper.

%%%%%%%%%%%%%%%%%%%%%%%%%%%%%%%%%%%%%%%%%%%%%%%%%%%%%%%%%%%%%%%%%%%%%%%%%%%%%%%
%%%%%%%%%%%%%%%%%%%%%%%%%%%%%%%%%%%%%%%%%%%%%%%%%%%%%%%%%%%%%%%%%%%%%%%%%%%%%%%
\subsection{Inductive power counting}
\label{sub-ind.1}

\begin{figure}
\begin{center}
\includegraphics[scale = 0.3]{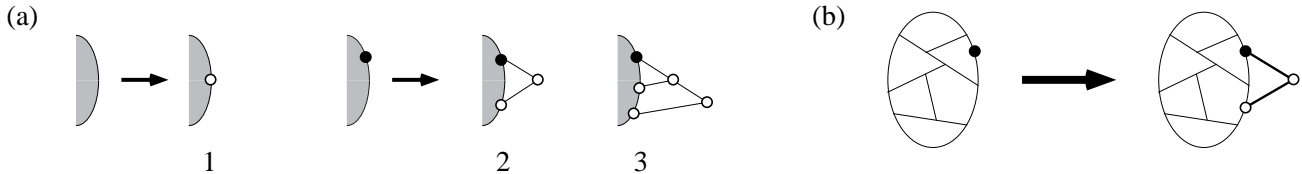} 
\end{center}
\caption{
(a) A portion of $G$ is shown together with  
the corresponding portion of $G^j$ resulting from construction~$j$.  
Solid dots denote vertices already
present in $G$, while open dots denote new vertices, present in $G^j$
but not in $G$.  Only those parts of $G$ which are changed are shown.  
(b)  An example of construction~2 is depicted.  
}
\label{fig-constructions}
\end{figure}

Our goal in this section 
is to provide a lemma which allows the infrared degree
of divergence of some of the Feynman diagrams appearing in this
paper to be accurately estimated in terms of the infrared degree of divergence
of related but simpler Feynman diagrams.   
This involves constructions in which a new Feynman diagram $G^j$ 
($j= 1, 2,3$) is
obtained from $G$ by the addition of new lines and/or vertices.
These constructions are illustrated in Figure~\ref{fig-constructions}.  
The following lemma gives bounds on $\underline{\deg}(G^{j})$ in 
terms of $\underline{\deg}(G)$.   In the lemma, either 
$\underline{\deg}_\mu(G)$ or $\underline{\deg}_0(G)$ can be used.

%%%%%FIGFIGFIGFIGFIGFIGFIGFIGFIGFIGFIGFIGFIGFIGFIGFIGFIGFIGFIGFIG
%%%%%FIGFIGFIGFIGFIGFIGFIGFIGFIGFIGFIGFIGFIGFIGFIGFIGFIGFIGFIGFIG
%%%%%FIGFIGFIGFIGFIGFIGFIGFIGFIGFIGFIGFIGFIGFIGFIGFIGFIGFIGFIGFIG
%%%%%FIGFIGFIGFIGFIGFIGFIGFIGFIGFIGFIGFIGFIGFIGFIGFIGFIGFIGFIGFIG

\begin{lemma}
\label{lem-Ginduction}
The infrared degree of divergence of $G^{j}$ is bounded in terms of
that of $G$ as follows:  
\eq 
\lbeq{Gind}
    	\underline{\deg}(G^{j})  \geq 
	\begin{cases}
		\underline{\deg}(G) -2  
			&	(j=1) \\
    		\underline{\deg}(G) + \min\{0 , d-6 \} 
		 	& 	(j=2) .
	\end{cases}
\en
\end{lemma}

Note that construction~3 results from two applications of construction~2,
and hence
\eq 
\lbeq{cor-Gind}
    	\underline{\deg}(G^{3})  \geq 
    		\underline{\deg}(G) + 2  \min\{0 , d-6 \}.
\en

\smallskip \noindent
{\bf Proof of Lemma~\ref{lem-Ginduction}.}
We begin with the observation that,
by definition,  $\degIR[\Hcal]$ depends only on the subspace spanned by
$\Hcal$.  Therefore, given $p' \nin \Hcal$ with $p'$ determined by $\Hcal$,
we can replace some vector $p \in \Hcal$ by $p'$ without changing 
$\degIR[\Hcal]$.  This fact was called {\em naturalness}\/ in \cite{Reis88a}. 

\begin{figure}
\begin{center}
\includegraphics[scale = 0.20]{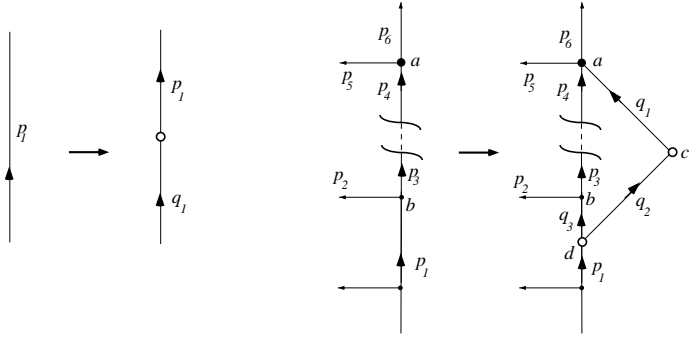} 
\end{center}
\caption{Constructions~1, 2 and labels of relevant momenta.
}
\label{fig-const-prf}
\end{figure}

Now we turn to the proof of \refeq{Gind} for Construction~1. 
In Construction~1, we add a new vertex to an existing line with 
momentum (say) $p_1$, thereby introducing a new line momentum $q_1$, as in
Figure~\ref{fig-const-prf}.  Let $V$ be the number of vertices of
$G$.  The momentum conservation
equations for $G$, which take the form $\sum_{j=1}^N\Lambda_{ij}p_j=0$
($i=1,\ldots,V-1$) for some integers $\Lambda_{ij}$,
are then supplemented with an additional equation $q_1=p_1$ for $G^1$.
Choose $\Hcal'$ such that the minimum in the definition of $\degIR[G^1]$ 
is attained at $\Hcal'$.  Either $q_1$ is determined by $\Hcal'$ or 
it is not.  If it is not, then $\Hcal'$ is a linearly independent subset
of the $p_j$ obeying $\sum_{j=1}^N\Lambda_{ij}p_j=0$, and
hence serves also as a subset $\Hcal=\Hcal'$
of possible loop momenta for $G$.  Therefore
\eq
	\degIR[G^1] = \degIR[\Hcal'] = \degIR[\Hcal] \geq \degIR[G] .
\en 
If, on the other hand, $q_1$ is determined by $\Hcal'$, then by naturalness, 
we may assume $p_1 \in \Hcal'$ and $q_1 \nin \Hcal'$.  
Let $\Hcal = \Hcal'$.
Then $\Hcal$ again
serves as a subset of possible loop momenta for $G$. 
Since $\Hcal$ determines one fewer line than $\Hcal'$ 
(due to the absence of $q_1$), we have 
\eq
	\degIR[G^1] = \degIR[\Hcal_1']  = \degIR[\Hcal] -2 \geq \degIR[G] -2.
\en 
This completes the proof of \refeq{Gind} for Construction~1.

Next, we turn to the proof of \refeq{Gind} for Construction~2. 
Momenta before and after Construction~2 are labelled as in 
Figure~\ref{fig-const-prf}.  Suppose $G$ has $V$ vertices.
The momentum conservation equations for $G$ can again be written as
\eq
\lbeq{Geqn0}
	\sum_{j=1}^N \Lambda_{ij} p_j = 0  \quad (i=1,\ldots,V-1).
\en 
For $G^2$, the corresponding equations can be written, for some integers
$C_i$, in the form
\eq
\lbeq{G2eq1}
	\sum_{j=1}^N \Lambda_{ij} p_j = C_{i} q_1 \quad (i=1,\ldots, V-1), 
	\qquad
	q_1 = q_2, \qquad q_3 = p_2 + p_3 .
\en 
The fact that the same coefficients $\Lambda_{ij}$ arise for $G$ and $G^2$
can easily be checked by comparing momentum conservation at vertices
$a,b$ for $G$ with that at $a,b,c,d$ for $G^2$.
Suppose the minimum in the definition of $\degIR[G^2]$ is attained 
by $\Hcal'$.  We again consider two cases, depending on whether 
$q_1$ is determined by $\Hcal'$ or not.

%%%%%FIGFIGFIGFIGFIGFIGFIGFIGFIGFIGFIGFIGFIGFIGFIGFIGFIGFIGFIGFIG
%%%%%FIGFIGFIGFIGFIGFIGFIGFIGFIGFIGFIGFIGFIGFIGFIGFIGFIGFIGFIGFIG
%%%%%FIGFIGFIGFIGFIGFIGFIGFIGFIGFIGFIGFIGFIGFIGFIGFIGFIGFIGFIGFIG
%%%%%FIGFIGFIGFIGFIGFIGFIGFIGFIGFIGFIGFIGFIGFIGFIGFIGFIGFIGFIGFIG

Suppose first that $q_1$ is determined by $\Hcal'$. 
By naturalness, we can assume $q_1 \in \Hcal'$. 
Also, we can assume $q_3 \nin \Hcal'$, 
because if $q_3 \in \Hcal'$ then $p_1$ is 
also determined (and would not be in $\Hcal'$ since
$p_1=q_1+q_3$), and thus by naturalness, we can take $p_1$ 
rather than $q_3$ as a member of $\Hcal'$. 
We therefore assume that $q_1 \in \Hcal'$ and $q_3 \nin \Hcal'$.  
Define $\Hcal = \Hcal' \backslash \{q_1\}$.  
Since $\Hcal'$ is a set of loop momenta corresponding to the momentum
constraints $\sum_{j=1}^N \Lambda_{ij} p_j = C_{i} q_1$, it follows
that $\Hcal$ is a set of loop momenta corresponding to 
$\sum_{j=1}^N \Lambda_{ij} p_j = 0$, and hence is a subset of loop momenta
for $G$.
Comparing \refeq{Geqn0} and 
\refeq{G2eq1} shows that the number of $G$'s line momenta $p_j$ 
determined by $\Hcal$ is the same as for $\Hcal'$. 
Since $\Hcal$ has one fewer momentum than $\Hcal'$, and since there are 
at most three more lines determined by $\Hcal'$ than by $\Hcal$
(namely $q_1,q_2,q_3$), we have
\eq
\lbeq{q1det}
	\degIR[G^2] = \degIR[\Hcal']  \geq \degIR[\Hcal] + d - 6 
	\geq \degIR[G] + d -6 .
\en 

It remains to consider the case where $q_1$ is not determined by $\Hcal'$.
In this case, at most one of the line momenta $p_1$ and $q_3$ is determined.  
However, the choice of labels $p_1$ and $q_3$ 
for these two lines was arbitrary,
and we can and do choose the labelling that guarantees that the line
momentum corresponding to $q_3$ is not determined.
Since none of $q_1, q_2, q_3$ is
determined, we have $q_1, q_2, q_3 \nin \Hcal'$.  
We now define $\Hcal = \Hcal'$. It is conceivable that $\Hcal$ may not be a
subset of a set of line momenta for $G$, since independence of line
momenta for $G^2$ (with its additional line momenta $q_i$)
does not necessarily imply independence for $G$.
To deal with this, we note that we can extend the definition \refeq{degIRHdef}
of $\degIR[\Hcal]$ also to dependent $\Hcal$, and still maintain 
\refeq{degIR.def1}, because a dependent $\Hcal$ will span a smaller space
than an independent $\Hcal$ consisting of the same number of vectors and
hence cannot give the minimum in \refeq{degIR.def1}. 
Thus we have 
\eq
\lbeq{q1notdet}
	\degIR[G^2] = 
	\underline{\rm deg}(\Hcal') 
	\geq 
	\underline{\rm deg}(\Hcal) 
	\geq 
	\degIR[G]
\en 
where the first inequality follows from the fact that $\Hcal$ may 
determine more momenta in $G$ than in $G^2$
(since $q_1$ provides an additional degree of freedom in \refeq{G2eq1}
compared to \refeq{Geqn0}), 
and the second inequality
makes use of the extended definition of $\degIR[\Hcal]$ described above.

Combining \refeq{q1det} and \refeq{q1notdet} completes the proof of 
\refeq{Gind} for Construction~2, and the lemma is proved.
\qed

%%%%%%%%%%%%%%%%%%%%%%%%%%%%%%%%%%%%%%%%%%%%%%%%%%%%%%%%%%%%%%%%%%%%%%%%%%%%%%%
%%%%%%%%%%%%%%%%%%%%%%%%%%%%%%%%%%%%%%%%%%%%%%%%%%%%%%%%%%%%%%%%%%%%%%%%%%%%%%%
%%%%%%%%%%%%%%%%%%%%%%%%%%%%%%%%%%%%%%%%%%%%%%%%%%%%%%%%%%%%%%%%%%%%%%%%%%%%%%%
%%%%%%%%%%%%%%%%%%%%%%%%%%%%%%%%%%%%%%%%%%%%%%%%%%%%%%%%%%%%%%%%%%%%%%%%%%%%%%%

\section*{Acknowledgements}
This work was supported in part by NSERC. 
It was carried out in part during extensive visits by both authors
to the University of British Columbia in 1997, and to
Microsoft Research and the Fields Institute in 1998.
The work of G.S.\ was also supported in part by an
Invitation Fellowship of the Japan Society for the Promotion of Science,
during a visit to Tokyo Institute of Technology in 1996.

%\bibliography{../bibdef/bib}    %For Gord
\bibliography{bib}             %For Takashi 
\bibliographystyle{plain}

\end{document}